\documentclass[12pt]{article}
\pdfoutput=1

\usepackage{float}

\usepackage[dvipsnames]{xcolor}
 \usepackage[a4paper,
   left=2.5cm, right=2.5cm,
   top= 3cm, bottom=4cm]{geometry}
\usepackage{amsmath,amssymb,amsfonts}

\usepackage{bbm,setspace} 
\usepackage{braket}
\usepackage{bm} 
\usepackage[toc,page]{appendix}

\usepackage{cite}
\usepackage{cancel}

\numberwithin{equation}{section}

\usepackage{makecell,multirow}
\usepackage{pgfplots}
\pgfplotsset{compat = newest}

\usepackage[linktocpage=true]{hyperref}
\hypersetup{
	colorlinks=true,
	citecolor=Maroon,
	linkcolor=Maroon,
	urlcolor=Blue,
	pdfauthor={},
	pdftitle={},
	pdfsubject={}
}

\newcommand{\bea}{\begin{eqnarray}}
\newcommand{\eea}{\end{eqnarray}}
\newcommand{\be}{\begin{equation}}
\newcommand{\ee}{\end{equation}}
\newcommand{\beq}{\begin{equation}}
\newcommand{\eeq}{\end{equation}}

\newcommand\os{{O(16)} \times {O(16)}}
\newcommand{\CE}{{\cal E}}
\newcommand\nn{\nonumber}
\newcommand{\nana}{\ket{\nabla}\bra{\nabla}}

\usetikzlibrary{arrows}
\pgfdeclarelayer{background}
\pgfsetlayers{background,main}
\usepackage{graphicx}
\usepackage{graphbox}

\def\AdS{{\rm AdS}}
\newcommand{\Z}{{\mathbb Z}}
\newcommand{\R}{{\mathbb R}}

\newcommand{\hn}{\hat{n}}

\newcommand{\C}[1]{$(\ref{#1})$}

\newcommand{\ul}{\underline}

\begin{document}

\begin{titlepage}

\vspace*{-2cm} 
\begin{flushright}
{\tt \phantom{xxx}  EFI-23-9} \qquad \qquad 
\end{flushright}

\vspace*{0.8cm} 
\begin{center}
{\LARGE Non-Supersymmetric Heterotic Strings on a Circle 
}\\

 \vspace*{1.5cm}
Bernardo Fraiman,$^{1}$ Mariana Gra\~na,$^2$  H\'ector Parra De Freitas$^2$ and Savdeep Sethi$^{3}$\\

 \vspace*{1.0cm} 
$^1$ {\it Theoretical Physics Department, CERN - 1211 Geneva 23 - Switzerland}\\[2mm]

$^2$ {\it Institut de Physique Th\'eorique, Universit\'e Paris Saclay, CEA, CNRS Orme des Merisiers, 91191 Gif-sur-Yvette CEDEX, France}\\[2mm]

$^3$ {\it Enrico Fermi Institute \& Kadanoff Center for Theoretical Physics \\ University of Chicago, Chicago, IL 60637, USA}\\

\vspace{1cm}
\small{\bf Abstract} \\[3mm]\end{center}

Motivated by a recent construction of non-supersymmetric $\AdS_3$, 
we revisit the $O(16) \times O(16)$ heterotic string compactified on a torus. The string one-loop potential energy has interesting dependence on the classical moduli; extrema of this potential include loci where the gauge symmetry is maximally enhanced. 
Focusing on the case of a circle, we use
lattice embeddings to find the maximal enhancement points together with their spectra of massless and tachyonic modes. We find an extended Dynkin diagram that encodes the global structure of the moduli space, as well as all symmetry enhancements and the
loci where they occur. 
We find $107$ points of maximal enhancement with $8$ that are free of tachyons. The tachyon-free points each have positive cosmological constant.
We determine the profile of the potential energy near each of these points and find that one is a maximum while three are saddle points. The remaining four live at the boundary of a tachyonic region in field space. In this way, we show that every point of maximal symmetry enhancement is unstable. We further find that the curvature of this stringy potential satisfies the de Sitter swampland conjecture. Finally, we discuss the implications for constructions of $\AdS_3$.

\end{titlepage}

\newpage

\renewcommand{\baselinestretch}{0.75}\normalsize
\tableofcontents
\renewcommand{\baselinestretch}{1.0}\normalsize

\newpage

\section{Introduction} \label{intro}

The absence of experimental signals of supersymmetry motivates a deeper exploration of models with no ultraviolet supersymmetry. 
The $\os$ heterotic string is one of a handful of ten-dimensional non-supersymmetric  theories of quantum gravity which can be analyzed perturbatively using worldsheet techniques and is free of tachyons at tree level~\cite{Dixon:1986iz, Alvarez-Gaume:1986ghj}. It is constructed using a different choice of GSO projection than its supersymmetric counterparts, which leads to mismatched bosonic and fermionic  degrees of freedom at each mass level. 
 At the massless level the projection keeps the graviton, B-field, dilaton and $SO(16)\times SO(16)$ gauge fields, while projecting out their supersymmetric partners.\footnote{For definiteness and compatibility with past literature we use $\os$ for the name of the non-supersymmetric heterotic string and $SO(16)\times SO(16)$ for the gauge symmetry. The actual gauge group is $\tfrac{Spin(16)\times Spin(16)
}{\mathbb{Z}_2}\rtimes \mathbb{Z}_2$. See Section~\ref{sss:lattice}.} There are also new massless fermionic states, including a bifundamental under the gauge group, which are not present in the supersymmetric theory. At the massless level there are more fermionic than bosonic degrees of freedom. However, this mismatch oscillates with increasing mass or level, realizing a very interesting structure called  `misaligned supersymmetry'~\cite{Dienes:1994np, Dienes:1994jt}. Many aspects of non-supersymmetric strings have been studied in recent work; see, for example,~\cite{Itoyama:2019yst,Itoyama:2020ifw,Basile:2020mpt, Basile:2020xwi, Faraggi:2020wld,Sagnotti:2021mxb,Cribiori:2021txm,Itoyama:2021itj,Acharya:2022shu,Koga:2022qch,Avalos:2023ldc,Angelantonj:2023egh,Raucci:2023xgx, BoyleSmith:2023xkd
} and references therein.

Besides the $\os$ theory, there are five more ten-dimensional non-supersymmetric heterotic theories with tachyons~\cite{Dixon:1986iz,SEIBERG1986272,KawaiLewellenTye}.
In the bosonic formulation of the heterotic string, all non-supersymmetric theories of rank 16 can be described as Scherk-Schwarz reductions  on the internal $T^{16}$ of the $E_8 \times E_8$ and $SO(32)$ theories~\cite{Scherk:1979zr,Ferrara:1987es}. As such, they preserve many of the properties of the parent supersymmetric theory at the classical level, like the local structure of the moduli space for the toroidally compactified theory.

Yet this breaking still gives rise to interesting phenomena like a non-trivial spacetime potential energy, which determines the spacetime cosmological constant at critical points of the potential. We will somewhat loosely refer to the spacetime potential as the cosmological constant on occasion. It is important to stress that this potential energy computed with a Minkowski spacetime is not even a continuous function of the moduli. As one enters a region with tachyons, it can jump discontinuously to minus infinity. There are more exotic pathologies of this type that we will encounter later.

As is typical for non-supersymmetric compactifications, Minkowski spacetime is not a stable solution because of the non-vanishing string one-loop vacuum energy, which serves as a potential for the dilaton. One can remedy this problem by introducing extra ingredients like fluxes which lead to $\AdS$ spacetimes with a stabilised dilaton but often with additional tachyons; see, for example,~\cite{Mourad:2016xbk,Basile:2020xwi, Raucci:2022bjw}. A recent construction uses the $\os$ heterotic string on $\AdS_3 \times S^3 \times S^3 \times S^1$~\cite{Baykara:2022cwj}. The dilaton is already stabilized at tree-level using fluxes and the string coupling can be made arbitrarily weak. The internal circle was chosen to be at the self-dual radius $R = \sqrt{\alpha'}$ with all Wilson lines turned off. This choice by itself extremizes the one-loop cosmological constant at a positive value~\cite{Ginsparg:1986wr}. In turn, the value of the cosmological constant for the $\AdS$ background is uplifted by this $1$-loop correction, although not enough to result in a de Sitter vacuum. 

In principle, there are many other critical points in the classical moduli space of $S^1$ compactifications which could be used in such constructions. This motivates us to tackle a basic string theory problem:  characterizing the classical moduli space of non-supersymmetric heterotic strings on $S^1$, and $T^d$ more generally, with a focus on extrema of the $1$-loop cosmological constant and their stability properties.

The main features of toroidal compactifications of non-supersymmetric heterotic strings have been understood for a long time \cite{Ginsparg:1986wr,Itoyama:1986ei} (see also \cite{Dienes:1990ij} for one of the original four-dimensional constructions in the free fermionic formulation). Similar to their supersymmetric cousins, they exhibit gauge symmetry enhancement at special points of the classical moduli space. In this case, however, such points have the added feature of extremizing the potential energy to all orders in string perturbation theory when the gauge enhancement is maximal.\footnote{By a point of maximal enhancement, we mean a point in moduli space that is fully fixed under reflections in the T-duality group. This will be discussed more fully in Section \ref{sec:explo}.} The task of finding extrema of the spacetime potential energy then includes finding points of maximal symmetry enhancement. There can also be points of non-maximal enhancement and even no enhancement that also extremize the potential energy. We will find examples of non-maximal critical points even on $S^1$. It  then remains to determine whether these critical points are minima, maxima or saddle points. This question has been addressed for a few examples in past literature~\cite{Ginsparg:1986wr,Koga:2022qch}, but so far no systematic study has been performed. 

As we will explain in the main text, the task of finding maximal symmetry enhancements on $T^d$ is formally the same for the supersymmetric and non-supersymmetric strings. For this reason, we can carry over techniques used for classifying maximal enhancements in the supersymmetric setting \cite{Fraiman:2018ebo,Font:2020rsk,Font:2021uyw,Fraiman:2021soq,Fraiman:2022aik}. In particular, for supersymmetric $S^1$ compactifications, there exists an extended Dynkin diagram (EDD), related  to the Coxeter-Dynkin diagrams that appear in mathematics, which encodes the global structure of the moduli space~\cite{Goddard-Olive,Cachazo:2000ey,Fraiman:2018ebo}.\footnote{This diagram also made a recent appearance in the study of 3D magnetic quivers \cite{DelZotto:2023myd,DelZotto:2023nrb}.} A main result of this paper is to construct an EDD for the non-supersymmetric case. From this diagram, we can easily obtain every non-Abelian symmetry enhancement and the values of the moduli where the enhancement occurs. We find that there are 107 points of maximal enhancements in total.  Studying the spectra of these enhancements we find that only 8 of them are free of tachyons.

We then compute the potential energy and its variation around these points of maximal enhancement, each of which is an extremum of the potential. We find that {\it none} of the 8 tachyon-free maximal enhancement points are minima.\footnote{This is different from the results of  \cite{Ginsparg:1986wr} where the enhancement to $SO(16)\times SO(16)\times SU(2)$ was thought to be a minimum of the potential. While this point is indeed a minimum with respect to changing the circle radius, we find that variations in Wilson line moduli decrease the value of the energy, in agreement with recent results in the literature~\cite{Koga:2022qch}; see Section \ref{maximal-saddle?}. The authors of \cite{Ginsparg:1986wr}, when contacted $36$ years after their work appeared, did not feel our conclusions are unreasonable. } Three of them are saddle points, one is a local maximum, while the other four, using the mountain-inspired language of \cite{Ginsparg:1986wr}, reside at `knife edges.' Namely, they lie at a boundary of a tachyonic region where the potential energy drops to minus infinity. The three saddle points are minima along the radial direction, but are unstable along the Wilson line directions. 
We also find that the minimum value of the Hessian of the one-loop potential at these points, divided by the value of the potential itself, is always smaller than approximately -0.64. At these very stringy critical points, which are far from the regime of validity of effective field theory, the refined de Sitter swampland conjecture appears to be satisfied.

As a byproduct, we also find four points which extremize the potential but \textit{do not} correspond to maximal symmetry enhancements, showing that maximal enhancement is a sufficient but not necessary condition for extrema. These four points are also unstable; each either a saddle point or a knife edge.  Again they are minima of the radial modulus but have instabilities along some of the Wilson line directions.

Although our focus is on $S^1$ compactifications, we set up and analyse in detail the case of generic $T^d$. We show, for example, how the massless and tachyonic spectra can be computed using lattice embedding techniques, and how these embeddings encode the topology of the gauge symmetry groups. We also determine the full T-duality group of the theory, extending the results of \cite{Itoyama:2021itj}; concretely we find that, as in the supersymmetric case, the group corresponds to the automorphism group of the charge lattice of the theory, with some dualities generated by charge vectors associated to tachyonic states. In this way we connect to recent results about tachyons and spin structures in non-supersymmetric heterotic worldsheets \cite{BoyleSmith:2023xkd}.

Lastly, we revisit the construction of $\AdS_3$ in light of our results about $S^1$. We find that any extremum of the potential energy on $\R^9 \times S^1$ can be used to build a perturbatively stable solution for some range of flux quantum numbers. However, there now appears to be a non-trivial quite fascinating constraint on the existence of intrinsically quantum $\AdS_3$ solutions with no electric $H_3$-flux threading the spacetime. These cases are particularly interesting as the likely endpoints of non-perturbative string nucleation processes that can discharge the electric flux. They are, perhaps, the best candidates for examples of non-supersymmetric AdS/CFT. Yet we see that their existence depends on the mass of the most tachyonic mode at a given critical point of the potential energy. This gives a more precise reason to know the exact value of the  constant that also appears in the refined de Sitter conjecture, where it is assumed to be of order one. 

The paper is organized as follows: 
to make our analysis self-contained, we start in Section \ref{sec:ferm} with a brief review of the supersymmetric $E_8\times E_8$ heterotic string. Its non-supersymmetric counterpart, the $\os$ heterotic string, is described in Section \ref{sec:os} initially using the fermionic formulation to showcase the difference in massless 
spectra. Passing to the bosonic formulation, we discuss 
torus compactifications  following \cite{Ginsparg:1986wr}, but with the introduction of further refinements based on lattice theory. In Section \ref{sec:explo} we construct the extended Dynkin diagram. This encodes all gauge groups, the loci in moduli space    where enhancements are realized, as well as the fundamental domain of the moduli space.   In Section \ref{s:cc} we compute the cosmological constant in the neighborhood of tachyon-free enhancements, showing that they are all unstable. In Section \ref{sec:AdS}, we examine the $\AdS_3$ construction. Finally in Section \ref{sec:discussion} we summarize our results.

\section{Supersymmetric heterotic strings on \texorpdfstring{$T^d$}{Td}} \label{sec:ferm}

Here we review  torus compactifications of the supersymmetric heterotic strings and their symmetry enhancements. The goal is to set the stage by introducing all the basic concepts that get generalized or modified in the non-supersymmetric case. For completeness we will start with a very brief review of the fermionic formulation of the $E_8 \times E_8$ heterotic string in ten dimensions, which can be later modified to give the non-supersymmetric $\os$ heterotic string, putting them on the same footing. We then move to the bosonic formulation, which is a more convenient  framework to study torus compactifications. 

\subsection{The \texorpdfstring{$E_8 \times E_8$}{E8 x E8} theory}

\subsubsection{Fermionic formulation}
In the fermionic formulation of the heterotic theory, the 16 extra left-moving bosons $X^I$ are replaced by $32$ left-moving worldsheet fermions $\lambda^{\tilde A}$, $\tilde A=1,...,32$. These fermions have the same mode expansion as the worldsheet fermions of the superstring, and their boundary conditions can also be periodic (Ramond) or anti-periodic (NS). In the $E_8 \times E_8$ heterotic theory that we will use as a starting point, these 32 fermions are split into two sets of $16$ and $16$, $\tilde A=(A,A')$, with possibly different boundary conditions for each set. One then has eight sectors, according to the two possible boundary conditions on the first 16, the last 16, or the 10 right-moving worldsheet fermions with a spacetime index. 

The GSO projection defining this supersymmetric theory leaves states that are even under the operators $(-1)^{F}$, $(-1)^{F'}$ and $(-1)^{\tilde F}$, where $F \, (F')$ and $\tilde F$ are the worldsheet fermion numbers for the first (last) 16 left-moving fermions and the right-moving fermions, respectively. This leaves the following eight sectors,\footnote{\label{foot:NSconvention}We use the conventions of \cite{Polchinski-book2}, where  the spacetime NS ground state is taken to have $\tilde F=1$, while the R ground states (either in the sixteen internal directions or in spacetime) have $(-1)^{F,F',{\tilde F}}=-1$, and the latter contain a chirality matrix (see (10.2.22) in \cite{Polchinski-book2}).}
\begin{equation} \label{sectorssusy}
	\begin{aligned}
		& \quad \   A  \quad  \quad \quad \ \ \ \, A'  \  \quad \quad  \quad \ \ \ \mu  \quad\quad \quad \\
		& \left\{  \begin{aligned} 
			&\text{NS}_+\\
			& R_-
		\end{aligned}   \right\}
		\otimes   \left\{  \begin{aligned} 
			&\text{NS}_+\\
			& R_-
		\end{aligned}   \right\}  \otimes  \left\{  \begin{aligned} 
			&\text{NS}_+\\
			& R_-
		\end{aligned}   \right\} \, , \\
	\end{aligned}
\end{equation}
where the sub-index $\pm$ denotes the required value of $(-1)^{F, F', \tilde{F}}$. 

Spacetime bosons belong to the spacetime NS sector i.e. the last column of \C{sectorssusy}, while fermions belong to the R sector. Since these two sectors appear on the same footing, the construction is spacetime supersymmetric. 

The zero point energies are $-1/2$ for every NS sector, $1/2$ for internal R sectors, and 0 for the spacetime R sector.    
Massless bosons, which are in the spacetime NS sector, can be constructed in two ways: either completely from the NS sector with $N+\tilde N=3/2$. The only possibility satisfying the level matching condition is then $N=1, \tilde N=1/2$. Or from one NS and one R sector with $N+\tilde N=1/2$ and level matching only allowing $N=0$, $\tilde N=1/2$. Thus we have the following states,
\begin{eqnarray} \label{massless10dsusy}
	\tilde \psi^{\mu}_{\tfrac12} \alpha^{\nu}_{-1} \ket{0,k}_{\rm NS-NS-NS} \ ,  & (1,1,35\oplus 28 \oplus 1) \nn \\
	\tilde \psi^{\mu}_{\tfrac12} \, \lambda^{A}_{\tfrac{1}{2}} \lambda^B_{\tfrac12} \ket{0,k}_{\rm NS-NS-NS} \ ,  & (120, 1, 8_v) \nn \\
	\tilde \psi^{\mu}_{\tfrac12} \, \lambda^{A'}_{\tfrac12} \lambda^{B'}_{\tfrac12} \ket{0,k}_{\rm NS-NS-NS} \ ,  & (1,120, 8_v)  \\
	\tilde \psi^{\mu}_{\tfrac12}  \ket{0,k}_{\rm R-NS-NS} \ , &(128, 1, 8_v) \nn \\
	\tilde \psi^{\mu}_{\tfrac12}  \ket{0,k}_{\rm NS-R-NS} \ , &(1,128, 8_v) \nn 
\end{eqnarray}
where we have listed the irreducible representations under $SO(16) \times SO(16) \times SO(8)$. The last factor is the little group in spacetime. 
The 128 representation appears because the R sector gives spinors of $Spin(16)$, which are chiral due to the GSO projection and thus have $2^{7}=128$ states. 
The first line corresponds to the {\it graviton, B-field} and {\it dilaton}, while the remaining lines are gauge fields. The second and third lines are $SO(16) \times SO(16)$ adjoint gauge fields, while the last two lines enhance this group, giving gauge fields in the adjoint of $E_8 \times E_8$. The spacetime R sector gives the superpartners of these fields. 

\subsubsection{Bosonic formulation and $T^d$ compactifications}\label{ss:susy}

In the bosonic formulation, the sixteen extra left-moving worldsheet bosons are not traded for fermions, but instead are compactified on a sixteen-dimensional  torus at a point of enhanced $E_8 \times E_8$ symmetry.  The oscillators along these directions are encoded in a 
sixteen-dimensional vector of this even self-dual lattice,
\beq
\pi^I \, , \,  I=1,...,16 \ , \quad \pi \in  \Gamma_8\oplus \Gamma_8\,.
\eeq
For compactifications on a torus, $T^d$, the momenta and winding numbers along $T^d$ combine with the internal torus to give a vector in the Narain lattice  
\beq
\Gamma_8\oplus \Gamma_8
\oplus \Gamma_{d,d}=\Gamma_{d+16,d}\,.
\eeq
The spectrum  is obtained from the mass formula and level matching condition, which take the form:\footnote{In the bosonic formulation, note that the left-moving R vacua in either of the 8 directions, which transform in the 128 spinor representation, are associated to vectors  $\pi$ with entries $\pm 1/2$ in all eight components, which have $\tfrac12 |\pi|^2=1$. On the other hand, in the fermionic formulation, the R vacua have no oscillations along these directions, thus giving no contribution to the mass. In order for the mass formulae to match, in the bosonic formulation one has to use $\hat a$, defined in Table \ref{ta:sectors}, for the left-moving sector. This is always 1 and therefore the zero point energy depends only on the right-moving sector to which the state belongs.}
\begin{equation} \label{mass}
	\frac14 m^2 = \frac12 P_L^2 + N - 1 = \frac12 p_R^2 + \tilde N - 
	\begin{cases}
		0 & \text{R sector}\\
		\frac12 & \text{NS sector}
	\end{cases}\,.
\end{equation}
Here NS and R refer to the right-moving sector,  
$(P_L,p_R)\equiv(p^I,p_L,p_R)$ are given by a vector in the Narain lattice polarized according to the values of the moduli $g_{ij},B_{ij}$ and the Wilson lines $A_i^I$ as follows:
\begin{subequations}
	\begin{align}
		p_{R}&= \frac1{\sqrt2} \hat e^{*i} \left[n_i-E_{ij}w^j-\pi \cdot A_i\right]\, , \label{pR} \\
		p_{L}&=\frac1{\sqrt2}\hat e^{*i}\left[n_i+\left(2g_{ij}-E_{ij}\right)w^j-\pi\cdot A_i\right]
		\, , \label{pL} \\[2mm]
		p^I &= \pi^I+A^I_iw^i\, , \label{pI}
	\end{align}\label{momentasusy}\end{subequations}
with $n_i$ and $w^i$  the integer momenta and winding numbers on the torus, $\hat e$ is the vielbein for $g^{-1}$ and $E$ is the following combination of metric, B-field and Wilson lines:
\beq \label{foot:E}
E_{ij}= g_{ij}+b_{ij}+\frac12A_i\cdot A_j \, .
\eeq
The massless bosons satisfying these conditions are given by
\begin{itemize}
    \item $N=1\, , \ \tilde N=\tfrac12\, , \ P_L=0 \, , \ p_R=0 \ \ \Rightarrow \ \ g_{\mu \nu} , \,B_{\mu \nu}  ,  \, g_{\mu i}, \, B_{\mu i} , \, A_\mu^I , \, g_{ij} , \, B_{ij} , \, A_i^I, \, \phi $
    \item $N=0\, , \ \tilde N=\tfrac12\, , \ P^2_L=2 \, , \ p_R=0 \ \ \Rightarrow \ \ A_{\mu}^\alpha $
\end{itemize}
where in the first line $g_{\mu i}\pm B_{\mu i}$ are the $U(1)_L^d \times U(1)_R^d$ gauge fields and $A_{\mu}^I$ the $U(1)^{16}_L$ gauge bosons, and in the second line $A_\mu^\alpha$ are extra massless gauge bosons that can appear at particular points in moduli space, as we discuss in detail later. Note that the massless condition on the right-moving sector can also be satisfied by $\tilde N=0$, $p_R^2=1$, but since the lattice of momenta is even, there are no vectors of length-squared one. This possibility will appear, however, in the non-supersymmetric case.

\subsection{Symmetry enhancement}

As mentioned, the $U(1)_L^{d+16} \times U(1)_R^d$ symmetry 
of the gauge fields in the first line is the generic symmetry in moduli space, arising from states with $n_i=w^i=0,\pi=0$.
The fact that the $E_8 \times E_8$ symmetry of the ten-dimensional theory is broken in compactifications at a generic point in moduli space is because of the Wilson lines: indeed, the condition $p_R=0$, for states with $w^i=0$ implies $\pi \cdot A_i=n_i \in {\mathbb Z}$, and thus for generic Wilson lines this is only satisfied for $\pi=0, \, n_i=0$. 

On the other hand, for Wilson lines $A_i \in \Gamma_8 \oplus \Gamma_8$ the condition $p_R=0$ can be satisfied at any $E_{ij}$ taking $n_i=\pi \cdot A_i$, $w^i=0$ with $\pi$ the roots of $E_8 \times E_8$, and one recovers the $E_8 \times E_8$ symmetry of the ten-dimensional theory along with an additional $U(1)_L^{d} \times U_R^d$. Additionally, at special values of 
$E_{ij}$ the $U(1)^d_L$ symmetry can also be enhanced by states with non-zero winding. For other (special) Wilson lines and values of $E_{ij}$, there can be vectors satisfying $P_L^2=2$ such that both $p_L$ and $p^I$ are non-zero, and thus the symmetry gets enhanced by states charged under both the torus and the 16 directions. Generically, the vectors with $ P_L^2 = 2$ (and $p_R=0$) then give the roots of the rank $16+d$ gauge algebra at that point (or surface) in moduli space. 

In terms of the Narain lattice, the condition $p_R = 0$ defines a primitive embedding of a rank $d+16$ positive definite lattice $W$ into $\Gamma_{d+16,d}$, provided the moduli are rational. The vectors in $W$ with $P_L^2 = 2$, to which we refer as roots, give massless states in the adjoint representation of the gauge algebra of the corresponding vacuum. In other words, the root sublattice of $W$ is exactly the root lattice of the physical gauge algebra. The overlattice of this root sublattice, which for maximal enhancements is simply $W$, is then the weight lattice of the gauge \textit{group} of the model.

In the case $d = 1$ every symmetry enhancement can be obtained by means of an extended Dynkin diagram (EDD) which encodes the global structure of the moduli space\cite{Goddard-Olive,Cachazo:2000ey}. We briefly review its construction in Section \ref{sss:eddsusy}, which we will in fact be able to generalize to the non-supersymmetric theory. For $d \geq 2$, however, such a diagram does not exist; one can use instead an algorithm  that explores the moduli space by ``jumping" along points of maximal enhancement \cite{Font:2020rsk}. This algorithm  has been successfully applied in $d = 2,3,4$, both in $T^d$ compactifications and asymmetric orbifolds of tori \cite{Font:2021uyw,Fraiman:2021soq,Fraiman:2021hma}. In the special case $d = 4$ every possible symmetry enhancement can be obtained from Niemeier lattices in a way similar to how one obtains symmetry enhancement from the EDD  \cite{Fraiman:2022aik}; Niemeier lattices similarly encode \textit{discrete} symmetry enhancements in this special case \cite{Gaberdiel:2011fg,Cheng:2016org}. In this paper we focus our attention primarily on the $d = 1$ case and the EDD, leaving the generalization of the other methods  to the non-supersymmetric theory for future work.

\section{Non-supersymmetric heterotic strings}
\label{sec:os}
We now generalize the discussion above to the non-supersymmetric heterotic string. We will also discuss the problem of classifying the ten-dimensional theories, the topology of the gauge groups and a relation between the presence of tachyons and chirality of the matter spectrum. 

\subsection{The \texorpdfstring{$\os$}{O(16) x O(16)} theory}
\subsubsection{Fermionic formulation}
The non-supersymmetric $\os$ theory is obtained in a similar manner to the $E_8\times E_8$ theory by imposing a GSO projection given instead by the operators
\beq \label{projections}
(-1)^{F+F'+\tilde F} \ , \quad (-1)^{F+\alpha'+\tilde \alpha} \ , \quad  (-1)^{F'+\alpha+\tilde \alpha} \ , \quad (-1)^{\tilde F+\alpha+\alpha'} \,.
\eeq
Here $F (F')$ and $\tilde F$ are, respectively, the worldsheet fermion numbers in the first (last) 16 left-movers  and the right-moving sector. For the R sector 
$\alpha =1 $  while  for the NS sector $\alpha=0$, and we use the same notation as for the $F$. Note that there is no independent projection $(-1)^{\tilde F}$, hence supersymmetry is broken; this will be clearly seen from the resulting spectrum.   

The eight sectors that survive the projections are given in Table \ref{ta:sectors}, where $a, \tilde a$ are minus the zero-point energies on the left and right-moving directions, and we have given the $SO(16) \times SO(16) \times SO(9,1)$ representations that appear in each sector\footnote{In the column labelled by $\mu$ we only give the right-moving part of the $SO(1,9)$, the left-moving part can be 0 or v in every sector}. In our conventions $s$ ($c$) is a spinor of negative
(positive) chirality. The first three sectors (counting 3a and 3b as a single sector) appear in the supersymmetric heterotic string (see eq. \eqref{sectorssusy}), while the last three, involving a R$_+$ sector and/or at least one NS$_-$ sector are new sectors which in the ordinary heterotic string are projected out by the GSO projection. On the other hand, sectors of the heterotic string like, for example, NS$_+$ NS$_+$ R$_-$, are projected out. Since the gravitino belongs to this sector, one sees again that supersymmetry is broken in this construction.  
\begin{table}
	\begin{center}
		\begin{tabular}{|c|ccc|ccc|ccc|} \hline
			$\#$ & A & A' & $\mu$ & A & A' & $\mu$ & $a$ & $\tilde a$  & $\hat a$ \\ \hline
			1 & NS$_+$ & NS$_+$ & NS$_+$ & 0 & 0 &    v  & 1 & 1/2 & 1  \\
			2 & R$_-$ & R$_-$ & NS$_+$ & s & s &    v & -1 & 1/2 & 1  \\
			3a & NS$_+$ & R$_-$ & R$_-$ & 0 & s &    s & 0 & 0 & 1 \\
			3b & R$_-$ & NS$_+$ & R$_-$ & s & 0 &    s & 0 & 0 & 1 \\
			4 & R$_+$ & R$_+$ & R$_+$ & c & c &    c & -1 & 0  & 1 \\
			5a & NS$_-$ & R$_+$ & NS$_-$ & v & c &     0 & 0 & 1/2 & 1  \\
			5b & R$_+$ & NS$_-$ & NS$_-$ & c & v &     0 & 0 & 1/2 & 1  \\
			6 & NS$_-$ & NS$_-$ & R$_+$ & v & v &     c & 1 & 0 & 1  \\ \hline
		\end{tabular}
		\caption{{\small States surviving the projections leading to the non-supersymmetric $\os$ theory. In the $\mu$ column we have given only the representation on the right-moving sector (the left-moving part can be in the scalar or vector representation).  The $a$ and $\tilde a$ are minus the zero point energy in the left and right-moving sectors respectively. We have introduced $\hat a\equiv a+\alpha+\alpha'$ for later convenience. }}
		\label{ta:sectors}
	\end{center}
\end{table}
We work out the ten-dimensional spectrum up to the massless level. Since by level matching condition the mass on the left-moving side has to be the same as that of the right-moving side, and the mass-squared has a term with $-a$ and $-\tilde a$, respectively, tachyons can only appear when both $(a, \tilde{a})$ are greater than zero. This happens only in the  NS$_+$  NS$_+$  NS$_+$ sector, but $\tilde N=0$, which is the only possibility that would give rise to a tachyon, is projected out (it is in the spacetime NS$_-$ sector, see footnote \ref{foot:NSconvention}). As a result, there are no tachyons in the spectrum.

Massless states can only arise when both $a$ and $\tilde a$ are greater than or equal to zero, which happens in four of the eight sectors.  Let us start with the NS$_+$ NS$_+$  NS$_+$ sector. The $+$ projections select states that have an even number of left-moving excitations along $A$ or $A'$ directions, and an odd number of right-moving excitations along spacetime. Since $a=1$, $\tilde a=1/2$, massless states  then have $N=1$, $\tilde N=1/2$, and are of the  form
\begin{eqnarray}
	\tilde \psi^{\mu}_{\tfrac12} \alpha^{\nu}_{-1} \ket{0,k}_{\rm NS-NS-NS} \, ,  & (1,1,35\oplus 28 \oplus 1) \, , \nn \\
	\tilde \psi^{\mu}_{\tfrac12} \, \lambda^{A}_{\tfrac{1}{2}} \lambda^B_{\tfrac12} \ket{0,k}_{\rm NS-NS-NS} \, ,  & (120, 1, 8_v) \, , \\
	\tilde \psi^{\mu}_{\tfrac12} \, \lambda^{A'}_{\tfrac12} \lambda^{B'}_{\tfrac12} \ket{0,k}_{\rm NS-NS-NS} \, ,  & (1,120, 8_v) \, . \nn
\end{eqnarray}
These states also appear in the supersymmetric theory (c.f. the first three lines of \eqref{massless10dsusy}), and correspond to the graviton, B-field and dilaton. The second and third line correspond to gauge fields in the adjoint of $SO(16) \times SO(16)$. 

The 3a and 3b sectors are spacetime fermions with positive (negative) chirality for an odd (even) number of right-moving excitations with integer oscillation number.  In the 2a sector, with NS$_+$ R$_-$  R$_-$, one has an even number of left-moving excitations with half-integer oscillation number along the first sixteen components, while the last 16 are spinors of $SO(16)$, and have an odd (even) number of excitations with integer oscillation number for positive (negative) chirality. In the 2b sector, the first 16 and the last ones are interchanged. Massless states have $N=\tilde N=0$, and transform under $SO(16) \times SO(16) \times SO(8)$ as
\beq
3a: \ \ (1,128_s,8_s) \ , \quad 3b:  \ \ (128_s,1,8_s) \ .
\eeq
In the supersymmetric theory, these states are superpartners to the gauge bosons in the fourth and fifth lines of \eqref{massless10dsusy}, which are projected out in this theory. Here instead they play the role of matter states transforming in the fundamental representation of each gauge group factor, giving a straightforward example of how supersymmetry breaking enriches the physics of the theory.

In the 5a and 5b sectors, there are no massless states because these states should have $N=0$ and $\tilde N=1/2$, but these are projected out of the spectrum by the NS$_-$ projections which require an odd number of left-moving excitations, incompatible with the former condition, and an even number of right-moving excitations, incompatible with the latter.  

Finally in sector 6 the massless states have $N=1, \tilde N=0$ and require an odd number of left-moving excitations in each of the 8 directions. They are therefore of the form
\beq
6: \ \ \lambda^A_{-\tfrac12} \lambda^{A'}_{-\tfrac12} \ket{0,k}_{R} \ , \quad (16,16,8_c) \ .
\eeq
These fermions transform in the bi-fundamental of $SO(16) \times SO(16)$ and do not appear in the supersymmetric heterotic theory. 

To recap, the massless bosonic states are the metric, the B-field, the dilaton and the gauge fields in the adjoint of $SO(16) \times SO(16)$. The massless fermionic states comprise two types of matter fields: those transforming in the spinor representation of one $SO(16)$ factor (and trivially under the other) and those transforming in the bi-fundamental representation of the full gauge group. 
\subsubsection{Bosonic formulation: orbifold construction}
\label{sec:comp}

As we saw in Section \ref{sec:os}, the non-supersymmetric $\os$ heterotic theory keeps some of the sectors of its supersymmetric counterpart, those numbered 1, 2 and 3 in Table \ref{ta:sectors}, while  sectors 4, 5 and 6 are new twisted sectors. The zero point energies $-\hat a$ and $-\tilde a$ only depend on the right-moving sector, both for the untwisted and the twisted sectors. The mass formula and level matching conditions are thus the same as in the supersymmetric version 
\beq \label{mass10d}
\frac14 m^2 = \frac12 \pi^2 + N - 1 = \tilde N - 
	\begin{cases}
		0 & \text{R sector}\\
		\frac12 & \text{NS sector}
	\end{cases}\, ,
\eeq
but now the momenta $\pi$ belong to a different lattice, as we now describe in detail.

In the following discussion, we give a heuristic description of the orbifold construction of the ten dimensional theory, which is easily extended to $T^d$ compactifications. We restrict to the facts relevant for computing the massless and tachyonic spectrum for different values of the moduli, and refer to the original work \cite{Ginsparg:1986wr} for more details. 

To break the $E_8 \times E_8$ gauge symmetry of the parent theory to $SO(16)\times SO(16)$, one can use a shift vector of the form
\begin{equation}
	\delta = (0^7,1;0^7,1) \,,
\end{equation}
which is of order two in the gauge lattice, $2\delta \in \Gamma_8 \oplus \Gamma_8$. This would leave only states with NS-NS or R-R boundary conditions in the $A,A'$ directions, i.e those in sectors 1 and 2 of Table \ref{ta:sectors}, together with their supersymmetric counterparts in the spacetime R sector. The latter should, however, be projected out from the spectrum.
On the other hand, the states with mixed NS-R and R-NS boundary conditions (again referring to the $A,A'$ directions) have $\pi \cdot \delta={\mathbb Z}+\tfrac12$, and in the supersymmetric compactification would be projected out, while here they appear in the spacetime fermions of sector 3. 
These two features are realized if we quotient the $E_8 \times E_8$ theory by, 
\beq
\beta = (-1)^{2\,   \pi \cdot \delta } (-1)^{\tilde \alpha} \, ,
\eeq
where $\tilde \alpha$, defined below eq. \eqref{projections}, is actually the spacetime fermion number. Projecting onto states that are even under this operation  leaves only sectors 
1, 2 and 3 as desired.

To describe the twisted sectors (4, 5 and 6), which involve R$_+$ and NS$_-$ instead of the R$_-$ and NS$_+$ appearing in the untwisted sectors, one can also use the vector $\delta$. For this, note that the elements of the original $\Gamma_8$ lattice are either in the NS$_+$ or R$_-$ sectors, since for the former the sum 
of its entries give an even number (thus they belong the conjugacy class 0, and for the latter the number of plus signs is even (in our conventions, this is the class s).
The NS$_-$ vectors on the other hand are in the class v, while the R$_+$ in the class c. These two types of vectors are precisely  of the form  $\pi+(0^7,1)$, with $\pi \in \Gamma_8$. The twisted sector thus contains vectors in $\Gamma_8 \oplus \Gamma_8 +\delta$.  Supersymmetry is also broken in this sector: there is no symmetry between the spacetime bosons, which have R$_+$-R$_+$ or NS$_-$-NS$_-$, and the fermions, which have mixed boundary conditions. In order to realize this feature, we again invoke  the spacetime fermion number.  The surviving states are those which are {\it odd} under $\beta$. 

The spectrum of the theory is thus separated into four conjugacy classes, which are conveniently described in terms of the subsets 
\begin{equation}
	\begin{split}
		\Gamma^+ &= \{ \pi \in \Gamma_8\oplus \Gamma_8 | \pi \cdot \delta \in \mathbb{Z} \}\,,\\
		\Gamma^- &= \{ \pi \in \Gamma_8\oplus \Gamma_8 | \pi \cdot \delta \in \mathbb{Z} + \tfrac 12 \}\,.\\
	\end{split}
\end{equation}
These distinguish between the NS-NS and R-R states (in $\Gamma^+$) and the NS-R and R-NS states that are in $\Gamma^-$. 
The untwisted sector where the (ten-dimensional) massless gauge bosons and spacetime fermions  then belong to the following conjugacy classes
\begin{equation}\label{untwisted}
\begin{split}
{\rm Sectors \, 1 \, and \, 2} : \ &(\Gamma^+;v)\,, \\ 
{\rm Sectors \, 3a \, and \, 3b}: \ & (\Gamma^-;s)\, ,
\end{split}
\end{equation}
where v and s denote their spacetime class. 
In the twisted sector the two conjugacy classes are
\begin{equation}\label{twisted}
\begin{split}
{\rm Sectors \, 5a \, and \, 5b}: \ &(\Gamma^- + \delta; 0)\, , \\ 
{\rm Sectors \, 4 \, and \, 6} : \ & (\Gamma^+ + \delta; c)\, .
\end{split}
\end{equation}
For convenience we let us rename
\begin{equation}\label{conjdef}
	\Gamma_v \equiv \Gamma^+\,, ~~~~~ \Gamma_s \equiv \Gamma^-\,, ~~~~~ \Gamma_c \equiv \Gamma^+ + \delta\,, ~~~~~ \Gamma_0 \equiv \Gamma^- + \delta\,.
\end{equation}

\subsubsection{Charge lattice and gauge group topology}\label{sss:lattice}
The charge lattice of the theory is the union of the sets corresponding to each sector as mentioned before. Let us denote it by
\begin{equation}\label{union}
	\Upsilon_{16} \equiv \Gamma_v \cup \Gamma_s \cup \Gamma_c \cup \Gamma_0\,.
\end{equation} 
It is not difficult to show that, in fact, this lattice is simply the dual of the vector conjugacy class lattice,
\begin{equation}\label{dual}
	\Upsilon_{16} = \Gamma_v^*\,;
\end{equation}
more details are provided in Section \ref{sec:others}. 

In the supersymmetric setting the heterotic string excitations produce the electrically charged particles in the spectrum, hence the perturbative electric charge lattice is complete. On the other hand, the allowed representations of the gauge group under which the particle content can transform is given by the gauge group's topology, and completeness of the spacetime spectrum implies that all of these representations  are realized in the lattice. It follows that the charge lattice encodes the topology of the gauge group, i.e. its fundamental group. For example, the $Spin(32)/\mathbb{Z}_2$ string has fundamental group $\mathbb{Z}_2$ in accord with the fact that there are only $Spin(32)$ spinors in the spectrum and no vectors or co-spinors, as encoded in the charge lattice $D_{16}^+$. 

Let us assume that, similarly, the $\os$ heterotic string excitations produce the full spectrum of electrically charged particles in the theory. The full gauge group $G$ can then be encoded in its so-called weight lattice, which is the dual of the lattice containing all the allowed representations on top of the adjoint of $G$. This latter lattice is nothing more than $\Upsilon_{16}$, and so from \eqref{dual} we see that $\Gamma_v$ is the weight lattice of $G$. This lattice comprises the root lattice of $D_8\oplus D_8$ together with the $(s,s)$ element  $\mu = (\tfrac12^{16})$ which satisfies $2\mu \in D_8\oplus D_8$, hence the fundamental group of $G$ is $\mathbb{Z}_2$. On the other hand there is an outer automorphism of the gauge group exchanging the $Spin(16)$ factors, just as in the $E_8\times E_8 \rtimes \mathbb{Z}_2$ heterotic string. The upshot is that the full gauge group takes the form
\begin{equation}
	G = \frac{Spin(16)\times Spin(16)}{\mathbb{Z}_2} \rtimes \mathbb{Z}_2 \,.
\end{equation}

\subsection{Other ten dimensional theories, tachyons and chirality}
\label{sec:others}
Apart from the $\os$ heterotic string, there are six other heterotic strings without supersymmetry in ten dimensions \cite{Dixon:1986iz,SEIBERG1986272,KawaiLewellenTye}. Although these theories are tachyonic, five of them are relevant to us because they become T-dual to the $\os$ string upon circle compactification. These have rank 16 charge lattices which we will denote $\Upsilon_{16}^{(p)}$, $p = 1,...,5$, with $p$ indicating the number of tachyons $2^p$; for the tachyon-free theory we will write $\Upsilon_{16}^{(0)} \equiv \Upsilon_{16}$, i.e. $p = 0$. These lattices are related to the gauge algebras as follows:
\begin{equation}
\label{10non-susy}
	\begin{split}
	\Upsilon_{16}^{(0)} &\sim \mathfrak{so}_{16}\oplus \mathfrak{so}_{16}\,, ~~~~ \Upsilon_{16}^{(1)} \sim \mathfrak{su}_{16}\oplus \mathfrak{u}_1\,, ~~~~ \Upsilon_{16}^{(2)} \sim 2(\mathfrak{e}_{7}\oplus \mathfrak{su}_2)\,,\\
	\Upsilon_{16}^{(3)} &\sim \mathfrak{so}_{8}\oplus \mathfrak{so}_{24}\,, ~~~~~\Upsilon_{16}^{(4)} \sim \mathfrak{e}_{8}\oplus \mathfrak{so}_{16}\,, ~~~~~ \Upsilon_{16}^{(5)} \sim \mathfrak{so}_{32}\,.
	\end{split}
\end{equation}
Other contributions to the lattice specify, in particular, the matter content and tachyons in the spectrum.

The spectrum of each of these theories is similarly divided into four sectors according to how the states transform under the spacetime symmetry group, and the charge lattice $\Upsilon_{16}^{(p)}$ is the union of the four conjugacy classes (cf. eq. \eqref{union}). The dual of this charge lattice is, as before, the lattice corresponding to the vector conjugacy class:
\begin{equation}\label{dualgen}
	\Upsilon_{16}^{(p)*} \simeq \Gamma_{v}^{(p)}\,.
\end{equation}
This is particularly easy to prove for $\Upsilon_{16}^{(4)}$; the theory is defined using a shift $\delta = (0^8)\times(0^7,1)$, hence $\Gamma_v^{(4)} \simeq E_8 \oplus D_8$, while the remaining three conjugacy classes are in one-to-one correspondence with the spinor, co-spinor and vector representations of $Spin(16)$ in such a way that $\Upsilon_{16}^{(4)} \simeq E_8 \oplus D_8^*$. The other cases can be similarly worked out. Heuristically, every $\Gamma_v^{(p)}$ has determinant 4, hence its discriminant group $\Gamma_v^{(p)*}/\Gamma_v^{(p)}$ has four elements each of which is associated to a spacetime conjugacy class and so we can identify $\Gamma_v^{(p)*} \simeq \Upsilon_{16}^{(p)}$.

\subsubsection{ Interpolation (T-duality) from lattice isomorphisms}

Without entering into the details of circle compactifications, we state here an important  consequence of extending the charge lattice by adding the hyperbolic lattice encoding the corresponding KK momentum and winding numbers on the circle: 
\begin{equation}\label{extension}
	\Upsilon_{16}^{(p)} \to \Gamma_{1,1} \oplus \Upsilon_{16}^{(p)} \,.
\end{equation} 
One can then check that the following lattice isomorphisms hold:
\begin{equation}\label{nonsusyiso}
	\Gamma_{1,1}\oplus\Upsilon_{16}^{(p)} \simeq \Gamma_{1,1}\oplus\Upsilon_{16}^{(q)}\,, ~~~~~ \forall ~p,q \,.
\end{equation}
In other words, the charge lattice is the same for the circle compactification of any of these theories, and we will denote this lattice by $\Upsilon_{17,1}$. This situation is exactly the same as the case of supersymmetric heterotic strings, where the isomorphism can be derived trivially by using uniqueness results for even self-dual lattices. We see that the six theories are T-dual on the circle, as already worked out in \cite{Ginsparg:1986wr}. 

What is perhaps more interesting is that the lattices $\Upsilon_{16}^{(p)}$ form a lattice genus. This statement can be understood better in terms of their duals $\Gamma_v^{(p)}$. For even lattices, a genus turns out to comprise exactly those lattices which become isomorphic upon extension by a $\Gamma_{1,1}$ factor just as above \cite{NikulinVV1980ISBF}; that these belong to the same genus is therefore trivial as seen by taking the dual of both sides in \eqref{nonsusyiso}, and it follows that the charge lattices themselves are also in the same genus. It is non-trivial however that the genus is complete, which can be checked by using the Smith-Minkowski-Siegel mass formula in the same way one classifies even self-dual lattices \cite{conway2013sphere}. This can be interpreted as a kind of classification of T-dual non-supersymmetric heterotic strings, or more accurately a classification of certain decompactification limits in the classical moduli space under consideration.

More generally, decompactification limits are characterized by orthogonal decompositions of the charge lattice $\Upsilon_{17,1}$ into a possibly scaled $\Gamma_{1,1}$ sublattice and another rank 16 sublattice. Each such decomposition assigns a real positive modulus to the $\Gamma_{1,1}$ part which is equivalent to the compactification radius from the point of view of the limiting ten dimensional theory; the rank 16 part is then associated to the heterotic gauge bundle. 

There are only two other possible orthogonal decompositions of $\Upsilon_{17,1}$ aside from those of \eqref{nonsusyiso}. Namely,
\begin{equation}\label{isosup}
	\Upsilon_{17,1} \simeq \Gamma_{1,1}(\tfrac12) \oplus E_8 \oplus E_8 \simeq \Gamma_{1,1}(\tfrac12)\oplus D_{16}^+\,,
\end{equation}
where $D_{16}^+$ is the weight lattice of $Spin(32)/\mathbb{Z}_2$. We recognize the two decompactification limits to the supersymmetric heterotic strings in ten dimensions. In total there are eight decompactification limits, matching the known classification of heterotic strings \cite{Dixon:1986iz,SEIBERG1986272,KawaiLewellenTye}. In particular one can consider curves in the moduli space which go from one such limit to another; these are known as ``interpolating models" in the literature \cite{Itoyama:1986ei,Blum:1997gw}. 

We note that given these lattice isomorphisms one can define a change of basis
\begin{equation}
	(n,w;\pi) \mapsto (n',w';\pi') \, ,
\end{equation}
with the first two integers describing the hyperbolic lattice factor and $\pi$, $\pi'$ describing the remaining rank 16 part of the charge lattice. This change of basis induces a transformation in the moduli, i.e. the compactification radius and the Wilson line, that maps from one compactified theory to another. This procedure is described in detail for the supersymmetric case in \cite{Font:2020rsk} and is straightforward to extend to the case at hand once the basis changes are known explicitly.

\subsubsection{Symmetries of the charge lattice and tachyons}\label{sss:latsym}

The automorphism group of the charge lattice in ten dimensions is a group of global symmetries of the heterotic worldsheet theory, hence gauge symmetries of the spacetime physics. For the $E_8\times E_8\rtimes \mathbb{Z}_2$ string, the inner automorphisms comprise the Weyl group of $E_8\times E_8$ while the outer automorphism exchanges the $E_8$ factors. For the $Spin(32)/\mathbb{Z}_2$ string, there are only inner automorphisms corresponding to the Weyl group of $Spin(32)$; the lack of $Spin(32)$ co-spinors means that the outer automorphism conjugating the spinors is not present. 

Except for the exchange of $E_8$'s, both automorphism groups are generated by reflections. Recall that a reflection $\sigma_v$ of a lattice $\Gamma$ is an involution defined by a vector $v \in \Gamma$ acting on every other vector $w \in \Gamma$ as
\begin{equation}
v \mapsto \sigma_v (w) = w - 2\frac{v\cdot w}{v^2}v\,.
\end{equation}
For self-dual lattices it turns out that only vectors with norm $v^2 = 2$, i.e. roots, can generate reflections, and in this case it is the set of simple roots which generates the full Weyl group. For more general lattices, such as the charge lattices $\Upsilon_{16}^{(p)}$, there can be in principle more possibilities for which we have to account.

We will start by considering the charge lattice $\Upsilon_{16}^{(4)} \simeq E_8 \oplus D_8^*$. A naive approach could consist in noting that this lattice is an extension of $E_8\oplus E_8$ by the co-spinor and vector classes in $D_8 \subset E_8$ associated to twisted states. This suggests that the automorphism group of the lattice contains, in particular, the Weyl reflections of $E_8 \times E_8$, but this is not the case; reflections generated by the spinor class, e.g.
\begin{equation}
	v = (\tfrac12,...,\tfrac12) \in E_8\,,
\end{equation}
act on an twisted state vector class element $w = (0,...,0,1)$ by
\begin{equation}
\label{refltachyon}
	\sigma_v(w) = w - \frac{v}{v^2} = \left( -\tfrac14,...,-\tfrac14,\tfrac34 \right) \notin D_8^*\,.
\end{equation}
In other words, the spinor class elements of $D_8$ are reflective in $E_8$ but \textit{not} in $D_8^*$. The upshot is that the automorphism group of $E_8\oplus E_8$ gets reduced when passing to $E_8\oplus D_8^*$ to that of the Weyl group of $E_8\times Spin(16)$. This is not the whole story, however, the vector class elements in fact turn out to be reflective in $D_8^*$. For an arbitrary vector $w \in D_8^*$ and $v = (0,...,0,1)$, we have
\begin{equation}
	\sigma_v(w) = w - 2 (v\cdot w) v\,,
\end{equation}
and the product $(v\cdot w)$ is always in $\tfrac12 \mathbb{Z}$; hence $\sigma_v(w) \in D_8^*$. Note that this reflection acts by changing the sign of the last component of $w$, which indeed is a symmetry of $D_8^*$ because \textit{both} spinor and co-spinor classes are present.\footnote{This reflection in $\Upsilon_{16}^{(4)} \simeq\Gamma_v^*$ is also an automorphism of the lattice $\Gamma_v$ since the automorphism group of a lattice and its dual are the same. In this latter case the automorphism is not a reflection since there are no norm 1 vectors in $\Gamma_v$ to generate it; it is instead an outer automorphism exchanging the two spinorial nodes in the Dynkin diagram of $D_8$.} Note that using instead the basis of the $\os$ string, the reflection is a bit more complicated because it has to exchange, for example, elements in the $(0,s)$ representation of $SO(16)\times SO(16)$ with elements in the $(v,v)$, which are associated to spacetime spinors and co-spinors, respectively.

Observe that the Weyl group of $E_8$ in general mixes the elements of the underlying $D_8$ with its spinor conjugacy class. This mixing is precisely due to the reflections generated by the latter, and that they are projected out in the non-supersymmetric theory is consistent with keeping spacetime vectors and spinors independent. In general, the reflection group of the underlying $E_8\oplus D_8 \subset \Upsilon_{16}^{(4)}$ will not mix any of the four conjugacy classes with another. On the other hand, the extra reflections generated by the vector class \textit{exchange} the spinor and co-spinor classes, as shown above, while acting trivially on the adjoint and vector classes. The presence of these reflective vectors is synonymous with the spectrum being \textit{non-chiral}. In turn, it is precisely these vectors which, because of their length, give rise to tree-level tachyons (see eq. \eqref{mass10d}). Hence we see that
\begin{equation}\boxed{
	\text{Matter is non-chiral} ~~~\Leftrightarrow~~~ \text{Spectrum has tachyons}}\,.
\end{equation}
This whole analysis generalizes to the other non-supersymmetric theories. Indeed, the $\os$ string is tachyon-free and its spectrum is chiral.\footnote{The same conclusion can be reached from the point of view of the string worldsheet \cite{BoyleSmith:2023xkd}, which also applies to the $E_8$ string. In this language, spacetime tachyons correspond to worldsheet free Majorana-Weyl fermions, which in turn imply invariance under changes in the spin structure induced by stacking Arf theories.}

The fact that certain Weyl reflections are projected out naturally by extending the charge lattice is also in accord with projecting out the corresponding gauge bosons. In other words, knowledge of the charge lattice itself and its reflection group is enough to determine the gauge algebra of the theory, without any reference to its microscopic construction. 

We conclude this discussion by noting that because of \eqref{dualgen}, the lattice $\Gamma_v^{(p)}$ corresponds in general to the weight lattice of the gauge group. Denoting the corresponding gauge groups by $G_p$, we see that
\begin{equation}
\begin{split}
	G_0 &= \frac{Spin(16)\times Spin(16)}{\mathbb{Z}_2} \rtimes \mathbb{Z}_2\,,  ~~~~~ ~~~~~ G_1 = \frac{SU(16)}{\mathbb{Z}_2}\times U(1) \rtimes \mathbb{Z}_2\,,\\\\
	G_2 &= \frac{(E_7 \times SU(2))^2}{\mathbb{Z}_2} \rtimes \mathbb{Z}_2\,, ~~~~~~~~~~~~~~~~~~ G_3 = \frac{Spin(8)\times Spin(24)}{\mathbb{Z}_2}\,,\\\\
	G_4 &= E_8 \times Spin(16)\,, ~~~~~~~~~~~~~~~~~~~~~~~~~~ G_5 = Spin(32)\,.
\end{split}
\end{equation}
For $G_1$, the outer automorphism acts by conjugation of the $SU(16)$ and $U(1)$ representations. The tachyons transform with unit charge with respect to $U(1)$, hence the fundamental group of the full gauge group is only a subgroup of the center of $SU(16)$, as presented. 

\subsection{\texorpdfstring{$T^d$}{Td} compactifications}

We now move to torus compactifications of the $\os$ string which, as discussed above, is equivalent to those of the other five non-supersymmetric heterotic strings. 

 The mass formula and level-matching conditions are the same as in the toroidal compactification of the supersymmetric heterotic string, given in \eqref{mass}. The momenta are also defined as in the supersymmetric setting  (eqs. \eqref{pR},\eqref{pL} and \eqref{pI}):
\begin{subequations}
	\begin{align}
		p_{R}&= \frac1{\sqrt2} \hat e^{*i} \left[n_i-E_{ij}w^j-\pi \cdot A_i\right]\, ,  \\
		p_{L}&=\frac1{\sqrt2}\hat e^{*i}\left[n_i+\left(2g_{ij}-E_{ij}\right)w^j-\pi\cdot A_i\right]
		\, , \\[2mm]
		p^I &= \pi^I+A^I_iw^i\, , 
	\end{align}
	\label{momenta}
\end{subequations}
where $E$ is defined in \eqref{foot:E} and now $\pi \in \Upsilon_{16}$. One defines
\begin{equation} \label{ketZ}
    \ket{Z}=\ket{w^i,n_i,\pi}
\end{equation}
with inner product given by 
\begin{equation} \label{innproduct}
    \braket{Z|\tilde Z} = n \tilde w + \tilde n w + \pi \cdot \tilde \pi \, .
\end{equation}
Note that
\beq
\braket{Z|Z}= P_L^2-p_R^2= 2 \Bigg[\tilde N-N+ \begin{cases}
		1 & \text{R sector}\\
		\frac12 & \text{NS sector}
	\end{cases} \Bigg]  \, , 
\eeq
where as before $P_L=(p_L,p^I)$.

To determine the lattice of the vectors $Z$ one can apply the procedure used in ten dimensions to the torus compactifications of the $E_8\times E_8$ string by restricting the action of the orbifold to the $E_8\oplus E_8$ part of the Narain lattice
\begin{equation}
	\Gamma_{d+16,d} = \Gamma_{d,d}\oplus E_8 \oplus E_8\,.
\end{equation}
The shift vector is now taken as $\delta = (0^{2d};0^7,1;0^7,1)$, and its action on $\Gamma_{d+16,d}$ defines the subsets
\begin{equation}
\begin{split}
\Gamma^+ &= \{ v \in \Gamma_{d+16,d} | v \cdot \delta \in \mathbb{Z} \}\,,\\
\Gamma^- &= \{ v \in \Gamma_{d+16,d} | v \cdot \delta \in \mathbb{Z} + \tfrac 12 \}\,,\\
\end{split}
\end{equation}
and the four sectors according to spacetime conjugacy class (cf. eqs. \eqref{untwisted} to \eqref{conjdef}). The subtlety now is that torus compactifications are non-chiral, and so the labels in $\Gamma_s$ and $\Gamma_c$ are now unphysical. We will keep this notation for clarity, however.
 
In this procedure, the sublattice $\Gamma_{d,d}$ of the Narain lattice is simply a spectator, and so each of the conjugacy classes is exactly as for the ten dimensional theory up to an orthogonal sum with $\Gamma_{d,d}$. The union of these classes gives, as before, the charge lattice of the theory, and so we see explicitly how the extension \eqref{extension} can be deduced. We write the charge lattice as
\begin{equation}
	\Upsilon_{d+16,d} \simeq \Gamma_{d,d}\oplus \Upsilon_{16}\,,
\end{equation}
with $\Upsilon_{16}$ the charge lattice of the ten dimensional theory. Alternatively we could choose $\delta$ such that $2\delta \in \Gamma_{1,1}$, e.g. with half an unit of winding and zero momentum along the associated circle, which corresponds to turning on a spacetime fermion number holonomy along this circle. This is nothing more than the Scherk-Schwarz mechanism and leads explicitly to the presentation
\begin{equation}
	\Upsilon_{d+16,d} \simeq \Gamma_{d-1,d-1}\oplus \Gamma_{1,1}(\tfrac12) \oplus E_8 \oplus E_8\,,
\end{equation}
in accord with the lattice isomorphism \eqref{isosup}.

 We will now show some properties of this lattice, while in section \ref{sec:enhancements} we discuss the massless states.

\subsubsection{T-duality}\label{ss:tduality}
As discussed in Section \ref{sss:latsym}, automorphisms of the charge lattice in ten dimensions are gauge symmetries of the theory. Upon torus compactification, the charge lattice gets promoted from an Euclidean lattice to a Lorentzian one, i.e. with time-like as well as space-like directions. This in turn promotes its automorphism group to one of \textit{dualities}. In the supersymmetric setting it is customary to refer to this automorphism group as the T-duality group, extending the usual notion of exchanging winding and KK momentum appearing in the bosonic string to operations involving the quantum numbers in the gauge lattice. In the non-supersymmetric theory we proceed in a similar manner, but further taking into account the subtleties involving mixing of conjugacy classes discussed above.

The way in which the T-duality group of the supersymmetric heterotic string changes on breaking supersymmetry was investigated at length in \cite{Itoyama:2021itj}. This change was determined to be a reduction to a congruence subgroup, which in practice consists of those elements of the original T-duality group which do not mix spacetime conjugacy classes. To illustrate this consider the case of Wilson line shifts in nine dimensions, which act on the moduli and quantum numbers as follows,
\begin{equation}
    \begin{split}
        A &\to A + Q\,, ~~~~~~~~~~~~~~~~~~~~~~~~ R \to R\,, \\\\
        n &\to n - \tfrac12 Q^2 w - Q \cdot \pi\,, ~~~~~~~~~~ w \to w\,, ~~~~~~~~~~ \pi \to \pi + Q\,,
    \end{split}
\end{equation}
with $Q$ an element in the gauge lattice $E_8 \oplus E_8$. A translation by a $D_8 \subset E_8$ spinor $Q = (\tfrac12^8)$ would mix elements in the classes $\Gamma_v$ and $\Gamma_s$, hence is projected out in the congruence subgroup. We must restrict to $Q \in \Upsilon_{16}^{(p)}$, e.g. $Q \in E_8 \oplus D_8$.

Similarly to the ten-dimensional case with $p = 4$, we find here that the presence of vector class elements in $D_8^*$ takes care of projecting out the undesired transformations. Indeed, for e.g. $Q = (0^8)\times (1,0^7)$, we have that $n \to n - \tfrac12 w - \pi_9$ which is generically in $\tfrac12 \mathbb{Z}$ and so the shift is not an automorphism of $\Upsilon_{17}$. But again by projecting out such transformations we do not get the full automorphism group. We should extend the congruence subgroup by including the transformation which exchanges the classes $\Gamma_s$ and $\Gamma_c$, as they are physically indistinguishable because of the lack of spacetime chirality in nine dimensions. In the basis with $p = 4$ used above, it consists in changing the sign of \textit{one} of the components of the Wilson line sitting in $D_8$, whereas in the original supersymmetric theory one is allowed to change the sign of a pair of components at a time.

The moral of the story is that the full duality group of the theory \textit{is} the automorphism group $O(\Upsilon_{17,1})$ and so the classical moduli space, now for arbitrary $d$, is of the familiar form
\begin{equation}
    \mathcal{M}_\text{non-susy}  \simeq \left[O(\Upsilon_{d+16,d})\backslash O(d+16,d)/O(d+16)\times O(d)\right] \times \mathbb{R}^+\,,
\end{equation}
with $\mathbb{R}^+$ the dilaton factor. Interestingly, since the automorphism group of a lattice and its dual are the same, one can describe the global structure of the moduli space in terms of $O(\Gamma_v)$ instead; the spacetime vector class already contains this global information.

\subsection{Symmetry enhancement}
\label{sec:enhancements}

We are interested in the appearance of extra massless states  at special points in moduli space. 
The mass formula and level matching conditions are the same as for compactifications of the $E_8 \times E_8$ theory, given in \eqref{mass}, where now $(n_i,w^i,\pi) \in \Upsilon_{d+16,d}$.

The massless bosons satisfy any one of the following conditions:
\begin{equation}\label{masslessBosons}
\begin{split}
    &\bullet N=1\, , \ \tilde N=\tfrac12\, , \ P_L=0 \, , \ p_R=0 \ \ \Rightarrow P =0 \, ,\\ 
    &\bullet N=0\, , \ \tilde N=\tfrac12\, , \ P^2_L=2 \, , \ p_R=0 \ \ \Rightarrow P \in \Gamma_v\, ,\\
    &\bullet N=0\, , \ \tilde N=0 \, , \ P_L^2=2 \, , \ p_R^2=1 \ \ \Rightarrow P \in \Gamma_0\, ,\\
    &\bullet N=1\, , \ \tilde N=0 \, , \ P_L=0 \, , \ p_R^2=1 \ \ \Rightarrow P \in \Gamma_0 \, .
\end{split}
\end{equation}
The last two lines are states that are not present in the supersymmetric theory. The third line gives scalars which are discussed in detail in the next section, while the last line gives right-moving vectors, to be discussed below. 

At generic points in the moduli space, i.e. in the bulk of the fundamental domain, the massless bosonic spectrum of the theory are given by states satisfying the conditions in the first line. These particles are the graviton, B-field, dilaton as well as $16+d$ gauge vectors and $d$ graviphotons. This is nothing but the bosonic content of the supersymmetric theory. Generically the gauge symmetry group is $U(1)^{16+d}\times U(1)^d$, while at special regions corresponding to finite distance boundaries in the fundamental domain the $U(1)^{d+16}$ factor undergoes enhancement to a non-Abelian group. 

Symmetry enhancement in the left-moving sector occurs when, as in the supersymmetric string, there are massless states of the type in the second line. The discussion in Section \ref{ss:susy} then carries over if we exchange the Narain lattice by the vector class lattice $\Gamma_v$. Massless gauge bosons have $p_R = 0$ and $P_L^2 = 2$ and define a root lattice whose overlattice $W$ primitively embedded into $\Gamma_v$ is the weight lattice of the gauge group $G$. To prove this last statement simply note that since $\Upsilon_{d+16,d}$ is the dual of $\Gamma_v$ and $W$ is primitively embedded into the latter, $\Upsilon_{d+16,d}$ has by construction every vector which is integrally paired with $W$ and so its projection onto the real span of $W$ yields the dual lattice $W^*$.\footnote{Alternatively one can use the general framework for computing the fundamental group of a gauge group when the charge lattice of the theory is not self-dual \cite{Cvetic:2021sjm}.
We look for embeddings of the \textit{co-root} lattice into the \textit{dual charge lattice} where the primitively embedded overlattice corresponds to the weight lattice of the gauge group. Since every root system here is simply-laced and the dual of the charge lattice is $\Gamma_v$ the result above follows.} 

At this point it is worth making a curious observation. The lattice $\Gamma_v$ always admits a primitive embedding into the Narain lattice corresponding to the supersymmetric theory with one more compact direction, 
\begin{equation}
    \Gamma_v \simeq \Gamma_{d,d}\oplus E_8 \oplus D_8 \hookrightarrow \Gamma_{d,d}\oplus \Gamma_{1,1} \oplus E_8 \oplus E_8\,,
\end{equation}
with orthogonal complement $\Gamma_{1,1}(2)$; to see this just embed $D_8$ primitively into $\Gamma_{1,1}\oplus E_8$. This implies that every weight lattice $W$ considered above admits a primitive embedding into $\Gamma_{d+16,d}\oplus \Gamma_{1,1}$ and so its associated gauge group is realized in the supersymmetric theory with one more compact dimension. This implies that for $d = 2$ every non-simply-connected gauge group $G$ satisfies the constraint derived in \cite{Cvetic:2020kuw} for $\mathcal{N}=1$ theories in eight dimensions to be free of 1-form center anomalies since every weight lattice embedded primitively into the Narain lattice for any $d$ trivially satisfies it \cite{Font:2021uyw}. Since the analysis of this reference is independent of the theory being supersymmetric, this result serves as an interesting verification extending the scope of the expected properties of charge lattices. 

We come now to the states in the last line of \eqref{masslessBosons}, which are absent in the supersymmetric theory. They give rise to extra right-moving vectors, enhancing the $U(1)^d$ part. The enhanced gauge algebra is given by the root sublattice of the orthogonal complement $T = W^\perp$ in $\Upsilon_{d+16,d}$, where roots have norm $-1$, and as such it can only involve $\mathfrak{su}_2$ factors. To see this, note that these roots live in the sublattice $\Gamma_v\cup \Gamma_0 \simeq \Gamma_{d,d}\oplus E_8 \oplus \mathbb{Z}^8$, which is manifestly integral. Hence the product of two norm $-1$ vectors cannot be $-1/2$ as required to form an $\mathfrak{su}_3$. Moreover, as suggested by the length of the roots, the associated worldsheet current algebras have level 2. 

\subsection{Matter content and tachyons}
\label{ss:content}

At special points in the moduli space the spectrum of the theory acquires {\it massless fermions} as well. The mass formula and level matching condition \eqref{mass} in the R sector gives massless fermions for\footnote{Another solution is $N=1\, ,  \tilde N=0\, , P_L^2=0 \, , p_R=0$, corresponding to the gravitino, but it is projected out of the spectrum; otherwise it would be in a  sector of the form NS$_+$ NS$_+$ R.} 
\begin{equation}
    \bullet N=0\, , \ \tilde N=0\, , \ P^2_L=2 \, , \ p_R=0 \, .
\end{equation}
These are spin 1/2 fermions corresponding to elements in the conjugacy classes $\Gamma_s$ and $\Gamma_c$ with the same $P_L^2,\, p_R$ as the spacetime vectors. Their charge vectors are  not reflective, however, and so their appearance is not tied to being at a boundary of the moduli space. On the other hand they do appear generically accompanying symmetry enhancements, and this is the situation we will be interested in.

The first thing to note is that the union of any of the spinor conjugacy classes with the vector class produces an even self-dual lattice. Indeed the union $\Gamma_v\cup \Gamma_s$ is nothing but the Narain lattice of the parent supersymmetric theory. Having at hand a primitive embedding $W \hookrightarrow \Gamma_v$ it may then admit an overlattice inside e.g. $\Gamma_v \cup \Gamma_s$ with extra elements satisfying $p_R = 0$; if some of them in turn satisfy $P_L^2 = 2$, they give rise to massless spinors. This is a generalization of the ten-dimensional situation where, for example, each $D_8$ has overlattice $E_8$ in the $\os$ heterotic string charge lattice. The situation is similar for the $\Gamma_c$ class, where an overlattice of $W$ in $\Gamma_v \cup \Gamma_c$ can lead to having massless co-spinors.

There are {\it massless scalars} accompanying any massless vector, since the states with $\tilde N=\tfrac12$ in the second line of \eqref{masslessBosons} are allowed to have this oscillation along the external space, giving rise to vectors, as well as on the internal space, giving rise to scalars. These scalars, which also appear in the supersymmetric theory, transform in the adjoint of the gauge group, and are massive in the vicinity of the points of enhancement.  
Similarly, the states in the last line have their (right-moving) scalar partners. 
On the other hand, for the scalars with $N=\tilde N=0$ the situation is subtler. On top of those becoming massless at special points (third line of \eqref{masslessBosons}), there are scalars with
\begin{equation}
\bullet N=0\, , \ \tilde N=0 \, , \ P_L^2=1 \, , \ p_R^2=0 \, .
\end{equation}
These are {\it tachyonic} with $m^2 = -2$. We can compute the overlattice of $W$ in $\Gamma_v\cup \Gamma_0$, which is an odd self-dual lattice, and for each element with norm 1 we have a tachyon. The subtlety lies in considering other tachyonic states with $p^2_R > 0$ up to $p_R^2 = 1$ at which point they become massless scalars. Such states are not encoded in overlattices of $W$, but as we explain shortly they are still amenable to computation using basic lattice embedding techniques.

The general situation in $\Gamma_0$ is that any vector in $\Upsilon_{d+16,d}$ with 
\begin{equation} \label{tachyons}
    0 \leq p_R^2 \leq 1\,, ~~~~~ P_L^2 = p_R^2 + 1\,, 
\end{equation}
gives a tachyonic or massless scalar with 
\begin{equation}
    m^2 = -2(1-p_R^2)\,,
\end{equation}
and to obtain them we proceed as follows: as already mentioned, the union $\Gamma_v \cup \Gamma_0$ is an odd self-dual lattice. Most importantly, however, it is integral. Every element $u$ in $\Gamma_0$ is integrally paired with those in $\Gamma_v$, and in particular with those in $W$ and its orthogonal complement $T \equiv W^\perp$ in $\Gamma_v$. This means that the projections $u_W$ and $u_T$ on these sublattices are respectively integrally paired with $W$ and $T$, hence they belong to their duals $W^*$ and $T^*$. It follows that every possible $u = (u_W,u_T)$ can be obtained by considering every possible pair of elements in these dual lattices and checking afterwards that the result indeed lies in $\Gamma_0$ i.e. that their components are well quantized. Since $u_W^2 = P_L^2$, we can filter at the outset the elements in $W^*$ by asking that $1 \leq u_W^2 \leq 2$, and the result is the full scalar conjugacy class spectrum up to massless level.

Among the tachyons that can appear for different enhancements in the theory, those with $m^2 = -2$, i.e. $p_R = 0$, play a special role. As discussed in Section \ref{ss:tduality}, their charge vectors are reflective, and these reflections become symmetries of the theory at the loci of the moduli space where $p_R = 0$ is satisfied. It follows that whenever there is a tachyon of this type in the spectrum, the content of the spinor and co-spinor conjugacy classes is exactly the same; this generalizes the situation in ten dimensions where tachyonic theories are non-chiral. We
will see in Section \ref{sec:maximal} in which representation of the gauge group these tachyons transform. Note that there can be tachyons with $p_R \neq 0$ in the compactified theory, which \textit{do not} imply this symmetry.

\section{Classical moduli space and symmetry enhancements in \texorpdfstring{$d = 1$}{d=1}}
\label{sec:explo}

We now specialize to circle compactifications, which is the main subject of this paper. Our interest is in obtaining the full list of maximal symmetry enhancements,\footnote{By maximal symmetry enhancement we mean rank 17 gauge groups without $U(1)$ factors (enhanced gauge symmetry completely fixed by reflections in the duality group). For example, $SU(18)$, ignoring the graviphoton, or enhancements with one $U(1)$ factor associated to a pair of tachyons such as the $SU(16)\times U(1)$ gauge symmetry in ten dimensions. In this definition we do not consider enhancements of the right-moving sector because these enhancements do not constrain the moduli in the same way and can be considered accidental.} which as shown in \cite{Ginsparg:1986wr} extremize the cosmological constant. Such enhancements can be obtained very easily in the supersymmetric case by means of an extended Dynkin diagram, which encodes the codimension one walls of a fundamental domain of the moduli space and how these walls intersect; equivalently it encodes the generators of the reflexive part of the T-duality group. This prompts the question of whether such a diagram exists in the non-supersymmetric case we are considering.

As we will see, such a diagram does exist, allowing us to find the symmetry enhancements in a manifestly exhaustive way. After constructing it and extracting this data we turn to characterizing each maximal enhancement in the rest of the section.

\subsection{Extended Dynkin diagram}

\subsubsection{Diagram for the supersymmetric theory}\label{sss:eddsusy}

In the supersymmetric case the extended Dynkin diagram (EDD) can be constructed by adding the affine node to both $E_8$ factors (0), (0'), giving them a momentum or winding charge with respect to the circle direction, and adding an extra node (\texttt{c}) connecting the two affine nodes:
\begin{eqnarray}
\begin{aligned}
\begin{tikzpicture}[scale = 1.2]
\draw(5.5,1);
\draw(0,0)--(7,0);
\draw(0.5,0)--(0.5,1);
\draw(6.5,0)--(6.5,1);
\draw[fill=white](0,0) circle (0.1)node[below=0.05in]{\scriptsize{1}};
\draw[fill=white](0.5,0) circle (0.1)node[below=0.05in]{\scriptsize{2}};
\draw[fill=white](1,0) circle (0.1)node[below=0.05in]{\scriptsize{3}};
\draw[fill=white](1.5,0) circle (0.1)node[below=0.05in]{\scriptsize{4}};
\draw[fill=white](2,0) circle (0.1)node[below=0.05in]{\scriptsize{5}};
\draw[fill=white](2.5,0) circle (0.1)node[below=0.05in]{\scriptsize{6}};
\draw[fill=Aquamarine](3,0) circle (0.1)node[below=0.05in]{\scriptsize{0}};
\draw[fill=white](0.5,0.5) circle (0.1)node[left=0.05in]{\scriptsize{7}};
\draw[fill=white](0.5,1) circle (0.1)node[left=0.05in]{\scriptsize{8}};
\draw[fill=Aquamarine](3.5,0) circle (0.1)node[below=0.05in]{\scriptsize{\texttt{c}}};
\draw[fill=Aquamarine](4,0) circle (0.1)node[below=0.05in]{\scriptsize{0'}};
\draw[fill=white](4.5,0) circle (0.1)node[below=0.05in]{\scriptsize{6'}};
\draw[fill=white](5,0) circle (0.1)node[below=0.05in]{\scriptsize{5'}};
\draw[fill=white](5.5,0) circle (0.1)node[below=0.05in]{\scriptsize{4'}};
\draw[fill=white](6,0) circle (0.1)node[below=0.05in]{\scriptsize{3'}};
\draw[fill=white](6.5,0) circle (0.1)node[below=0.05in]{\scriptsize{2'}};
\draw[fill=white](7,0) circle (0.1)node[below=0.05in]{\scriptsize{1'}};
\draw[fill=white](6.5,0.5) circle (0.1)node[right=0.05in]{\scriptsize{7'}};
\draw[fill=white](6.5,1) circle (0.1)node[right=0.05in]{\scriptsize{8'}};
\end{tikzpicture}
\end{aligned}
\end{eqnarray}
where we colored the nodes that have momentum and/or winding charge.

Alternatively, one can affinize the Dynkin diagram of $D_{16}$ and add two nodes (\texttt{c}) and (\texttt{w}) which respectively account for the extra root obtained from the $\Gamma_{1,1}$ lattice and the spinor which extends $D_{16}$ to the charge lattice $D_{16}^+$: 
\begin{eqnarray}
\begin{aligned}
\begin{tikzpicture}[scale = 1.2]
\draw(5.5,1);
\draw(0,0)--(7,0);
\draw(0.5,0)--(0.5,1);
\draw(6.5,0)--(6.5,1);
\draw[fill=white](0,0) circle (0.1)node[below=0.05in]{\scriptsize{1}};
\draw[fill=white](0.5,0) circle (0.1)node[below=0.05in]{\scriptsize{2}};
\draw[fill=white](1,0) circle (0.1)node[below=0.05in]{\scriptsize{3}};
\draw[fill=white](1.5,0) circle (0.1)node[below=0.05in]{\scriptsize{4}};
\draw[fill=white](2,0) circle (0.1)node[below=0.05in]{\scriptsize{5}};
\draw[fill=white](2.5,0) circle (0.1)node[below=0.05in]{\scriptsize{6}};
\draw[fill=white](3,0) circle (0.1)node[below=0.05in]{\scriptsize{7}};
\draw[fill=white](0.5,0.5) circle (0.1)node[left=0.05in]{\scriptsize{16}};
\draw[fill=Aquamarine](0.5,1) circle (0.1)node[left=0.05in]{\scriptsize{\texttt{w}}};
\draw[fill=white](3.5,0) circle (0.1)node[below=0.05in]{\scriptsize{8}};
\draw[fill=white](4,0) circle (0.1)node[below=0.05in]{\scriptsize{9}};
\draw[fill=white](4.5,0) circle (0.1)node[below=0.05in]{\scriptsize{10}};
\draw[fill=white](5,0) circle (0.1)node[below=0.05in]{\scriptsize{11}};
\draw[fill=white](5.5,0) circle (0.1)node[below=0.05in]{\scriptsize{12}};
\draw[fill=white](6,0) circle (0.1)node[below=0.05in]{\scriptsize{13}};
\draw[fill=white](6.5,0) circle (0.1)node[below=0.05in]{\scriptsize{14}};
\draw[fill=white](7,0) circle (0.1)node[below=0.05in]{\scriptsize{15}};
\draw[fill=Aquamarine](6.5,0.5) circle (0.1)node[right=0.05in]{\scriptsize{0}};
\draw[fill=Aquamarine](6.5,1) circle (0.1)node[right=0.05in]{\scriptsize{\texttt{c}}};
\end{tikzpicture}
\end{aligned}
\end{eqnarray}
As one can see, both procedures lead to the same diagram. This diagram was first considered in the context of circle compactifications of heterotic strings in \cite{Ginsparg:1986bx}, but had made appearances before in other contexts \cite{Goddard-Olive,Vinberg}. 

From the extended Dynkin diagram one can obtain  all the maximal enhancements in the theory by deleting two nodes such that what remains is of ADE type \cite{Cachazo:2000ey}. Each node gives a co-dimension one surface in moduli space, and the intersection of seventeen of them defines the point where the maximal enhancement occurs. More generally, all enhancements of rank $k$, and the $17-k$-dimensional surfaces in moduli space where they occur, can be obtained by deleting $19-k$ nodes. Extensions of these ideas to compactifications to lower dimensions were explored in \cite{Font:2020rsk}.

\subsubsection{Diagram for the non-supersymmetric theory}

We now apply the procedure of affinization and extension outlined above to get the extended Dynkin diagram for the non-supersymmetric theory. In principle we can take as a starting point any of the six non-supersymmetric theories, since they are dual upon circle compactification. However, the absence of tachyons in the $\os$ string makes the process considerably clearer. We will proceed in analogy with the supersymmetric case by adding nodes to the affine diagrams of both $SO(16) \sim D_8$ gauge factors, accounting for the compactification circle and the spinor-spinor element in the gauge lattice $[2\, D_8]^+$.  

We work in the basis of $\Gamma_v \simeq \Upsilon_{17,1}^*$ where charge vectors are written as 
\begin{equation}
    \ket{Z} = \ket{w,n;\pi}\,, ~~~~~ w,n\in \mathbb{Z}\,, ~~~ \pi \in [2\,D_8]^+\,,
\end{equation}
with inner product given by \eqref{innproduct}. The simple roots and lowest root of the $D_8$ sublattices are embedded into $\Gamma_v$ as follows
\begin{equation}\label{nodes1}
    \begin{split}
        \varphi_i &= \ket{0,0;\alpha_i,0^8}\,, ~~~~~~~ \varphi_i' = \ket{0,0;0^8,\alpha_i}\,, ~~~~~~ i = 1,...,8\,,\\\\
        \varphi_0 &= \ket{0,1;\alpha_0,0^8}\,, ~~~~~ \varphi_0' = \ket{0,1;0^8,\alpha_0}\,,
    \end{split}
\end{equation}
where 
\begin{equation}
    \begin{split}
        \alpha_1 = (1,-1,0^6)\,, ~~~...\,, ~~~\alpha_7 = (0^6,1,-1)\,,~~~ \alpha_8 = (-1,-1,0^6)\,, ~~~ \alpha_0 = (0^6,1,1)\, .
    \end{split}
\end{equation}
Note that the nodes corresponding to the lowest roots have momentum charge on the circle, making them linearly independent from the simple roots. 

We extend the root system by adding two more nodes. The first one corresponds to the circle direction and reads
\begin{equation}\label{nodec}
    \varphi_{\texttt{c}} = \ket{-1,-1;0^8,0^8}\,.
\end{equation}
The second one is constructed in analogy with the $SO(32)$ heterotic string: we take the spinor-spinor element $(\tfrac12^{16})$ and add winding $w = -1$ and momentum $n = 1$ such that its norm goes from 4 to 2,
\begin{equation}\label{nodew}
\varphi_{\texttt{w}} = \ket{-1,1;\tfrac12^8,\tfrac12^8}\,.
\end{equation}
The way in which the extra nodes are linked to the affine $2\,D_8$ turns out to be very simple, leading to a very symmetric extended diagram:
\begin{eqnarray}\label{diag1}
\begin{aligned}
\begin{tikzpicture}[scale = 1.2]
\draw(-3,1)node{};
\draw(5.5,1);
\draw(0.5,0)--(2.5,0)--(2.5,2)--(0.5,2)--(0.5,0);
\draw(0.5,0)--(0,-0.5);
\draw(2.5,0)--(3,-0.5);
\draw(0.5,2)--(0,2.5);
\draw(2.5,2)--(3,2.5);
\draw[fill=white](0.5,0) circle (0.1)node[below=0.1in]{$\varphi_2$};
\draw[fill=white](1,0) circle (0.1)node[below=0.1in]{$\varphi_3$};
\draw[fill=white](1.5,0) circle (0.1)node[below=0.1in]{$\varphi_4$};
\draw[fill=white](2,0) circle (0.1)node[below=0.1in]{$\varphi_5$};
\draw[fill=white](2.5,0) circle (0.1)node[below=0.1in]{$\varphi_6$};

\draw[fill=white](0.5,2) circle (0.1)node[above=0.1in]{$\varphi_2'$};
\draw[fill=white](1,2) circle (0.1)node[above=0.1in]{$\varphi_3'$};
\draw[fill=white](1.5,2) circle (0.1)node[above=0.1in]{$\varphi_4'$};
\draw[fill=white](2,2) circle (0.1)node[above=0.1in]{$\varphi_5'$};
\draw[fill=white](2.5,2) circle (0.1)node[above=0.1in]{$\varphi_6'$};

\draw[fill=white](0.5,0.5) circle (0.1)node[left=0.1in]{$\varphi_8$};
\draw[fill=Aquamarine](0.5,1) circle (0.1)node[left=0.1in]{$\varphi_\texttt{w}$};
\draw[fill=white](0.5,1.5) circle (0.1)node[left=0.1in]{$\varphi_8'$};

\draw[fill=Aquamarine](2.5,0.5) circle (0.1)node[right=0.1in]{$\varphi_0$};
\draw[fill=Aquamarine](2.5,1) circle (0.1)node[right=0.1in]{$\varphi_\texttt{c}$};
\draw[fill=Aquamarine](2.5,1.5) circle (0.1)node[right=0.1in]{$\varphi_0'$};

\draw[fill=White](0,-0.5) circle (0.1)node[below=0.1in]{$\varphi_1$};
\draw[fill=White](3,-0.5) circle (0.1)node[below=0.1in]{$\varphi_7$};
\draw[fill=White](3,2.5) circle (0.1)node[above=0.1in]{$\varphi_7'$};
\draw[fill=White](0,2.5) circle (0.1)node[above=0.1in]{$\varphi_1'$};
\end{tikzpicture}
\end{aligned}
\end{eqnarray}
As in the supersymmetric case reviewed above, we have colored the nodes extending the original (non-affine) $2\,D_8$ diagram, i.e. those which have non-zero momentum and/or winding.

The diagram \eqref{diag1} looks quite promising and indeed we will see that it plays a special role. Strictly speaking, however, it does not tell the full story. An extended diagram is supposed to encode the reflective part of the duality group, and as we have seen in Section \ref{ss:tduality}, there are reflections generated by vectors with norm 1 which are not accounted for so far. This suggests that the diagram has to be extended by adding ``short roots."\footnote{The reason for the quotation marks is that in the nine-dimensional theory these correspond to scalars rather than vectors, so they are not part of the gauge group.} Before proceeding, however, note that any extra node we add must correspond to a vector in the charge lattice $\Upsilon_{17,1}$, and that there are already 20 nodes in our diagram. For this reason, a generic extra node will have non-trivial links with many of those already in the diagram --- constructing a ``simple" extension of \eqref{diag1} is a highly constrained problem. 

\subsubsection{Tachyon contributions and infinite distance limits}

With this constraint in mind we look for extensions of \eqref{diag1} by solving a system of linear equations which determines a norm 1 vector in $\Upsilon_{17,1}$ linked as simply as possible with those given in \eqref{nodes1}, \eqref{nodec} and \eqref{nodew}. Remarkably, there is a solution with only two links, given by
\begin{equation}
     \varphi_\texttt{t} = \ket{-1,1;-\tfrac12,\tfrac12^7,0^7,1}\,,
\end{equation}
and the resulting diagram takes the form
\begin{eqnarray}\label{diag2}
\begin{aligned}
\begin{tikzpicture}[scale = 1.3]
\draw(-3,1)node{};
\draw(5.5,1);
\draw(0.5,0)--(2.5,0)--(2.5,2)--(0.5,2)--(0.5,0);
\draw(0.5,0)--(1,0.5);
\draw(2.5,0)--(3,-0.5);
\draw(0.5,2)--(0,2.5);
\draw(2.5,2)--(2,1.5);
\draw(0.95,0.55)--(1.95,1.55);
\draw(1.05,0.45)--(2.05,1.45);
\draw(1.3,0.8)--(1.1,0.8);
\draw(1.3,0.8)--(1.3,0.6);
\draw(1.7,1.2)--(1.9,1.2);
\draw(1.7,1.2)--(1.7,1.4);
\draw[fill=white](0.5,0) circle (0.1)node[below=0.1in]{$2$};
\draw[fill=white](1,0) circle (0.1)node[below=0.1in]{$3$};
\draw[fill=white](1.5,0) circle (0.1)node[below=0.1in]{$4$};
\draw[fill=white](2,0) circle (0.1)node[below=0.1in]{$5$};
\draw[fill=white](2.5,0) circle (0.1)node[below=0.1in]{$6$};

\draw[fill=white](0.5,2) circle (0.1)node[above=0.1in]{$2'$};
\draw[fill=white](1,2) circle (0.1)node[above=0.1in]{$3'$};
\draw[fill=white](1.5,2) circle (0.1)node[above=0.1in]{$4'$};
\draw[fill=white](2,2) circle (0.1)node[above=0.1in]{$5'$};
\draw[fill=white](2.5,2) circle (0.1)node[above=0.1in]{$6'$};

\draw[fill=white](0.5,0.5) circle (0.1)node[left=0.1in]{$8$};
\draw[fill=Aquamarine](0.5,1) circle (0.1)node[left=0.1in]{$\texttt{w}$};
\draw[fill=white](0.5,1.5) circle (0.1)node[left=0.1in]{$8'$};

\draw[fill=Aquamarine](2.5,0.5) circle (0.1)node[right=0.1in]{$0$};
\draw[fill=Aquamarine](2.5,1) circle (0.1)node[right=0.1in]{$\texttt{c}$};
\draw[fill=Aquamarine](2.5,1.5) circle (0.1)node[right=0.1in]{$0'$};

\draw[fill=White](1,0.5) circle (0.1)node[above=0.1in]{$1$};
\draw[fill=White](3,-0.5) circle (0.1)node[below=0.1in]{$7$};
\draw[fill=White](2,1.5) circle (0.1)node[below=0.1in]{$7'$};
\draw[fill=White](0,2.5) circle (0.1)node[above=0.1in]{$1'$};

\draw[fill=Yellow](1.5,1) circle (0.1)node[above=0.1in]{$\texttt{t}$};
\end{tikzpicture}
\end{aligned}
\end{eqnarray}
where we have dropped the $\varphi$'s for clarity. As already discussed, the short roots are associated to tachyons, hence the subscript \texttt{t}. 

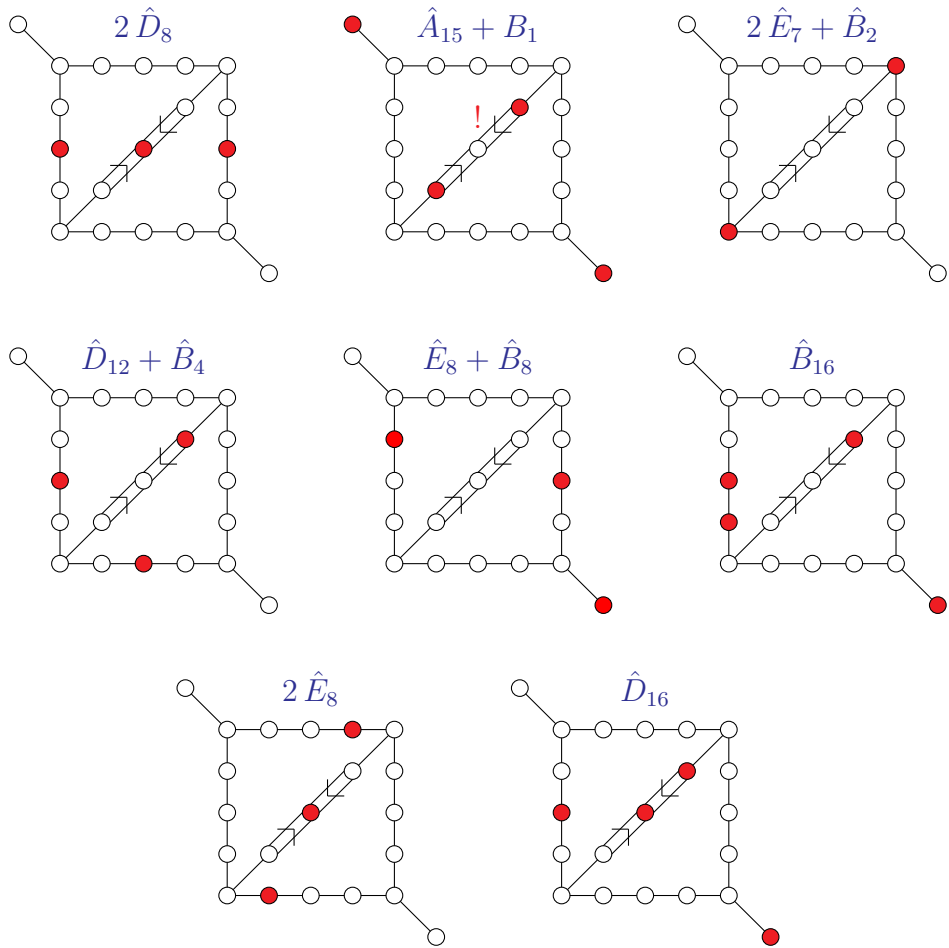
\begin{figure}
\begin{center}
\begin{tikzpicture}[scale = 1.1]
\draw[Blue](1.5,2.5)node{$2\,\hat D_8$};
\draw(0.5,0)--(2.5,0)--(2.5,2)--(0.5,2)--(0.5,0);
\draw(0.5,0)--(1,0.5);
\draw(2.5,0)--(3,-0.5);
\draw(0.5,2)--(0,2.5);
\draw(2.5,2)--(2,1.5);
\draw(0.95,0.55)--(1.45,1.05);
\draw(1.95,1.55)--(1.45,1.05);
\draw(1.05,0.45)--(1.55,0.95);
\draw(2.05,1.45)--(1.55,0.95);
\draw(1.3,0.8)--(1.1,0.8);
\draw(1.3,0.8)--(1.3,0.6);
\draw(1.7,1.2)--(1.9,1.2);
\draw(1.7,1.2)--(1.7,1.4);

\draw[fill=White](0.5,0) circle (0.1);
\draw[fill=white](1,0) circle (0.1);
\draw[fill=white](1.5,0) circle (0.1);
\draw[fill=white](2,0) circle (0.1);
\draw[fill=white](2.5,0) circle (0.1);

\draw[fill=white](0.5,2) circle (0.1);
\draw[fill=white](1,2) circle (0.1);
\draw[fill=white](1.5,2) circle (0.1);
\draw[fill=white](2,2) circle (0.1);
\draw[fill=White](2.5,2) circle (0.1);

\draw[fill=white](0.5,0.5) circle (0.1);
\draw[fill=Red](0.5,1) circle (0.1);
\draw[fill=white](0.5,1.5) circle (0.1);

\draw[fill=white](2.5,0.5) circle (0.1);
\draw[fill=Red](2.5,1) circle (0.1);
\draw[fill=white](2.5,1.5) circle (0.1);

\draw[fill=White](1,0.5) circle (0.1);
\draw[fill=White](3,-0.5) circle (0.1);
\draw[fill=White](2,1.5) circle (0.1);
\draw[fill=White](0,2.5) circle (0.1);

\draw[fill=Red](1.5,1) circle (0.1);
\begin{scope}[shift={(4,0)}]
\draw[Blue](1.5,2.5)node{$\hat A_{15} + B_1$};
\draw(0.5,0)--(2.5,0)--(2.5,2)--(0.5,2)--(0.5,0);
\draw(0.5,0)--(1,0.5);
\draw(2.5,0)--(3,-0.5);
\draw(0.5,2)--(0,2.5);
\draw(2.5,2)--(2,1.5);
\draw(0.95,0.55)--(1.95,1.55);
\draw(1.05,0.45)--(2.05,1.45);
\draw(1.3,0.8)--(1.1,0.8);
\draw(1.3,0.8)--(1.3,0.6);
\draw(1.7,1.2)--(1.9,1.2);
\draw(1.7,1.2)--(1.7,1.4);

\draw[fill=white](0.5,0) circle (0.1);
\draw[fill=white](1,0) circle (0.1);
\draw[fill=white](1.5,0) circle (0.1);
\draw[fill=white](2,0) circle (0.1);
\draw[fill=white](2.5,0) circle (0.1);

\draw[fill=white](0.5,2) circle (0.1);
\draw[fill=white](1,2) circle (0.1);
\draw[fill=white](1.5,2) circle (0.1);
\draw[fill=white](2,2) circle (0.1);
\draw[fill=white](2.5,2) circle (0.1);

\draw[fill=white](0.5,0.5) circle (0.1);
\draw[fill=white](0.5,1) circle (0.1);
\draw[fill=white](0.5,1.5) circle (0.1);

\draw[fill=white](2.5,0.5) circle (0.1);
\draw[fill=white](2.5,1) circle (0.1);
\draw[fill=white](2.5,1.5) circle (0.1);

\draw[fill=Red](1,0.5) circle (0.1);
\draw[fill=Red](3,-0.5) circle (0.1);
\draw[fill=Red](2,1.5) circle (0.1);
\draw[fill=Red](0,2.5) circle (0.1);

\draw[fill=white](1.5,1) circle (0.1)node[above=0.05in]{\textcolor{Red}{!}};
\end{scope}
\begin{scope}[shift={(8,0)}]
\draw[Blue](1.5,2.5)node{$2\, \hat  E_7+\hat B_2$};
\draw(0.5,0)--(2.5,0)--(2.5,2)--(0.5,2)--(0.5,0);
\draw(0.5,0)--(1,0.5);
\draw(2.5,0)--(3,-0.5);
\draw(0.5,2)--(0,2.5);
\draw(2.5,2)--(2,1.5);
\draw(0.95,0.55)--(1.95,1.55);
\draw(1.05,0.45)--(2.05,1.45);
\draw(1.3,0.8)--(1.1,0.8);
\draw(1.3,0.8)--(1.3,0.6);
\draw(1.7,1.2)--(1.9,1.2);
\draw(1.7,1.2)--(1.7,1.4);
\draw[fill=Red](0.5,0) circle (0.1);
\draw[fill=white](1,0) circle (0.1);
\draw[fill=white](1.5,0) circle (0.1);
\draw[fill=white](2,0) circle (0.1);
\draw[fill=white](2.5,0) circle (0.1);

\draw[fill=white](0.5,2) circle (0.1);
\draw[fill=white](1,2) circle (0.1);
\draw[fill=white](1.5,2) circle (0.1);
\draw[fill=white](2,2) circle (0.1);
\draw[fill=Red](2.5,2) circle (0.1);

\draw[fill=white](0.5,0.5) circle (0.1);
\draw[fill=white](0.5,1) circle (0.1);
\draw[fill=white](0.5,1.5) circle (0.1);

\draw[fill=white](2.5,0.5) circle (0.1);
\draw[fill=white](2.5,1) circle (0.1);
\draw[fill=white](2.5,1.5) circle (0.1);

\draw[fill=White](1,0.5) circle (0.1);
\draw[fill=White](3,-0.5) circle (0.1);
\draw[fill=White](2,1.5) circle (0.1);
\draw[fill=White](0,2.5) circle (0.1);

\draw[fill=white](1.5,1) circle (0.1);
\end{scope}

\begin{scope}[shift={(0,-4)}]
\draw[Blue](1.5,2.5)node{$\hat  D_{12}+\hat  B_4$};
\draw(0.5,0)--(2.5,0)--(2.5,2)--(0.5,2)--(0.5,0);
\draw(0.5,0)--(1,0.5);
\draw(2.5,0)--(3,-0.5);
\draw(0.5,2)--(0,2.5);
\draw(2.5,2)--(2,1.5);
\draw(0.95,0.55)--(1.95,1.55);
\draw(1.05,0.45)--(2.05,1.45);
\draw(1.3,0.8)--(1.1,0.8);
\draw(1.3,0.8)--(1.3,0.6);
\draw(1.7,1.2)--(1.9,1.2);
\draw(1.7,1.2)--(1.7,1.4);
\draw[fill=white](0.5,0) circle (0.1);
\draw[fill=white](1,0) circle (0.1);
\draw[fill=Red](1.5,0) circle (0.1);
\draw[fill=white](2,0) circle (0.1);
\draw[fill=white](2.5,0) circle (0.1);

\draw[fill=white](0.5,2) circle (0.1);
\draw[fill=white](1,2) circle (0.1);
\draw[fill=white](1.5,2) circle (0.1);
\draw[fill=white](2,2) circle (0.1);
\draw[fill=white](2.5,2) circle (0.1);

\draw[fill=white](0.5,0.5) circle (0.1);
\draw[fill=Red](0.5,1) circle (0.1);
\draw[fill=white](0.5,1.5) circle (0.1);

\draw[fill=white](2.5,0.5) circle (0.1);
\draw[fill=white](2.5,1) circle (0.1);
\draw[fill=white](2.5,1.5) circle (0.1);

\draw[fill=White](1,0.5) circle (0.1);
\draw[fill=White](3,-0.5) circle (0.1);
\draw[fill=Red](2,1.5) circle (0.1);
\draw[fill=White](0,2.5) circle (0.1);

\draw[fill=white](1.5,1) circle (0.1);
\end{scope}
\begin{scope}[shift={(4,-4)}]
\draw[Blue](1.5,2.5)node{$\hat  E_8+\hat  B_8$};
\draw(0.5,0)--(2.5,0)--(2.5,2)--(0.5,2)--(0.5,0);
\draw(0.5,0)--(1,0.5);
\draw(2.5,0)--(3,-0.5);
\draw(0.5,2)--(0,2.5);
\draw(2.5,2)--(2,1.5);
\draw(0.95,0.55)--(1.95,1.55);
\draw(1.05,0.45)--(2.05,1.45);
\draw(1.3,0.8)--(1.1,0.8);
\draw(1.3,0.8)--(1.3,0.6);
\draw(1.7,1.2)--(1.9,1.2);
\draw(1.7,1.2)--(1.7,1.4);
\draw[fill=white](0.5,0) circle (0.1);
\draw[fill=white](1,0) circle (0.1);
\draw[fill=white](1.5,0) circle (0.1);
\draw[fill=white](2,0) circle (0.1);
\draw[fill=white](2.5,0) circle (0.1);

\draw[fill=white](0.5,2) circle (0.1);
\draw[fill=white](1,2) circle (0.1);
\draw[fill=white](1.5,2) circle (0.1);
\draw[fill=white](2,2) circle (0.1);
\draw[fill=white](2.5,2) circle (0.1);

\draw[fill=Red](0.5,0.5) circle (0.1);
\draw[fill=white](0.5,1) circle (0.1);
\draw[fill=white](0.5,1.5) circle (0.1);

\draw[fill=white](2.5,0.5) circle (0.1);
\draw[fill=Red](2.5,1) circle (0.1);
\draw[fill=white](2.5,1.5) circle (0.1);

\draw[fill=White](1,0.5) circle (0.1);
\draw[fill=White](3,-0.5) circle (0.1);
\draw[fill=Red](2,1.5) circle (0.1);
\draw[fill=White](0,2.5) circle (0.1);

\draw[fill=white](1.5,1) circle (0.1);
\end{scope}

\begin{scope}[shift={(8,-4)}]
\draw[Blue](1.5,2.5)node{$\hat  B_{16}$};
\draw(0.5,0)--(2.5,0)--(2.5,2)--(0.5,2)--(0.5,0);
\draw(0.5,0)--(1,0.5);
\draw(2.5,0)--(3,-0.5);
\draw(0.5,2)--(0,2.5);
\draw(2.5,2)--(2,1.5);
\draw(0.95,0.55)--(1.95,1.55);
\draw(1.05,0.45)--(2.05,1.45);
\draw(1.3,0.8)--(1.1,0.8);
\draw(1.3,0.8)--(1.3,0.6);
\draw(1.7,1.2)--(1.9,1.2);
\draw(1.7,1.2)--(1.7,1.4);
\draw[fill=White](0.5,0) circle (0.1);
\draw[fill=white](1,0) circle (0.1);
\draw[fill=white](1.5,0) circle (0.1);
\draw[fill=white](2,0) circle (0.1);
\draw[fill=white](2.5,0) circle (0.1);

\draw[fill=white](0.5,2) circle (0.1);
\draw[fill=white](1,2) circle (0.1);
\draw[fill=white](1.5,2) circle (0.1);
\draw[fill=white](2,2) circle (0.1);
\draw[fill=White](2.5,2) circle (0.1);

\draw[fill=Red](0.5,0.5) circle (0.1);
\draw[fill=Red](0.5,1) circle (0.1);
\draw[fill=white](0.5,1.5) circle (0.1);

\draw[fill=white](2.5,0.5) circle (0.1);
\draw[fill=White](2.5,1) circle (0.1);
\draw[fill=white](2.5,1.5) circle (0.1);

\draw[fill=White](1,0.5) circle (0.1);
\draw[fill=Red](3,-0.5) circle (0.1);
\draw[fill=Red](2,1.5) circle (0.1);
\draw[fill=White](0,2.5) circle (0.1);

\draw[fill=white](1.5,1) circle (0.1);
\end{scope}

\begin{scope}[shift={(2,-8)}]
\draw[Blue](1.5,2.5)node{$2\,\hat  E_8$};
\draw(0.5,0)--(2.5,0)--(2.5,2)--(0.5,2)--(0.5,0);
\draw(0.5,0)--(1,0.5);
\draw(2.5,0)--(3,-0.5);
\draw(0.5,2)--(0,2.5);
\draw(2.5,2)--(2,1.5);
\draw(0.95,0.55)--(1.95,1.55);
\draw(1.05,0.45)--(2.05,1.45);
\draw(1.3,0.8)--(1.1,0.8);
\draw(1.3,0.8)--(1.3,0.6);
\draw(1.7,1.2)--(1.9,1.2);
\draw(1.7,1.2)--(1.7,1.4);
\draw[fill=White](0.5,0) circle (0.1);
\draw[fill=Red](1,0) circle (0.1);
\draw[fill=white](1.5,0) circle (0.1);
\draw[fill=white](2,0) circle (0.1);
\draw[fill=white](2.5,0) circle (0.1);

\draw[fill=white](0.5,2) circle (0.1);
\draw[fill=white](1,2) circle (0.1);
\draw[fill=white](1.5,2) circle (0.1);
\draw[fill=Red](2,2) circle (0.1);
\draw[fill=White](2.5,2) circle (0.1);

\draw[fill=white](0.5,0.5) circle (0.1);
\draw[fill=white](0.5,1) circle (0.1);
\draw[fill=white](0.5,1.5) circle (0.1);

\draw[fill=white](2.5,0.5) circle (0.1);
\draw[fill=white](2.5,1) circle (0.1);
\draw[fill=white](2.5,1.5) circle (0.1);

\draw[fill=White](1,0.5) circle (0.1);
\draw[fill=White](3,-0.5) circle (0.1);
\draw[fill=White](2,1.5) circle (0.1);
\draw[fill=White](0,2.5) circle (0.1);

\draw[fill=Red](1.5,1) circle (0.1);
\end{scope}

\begin{scope}[shift={(6,-8)}]
\draw[Blue](1.5,2.5)node{$\hat  D_{16}$};
\draw(0.5,0)--(2.5,0)--(2.5,2)--(0.5,2)--(0.5,0);
\draw(0.5,0)--(1,0.5);
\draw(2.5,0)--(3,-0.5);
\draw(0.5,2)--(0,2.5);
\draw(2.5,2)--(2,1.5);
\draw(0.95,0.55)--(1.95,1.55);
\draw(1.05,0.45)--(2.05,1.45);
\draw(1.3,0.8)--(1.1,0.8);
\draw(1.3,0.8)--(1.3,0.6);
\draw(1.7,1.2)--(1.9,1.2);
\draw(1.7,1.2)--(1.7,1.4);
\draw[fill=White](0.5,0) circle (0.1);
\draw[fill=white](1,0) circle (0.1);
\draw[fill=white](1.5,0) circle (0.1);
\draw[fill=white](2,0) circle (0.1);
\draw[fill=white](2.5,0) circle (0.1);

\draw[fill=white](0.5,2) circle (0.1);
\draw[fill=white](1,2) circle (0.1);
\draw[fill=white](1.5,2) circle (0.1);
\draw[fill=white](2,2) circle (0.1);
\draw[fill=White](2.5,2) circle (0.1);

\draw[fill=white](0.5,0.5) circle (0.1);
\draw[fill=Red](0.5,1) circle (0.1);
\draw[fill=white](0.5,1.5) circle (0.1);

\draw[fill=white](2.5,0.5) circle (0.1);
\draw[fill=white](2.5,1) circle (0.1);
\draw[fill=white](2.5,1.5) circle (0.1);

\draw[fill=White](1,0.5) circle (0.1);
\draw[fill=Red](3,-0.5) circle (0.1);
\draw[fill=Red](2,1.5) circle (0.1);
\draw[fill=White](0,2.5) circle (0.1);

\draw[fill=Red](1.5,1) circle (0.1);
\end{scope}
\end{tikzpicture}
\end{center}
\caption{Affine Dynkin diagrams encoded in the almost complete extended diagram of the $\os$ heterotic string on $S^1$. Red nodes are to be removed. The first six cases correspond to the ten dimensional non-supersymmetric heterotic strings while the last two cases correspond to the supersymmetric ones. The second case, corresponding in the decompactification limit to the $SU(16)\times U(1)$ string, shows that an extra node is required in order to have affine $B_1$, but it is not necessary to obtain every possible finite distance gauge symmetry enhancement.}\label{fig:infdist}
\end{figure}

To see if the diagram just obtained completes the picture it is instructive to put it through another test. In the supersymmetric case, the extended diagram not only encodes every symmetry enhancement at finite distance. It also encodes infinite distance limits, where one gets the affine version of  the two ten-dimensional gauge symmetries \cite{Collazuol:2022jiy}. Indeed, the affine subdiagrams are the starting point of the construction. Naturally we expect the same to occur in the non-supersymmetric setting, although here the extended diagram is required to encode eight different decompactification limits instead of two (six corresponding to the non-supersymmetric theories given in \eqref{10non-susy} and two to the supersymmetric ones). Remarkably, the diagram \eqref{diag2} \textit{almost} passes this test. By appropriately deleting nodes we find the affine Dynkin diagrams of all the ten-dimensional limiting gauge groups except $SU(16)\times U(1)$ (see Figure \ref{fig:infdist}); in this latter case, an affine node is missing. Note that to properly read the diagrams we must take into account the fact that the charge vectors of tachyons are strictly seen as roots by the theory and that they transform as vectors of the gauge group $Spin(2n)$. The root system of $Spin(2n)$ combined with the weights sitting in the vector representation is isomorphic to the root system of $Spin(2n+1)$, hence the subdiagram we look for is \textit{not} $\hat D_n$ but rather $\hat B_n$.   

To correct the issue of the missing affine subdiagram, note that because of the symmetry of the diagram \eqref{diag1},  when we added the node (\texttt{t}) connecting ($1$) and ($7'$) we might as well have added instead a node ($\texttt{t}'$) connecting ($1'$) and ($7$). We must in fact add both nodes at the same time. The node ($\texttt{t}'$) is given by the vector
\begin{equation}
    \varphi_{\texttt{t}'} = \ket{-1,1;0^7,1,-\tfrac12,\tfrac12^7}\,,
\end{equation}
which has inner product $-1$ with $\varphi_{\texttt{t}}$; the two extra nodes are then linked by a double line and furnish the affine Dynkin diagram of $B_1$. The resulting diagram can be shown to account for all the walls in the fundamental domain of the moduli space, and so is complete. It takes the form:
\begin{eqnarray}\label{diagcomplete}
\begin{aligned}
\begin{tikzpicture}[scale=2]
\usetikzlibrary{decorations.markings}
\draw(0.5,0,0)--(2.5,0,0);
\draw[double](1,0,1.5)--(1.5,-1,1);
\draw[double](1.5,-1,1)--(2,0,0.5);
\draw[double](1.5,-1,1)--(1.5,1,1);
\draw(2.5,0,0)--(2.5,0,2)--(0.5,0,2)--(0.5,0,0);

\draw(0.5,0,0)--(1,0,0.5);
\draw(2.5,0,0)--(2,0,0.5);
\draw(0.5,0,2)--(1,0,1.5);
\draw(2.5,0,2)--(2,0,1.5);

\draw[fill=White](0.5,0,0) circle (0.06);
\draw[fill=white](1,0,0) circle (0.06);
\draw[fill=white](1.5,0,0) circle (0.06);
\draw[fill=white](2,0,0) circle (0.06);
\draw[fill=white](2.5,0,0) circle (0.06);

\draw[double](1,0,0.5)--(1.5,1,1);
\draw[double](2,0,1.5)--(1.5,1,1);

\draw[fill=white](0.5,0,2) circle (0.06);
\draw[fill=white](1,0,2) circle (0.06);
\draw[fill=white](1.5,0,2) circle (0.06);
\draw[fill=white](2,0,2) circle (0.06);
\draw[fill=White](2.5,0,2) circle (0.06);

\draw[fill=white](0.5,0,0.5) circle (0.06);
\draw[fill=Aquamarine](0.5,0,1) circle (0.06);
\draw[fill=white](0.5,0,1.5) circle (0.06);

\draw[fill=Aquamarine](2.5,0,0.5) circle (0.06);
\draw[fill=Aquamarine](2.5,0,1) circle (0.06);
\draw[fill=Aquamarine](2.5,0,1.5) circle (0.06);

\draw[fill=White](1,0,0.5) circle (0.06);
\draw[fill=White](2,0,0.5) circle (0.06);
\draw[fill=White](2,0,1.5) circle (0.06);
\draw[fill=White](1,0,1.5) circle (0.06);

\draw[fill=Yellow](1.5,1,1) circle (0.06);
\draw[fill=Yellow](1.5,-1,1) circle (0.06);
\end{tikzpicture}
\end{aligned}
\end{eqnarray}
Here we omit labels and arrows for clarity, but note that the double link between yellow nodes does not involve an arrow as the corresponding vectors have the same length. Remarkably, this same diagram was found by Vinberg in his classification of hyperbolic Coxeter groups \cite{Vinberg}, where he also found the one for the lattice $\Gamma_{1,17}$ corresponding to circle compactifications of the supersymmetric heterotic strings. This reflects the fact that the fundamental domains of the classical moduli space of both supersymmetric and non-supersymmetric heterotic strings on a circle (with rank 16) are Coxeter polytopes. 
\subsubsection{Fundamental domain}

The extended Dynkin diagram allows us to determine the fundamental region in moduli space, whose walls are the fixed points of the symmetry corresponding to each node of the diagram. In the supersymmetric $Spin(32)/\mathbb{Z}_2$ heterotic string the fundamental domain has the form of a chimney\footnote{Very much like in the $SL(2,\mathbb Z)$ case, where the vertical ``walls" (lines in this case) are at a fixed value of the real part of $\tau$, and there is a circular boundary at the bottom given by $|\tau|=1$.} \cite{Cachazo:2000ey} whose vertical walls are given by constant values of the Wilson lines, and the horizontal ones are two spherical walls corresponding to the nodes $\texttt{c}$ and $\texttt{w}$ in the extended Dynkin diagram (in the $E_8\times E_8$ string description there is only one spherical boundary due to the node $\texttt{c}$ because there is no node $\texttt{w}$). Here we determine the fundamental domain for the non-supersymmetric $\os$ string.

The boundaries of the fundamental domain are given by the co-dimension one walls in moduli space given by the moduli that satisfy
\begin{equation}
    p_R = 0 ~~~~~  \Leftrightarrow ~~~~~  n - Ew - \pi \cdot A = 0
\end{equation}
for a given charge vector $\ket{w,n;\pi}$ associated to each node in the extended Dynkin diagram; recall this is one of the conditions for the respective gauge bosons to become massless. Each wall then defines two T-dual regions of the moduli space such that only one of them is kept in the fundamental domain, e.g. the one that satisfies the inequality 
$n - Ew - \pi \cdot A \le 0$. The 22 nodes of the extended Dynkin diagram then give a set of 22 inequalities that define the fundamental region of moduli space. These are recorded in Table \ref{tab:boundaries} for the charge vectors in \eqref{nodes1}, \eqref{nodec} and \eqref{nodew}.

\begin{table}
\centering
\begin{tabular}
			{|>{$}c<{$}||
				>{$}c<{$}||
				>{$}c<{$}|} \hline
    \text{Node} & \text{Generator} & \text{Fundamental region} \\ \hline\hline
 \varphi_{i=1,\text{...},7} & \ket{0,0;\alpha _i,0^8} & A_i\leq A_{i+1} \\
 \varphi _8 & \ket{0,0;\alpha _8,0^8} & -A_2\leq A_1 \\
 \varphi _0 & \ket{0,1;\alpha _0,0^8} &  A_8\leq 1-A_7 \\
 \varphi '_{i=1,\text{...},7} & \ket{0,0;0^8,\alpha _i} & A_{i+8}\leq A_{i+9} \\
 \varphi '_8 & \ket{0,0;0^8,\alpha _8} &  -A_{10}\leq A_9 \\
 \varphi '_0 & \ket{0,1;0^8,\alpha _0} & A_{16}\leq 1-A_{15} \\
 \varphi _{\texttt{w}} & \ket{-1,1;\frac{1}{2}^8,\frac{1}{2}^8} & 
 2R^2\geq 2-\left(A-\texttt{w} \right)^2  \\
 \varphi _c & \ket{-1,-1;0^8,0^8} & 
 2R^2\geq 2-  A^2 \\
 \varphi_{\texttt{t}} & \ket{-1,1;-\frac{1}{2},\frac{1}{2}^7,0^7,1} & 2R^2 \geq 1-\left(A-\texttt{t}\right)^2 \\
 \varphi_{\texttt{t'}} & \ket{-1,1;0^7,1,-\frac{1}{2},\frac{1}{2}^7} & 2R^2 \geq 1 -\left(A-\texttt{t'}\right)^2 \\ \hline
\end{tabular}
\caption{Fundamental region of the $\os$ string compactified on a circle. Here $\texttt{w},\texttt{t}$ and $\texttt{t'}$ denote the 16-dimensional piece of the corresponding node. The boundaries are given by saturating the inequalities.}\label{tab:boundaries}
\end{table}

\begin{figure}
\centering
\includegraphics[width=0.5\textwidth]{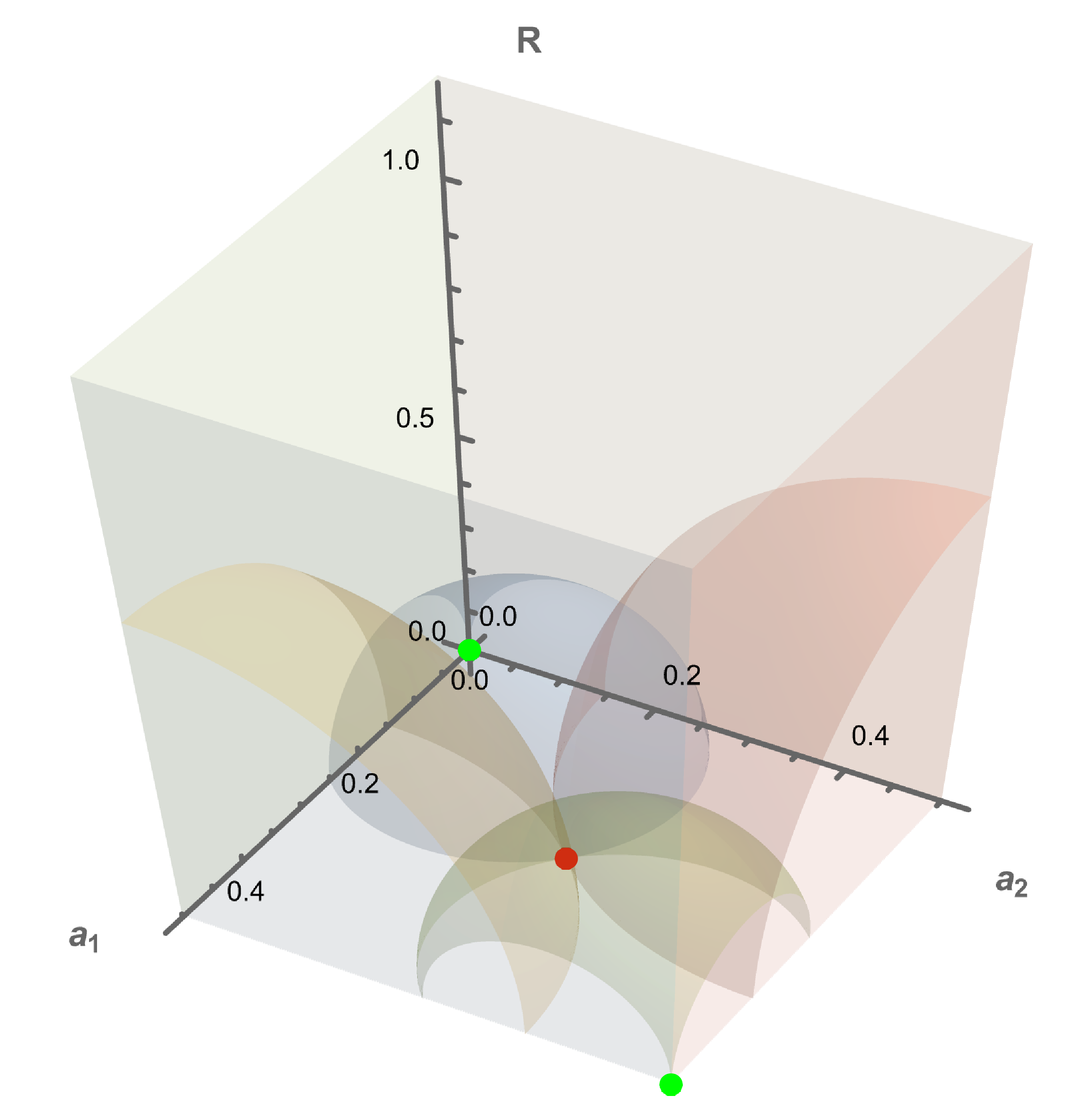}
\caption{Slice of moduli space with Wilson line $A = \left(-a_1,(a_1)^6,1-a_1,-a_2,(a_2)^6,1-a_2\right)$, showing the four spherical boundaries of the fundamental domain. The red and green dots are, respectively, associated to the $\hat A_{15}+\hat B_1$ and $\hat D_{16}$ decompactification limits (cf. Figure \ref{fig:infdist}).}\label{fig:chimney}
\end{figure}

The constraints associated to the nodes $(\texttt{c})$, $(\texttt{w})$, $(\texttt{t})$ and $(\texttt{t}')$ all involve the radius $R$, and give rise to four spherical boundaries at the bottom (small $R$) of the moduli space. In Figure \ref{fig:chimney} we plot these boundaries in a 3-dimensional slice given by an appropriate parametrization of the Wilson line moduli $A_i$. The two larger walls, absent in the supersymmetric case, correspond to the tachyon nodes which, as explained above, are doubly linked. They therefore intersect at infinite distance $R \to 0$ (red dot)  where the theory decompactifies\footnote{This point, as well as the green dots, all located at $R=0$, are actually T-dual to the usual decompactification limits $R\to \infty$. } to the $SU(16)\times U(1)$ string. The significance of this is that every finite distance intersection cannot involve both of these walls simultaneously, and so the full set of symmetry enhancements is encoded already in the subdiagram \eqref{diag2}. For our purposes, then, the full non-planar diagram \eqref{diagcomplete} will not be necessary.

\subsection{Maximal enhancements}
\label{sec:maximal}

Maximal symmetry enhancements are obtained from the EDD \eqref{diag2} by simply deleting four nodes such that what remains is the Dynkin diagram of a finite-dimensional Lie algebra. The presence of the tachyon node $(\texttt{t})$ leads to a $B_n$ component, which signals a $\mathfrak{spin}_{2n}$ gauge algebra at level 1 together with a tachyon of mass $m^2 = -2$ sitting in its vector representation $\mathbf{2n}$.  The remaining pieces of the diagram then correspond  to various ADE factors completing the rank of the gauge algebra to 17. The reader is encouraged to verify this for the trivial $SU(2)$ enhancements at the self-dual radius for the five tachyonic heterotic strings; for example, see the diagram: 
\begin{eqnarray}
\begin{aligned}
\begin{tikzpicture}
\draw[Blue](1.5,2.5)node{$2\, E_7+ B_2 + A_1$};
\draw(0.5,0)--(2.5,0)--(2.5,2)--(0.5,2)--(0.5,0);
\draw(0.5,0)--(1,0.5);
\draw(2.5,0)--(3,-0.5);
\draw(0.5,2)--(0,2.5);
\draw(2.5,2)--(2,1.5);
\draw(0.95,0.55)--(1.95,1.55);
\draw(1.05,0.45)--(2.05,1.45);
\draw(1.3,0.8)--(1.1,0.8);
\draw(1.3,0.8)--(1.3,0.6);
\draw(1.7,1.2)--(1.9,1.2);
\draw(1.7,1.2)--(1.7,1.4);
\draw[fill=Red](0.5,0) circle (0.1);
\draw[fill=white](1,0) circle (0.1);
\draw[fill=white](1.5,0) circle (0.1);
\draw[fill=white](2,0) circle (0.1);
\draw[fill=white](2.5,0) circle (0.1);

\draw[fill=white](0.5,2) circle (0.1);
\draw[fill=white](1,2) circle (0.1);
\draw[fill=white](1.5,2) circle (0.1);
\draw[fill=Red](2,2) circle (0.1);
\draw[fill=white](2.5,2) circle (0.1);

\draw[fill=white](0.5,0.5) circle (0.1);
\draw[fill=white](0.5,1) circle (0.1);
\draw[fill=white](0.5,1.5) circle (0.1);

\draw[fill=white](2.5,0.5) circle (0.1);
\draw[fill=white](2.5,1) circle (0.1);
\draw[fill=Red](2.5,1.5) circle (0.1);

\draw[fill=White](1,0.5) circle (0.1);
\draw[fill=White](3,-0.5) circle (0.1);
\draw[fill=Red](2,1.5) circle (0.1);
\draw[fill=White](0,2.5) circle (0.1);

\draw[fill=white](1.5,1) circle (0.1);
\end{tikzpicture} 
\end{aligned} 
\end{eqnarray}
The values of the moduli are then obtained by saturating the corresponding inequalities in Table \ref{tab:boundaries}, yielding in this case
\begin{equation}
    R = \tfrac14\, , ~~~ A = (0^5, \tfrac14^2, \tfrac12,0^2, \tfrac12^6) \, .
\end{equation}
Finally one can determine the rest of the spectrum up to the massless level as explained in Section \ref{ss:content}. This requires the embedding data of the root lattice into $\Upsilon_{17,1}$, which is simply given by the collection of charge vectors associated to the nodes in the left-over Dynkin diagram. 

The full list of maximal enhancements, including those with the tachyonic node in the EDD, comprises 107 distinct cases. We record the relevant data in Table \ref{tab:9Dlist} of Appendix \ref{app:results}. Out of these cases, there are 12 enhancements corresponding to deletions of the nodes $(1)$ and $(7')$ in the EDD \eqref{diag2}, which means they have a $U(1)$ factor in their gauge group. We take these enhancements to be on the same footing as those without $U(1)$ factors, since they are also obtained by deleting only four nodes, and thus appear at points in moduli space which are fixed points under 17 reflection symmetries. On the other hand, there are rank 16 non-maximal enhancements that occur at lines in moduli space, fixed under 16 reflection symmetries, and are therefore  not on the same footing as the preceding case.   

In general, enhancements without tachyons of mass $m^2 = -2$ are those encoded in the subdiagram \eqref{diag1}. There are in total $22$ of these which are maximal, up to symmetries of the diagram which are symmetries of the theory, but $14$ of them \textit{do} contain tachyons with mass $-2 < m^2 < 0$. Only the remaining $8$, recorded in Table \ref{tab:9Dtachyonfree}, are candidates for tachyon-free critical points of the cosmological constant.  As will be explained shortly, however, four of these have scalars in the conjugacy class $\Gamma_0$ which makes them automatically unstable.  

\begin{table}[H]
	\begin{center}\footnotesize
		\renewcommand{\arraystretch}{1.2}
		\begin{tabular}
  {|
  @{}>{ }c<{ }@{}|
  @{}>{ }c<{ }@{}|
  @{}>{ }l<{ }@{}|
  @{}>{ }c<{ }@{}|
  @{}>{ }c<{ }@{}|
  @{}>{ }c<{ }@{}|} \hline
\#& WL &
\ \ \ \ \ \quad \quad \quad v &  
s &   
c &  
0  \\ \hline
41 & $\left( 0^{16}\right)$ & $[A_1+2D_8\,;\mathbb{Z}_2]$& $\begin{matrix}({\bf 1},{\bf 128},{\bf 1}) \\ ({\bf 1},\textbf{1},\textbf{128}) \end{matrix}$ &$(\textbf{1},\textbf{16},\textbf{16})$&  none \\ \hline
40& $\left( \tfrac12^2, 0^{14}\right)$ & $[A_1+A_2+D_6+D_8\,;\mathbb{Z}_2]$& $\begin{matrix} (\textbf{2},\textbf{1},\textbf{32},\textbf{1})\\(\textbf{1},\textbf{1},\textbf{1},\textbf{128})\end{matrix}$ &$(\textbf{1},\textbf{1},\textbf{12},\textbf{16})$&  none \\ \hline
36& $\left( \tfrac12^3, 0^{13}\right)$ & $A_4+D_5+D_8$ & $(\textbf{1},\textbf{1},\textbf{128})$ & $(\textbf{1},\textbf{10},\textbf{16}$)&  none \\ \hline
17 &  $\left( \tfrac12^3, 0^5, \tfrac12^3, 0^5\right)$ &
$[A_7+2D_5\,; \mathbb{Z}_4]$  & none &
$\begin{matrix}(\textbf{1},\textbf{10},\textbf{10}) \\ (\textbf{70},\textbf{1},\textbf{1})\end{matrix}$
&   none\\ \hline
51 & $\left( 1, 0^{15}\right)$ & $D_8 + D_9$ & $(\textbf{128},\textbf{1})$ &$(\textbf{16},\textbf{18})$&  $(\textbf{128},\textbf{1})\times 2$ \\ \hline
37& $\left( \tfrac12^4, 0^{12}\right)$ & $[D_4+D_5+D_8\,;\mathbb{Z}_2]$&
$\begin{matrix}(\textbf{1},\textbf{1},\textbf{128}) \\ (\textbf{8},\textbf{10},\textbf{1})\end{matrix}$ &
$(\textbf{8},\textbf{1},\textbf{16})$ &  $(\textbf{8},\textbf{16},\textbf{1}) \times 2$ \\ \hline
28& $\left( \tfrac12^2, 0^6, \tfrac12^2, 0^6\right)$ & $[2A_1+A_3+2D_6\,; \mathbb{Z}_2^2]$&
$\begin{matrix}(\textbf{2},\textbf{1},\textbf{1},\textbf{32},\textbf{1})\\(\textbf{1},\textbf{2},\textbf{1},\textbf{1},\textbf{32})\end{matrix}$ & $\begin{matrix}(\textbf{1},\textbf{1},\textbf{1},\textbf{12},\textbf{12})\\(\textbf{2},\textbf{2},\textbf{6},\textbf{1},\textbf{1})
\end{matrix}$&
$\begin{matrix} (\textbf{2},\textbf{1},\textbf{1},\textbf{32},\textbf{1})\times 2 \\  (\textbf{1},\textbf{2},\textbf{1},\textbf{1},\textbf{32}) \times 2 \end{matrix}$  
\\ \hline
73& $\left( \tfrac12^5,0^3,\tfrac14^7,\text{-}\tfrac14\right)$ &
$[A_{11} + E_6 \,;\mathbb{Z}_3]$ &   none &    none  & $\begin{matrix} (\textbf{143},\textbf{1})\times 2 \\ (\textbf{1},\textbf{78}) \times 2 \end{matrix}$
\\ \hline
  	\end{tabular} \caption{The eight tachyon-free maximal enhancements, all realized at $R^2=1-\tfrac12 A^2$, their Wilson lines and massless spectra.  The numbers indicate their location in the complete list of maximal enhancements given in Table \ref{tab:9Dlist}.} \label{tab:9Dtachyonfree}.  \end{center}
\end{table}

\subsubsection{Scalars and knife edges}
\label{sec:knifeedges}

Consider the maximal enhancement obtained by turning on a Wilson line
\begin{equation}
    A = (1,0^7,0^8)
\end{equation}
and setting the critical radius $R^2 = 1/2$. The effect of the Wilson line is to project out the element $(\tfrac12^8)$ in the gauge lattice $[2\, D_8]^+$ making the gauge group simply connected,
\begin{equation}
    A: ~~~~~ G = \frac{Spin(16)\times Spin(16)}{\mathbb{Z}_2} ~\to~ Spin(16)\times Spin(16)\,.
\end{equation}
Upon tuning the radius to $1/\sqrt{2}$, the $Spin(16)$ factor in which the Wilson line sits is subsequently enhanced to $Spin(18)$, and so the maximal enhancement has gauge group $G = Spin(18)\times Spin(16)$. This model has no tachyons and correspondingly it can be obtained from the extended diagram \eqref{diag1}:
\begin{eqnarray}
\begin{aligned}
\begin{tikzpicture}[scale = 1.2]
\draw(-3,1)node{};
\draw(5.5,1);
\draw(0.5,0)--(2.5,0)--(2.5,2)--(0.5,2)--(0.5,0);
\draw(0.5,0)--(0,-0.5);
\draw(2.5,0)--(3,-0.5);
\draw(0.5,2)--(0,2.5);
\draw(2.5,2)--(3,2.5);
\draw[fill=white](0.5,0) circle (0.1);
\draw[fill=white](1,0) circle (0.1);
\draw[fill=white](1.5,0) circle (0.1)node[below=0.1in]{$D_9$};
\draw[fill=white](2,0) circle (0.1);
\draw[fill=white](2.5,0) circle (0.1);

\draw[fill=white](0.5,2) circle (0.1);
\draw[fill=white](1,2) circle (0.1);
\draw[fill=white](1.5,2) circle (0.1)node[above=0.1in]{$D_8$};
\draw[fill=white](2,2) circle (0.1);
\draw[fill=white](2.5,2) circle (0.1);

\draw[fill=white](0.5,0.5) circle (0.1);
\draw[fill=Red](0.5,1) circle (0.1);
\draw[fill=white](0.5,1.5) circle (0.1);

\draw[fill=white](2.5,0.5) circle (0.1);
\draw[fill=white](2.5,1) circle (0.1);
\draw[fill=Red](2.5,1.5) circle (0.1);

\draw[fill=White](0,-0.5) circle (0.1);
\draw[fill=Red](3,-0.5) circle (0.1);
\draw[fill=White](3,2.5) circle (0.1);
\draw[fill=White](0,2.5) circle (0.1);
\end{tikzpicture}
\end{aligned}
\end{eqnarray}
As observed in \cite{Ginsparg:1986wr}, however, there are massless scalars in the spectrum which do become tachyonic as soon as any of the Wilson line moduli associated to the $D_8$ factor are turned on. The cosmological constant goes to $-\infty$ for any infinitesimal variation along any of eight directions in moduli space, and so the maximal enhancement point sits on a ``knife edge" making it unstable. We discuss this in detail in Section \ref{sec:Hessianmaxenh}.

To see how these scalars are obtained, note first that the weight lattice of the gauge group $G$ is $W = D_8\oplus D_9$, as $G$ is simply connected (this and other lattice embedding statements can be checked explicitly by using the embedding data of the roots obtained from the EDD). The dual lattice $W^*$ therefore contains the spinor class of $D_8$ which comprises vectors with norm $2$. On the other hand, the orthogonal complement of $W$ in $\Gamma_v$ can be shown to be $T = A_1(-2)$, whose dual $T^*$ contains two vectors with norm $-1$. Combining these sets of dual vectors we obtain a total of $128\times 2 = 256$ vectors in $\Gamma_0$ with $p_L^2 = 2$ and $p_R^2 = 1$. As explained in Section \ref{ss:content}, these give rise to scalars in the spacetime spectrum. A pair of such vectors takes the form 
\begin{equation}
    \ket{\pm1,\mp1;\mp1,0^7,\text{-}\tfrac12,\tfrac12^7}\,.
\end{equation}
Turning on a Wilson line $\epsilon$ in the $D_8$ factor, say $A = (1,0^7,\epsilon,0^7)$, the value of $p_R^2$ for these vectors is then
\begin{equation}
    p_R^2 = \frac{1}{2 R^2}(\mp 1 \mp E - A\cdot \pi)^2 = (\mp 1 + \tfrac12 \epsilon)^2\,,
\end{equation}
so that one vector has $p_R^2 > 1$ while the other has $p_R^2 < 1$. The scalar associated to the first vector thus becomes massive while that associated to the second becomes tachyonic. In general, 128 out of the 256 scalars become massive while the other 128 become tachyonic, for any value of $\epsilon$. 

The general lesson here is that whenever there are 
massless scalars transforming in a non-trivial representation of a gauge group factor, varying these moduli such that this gauge symmetry is broken will always produce tachyons. The presence of scalars of this type always signals an instability in the theory. 

\subsubsection{Scalars in the adjoint and right-moving gauge bosons}
Let us now consider the tachyon-free maximal enhancement with gauge group $E_6\times SU(12)/\mathbb{Z}_3$. In the supersymmetric heterotic string on $S^1$ there is also a maximal enhancement with this gauge group, and it turns out that its weight lattice $W$ has orthogonal complement $T = A_1(-2)$ in the Narain lattice $\Gamma_{17,1}$. We can therefore break supersymmetry at this point by using a norm $-1$ shift vector along $T$ given by its generator divided by 2. Since this shift is orthogonal to $W$, the gauge group is preserved, and the result is the aforementioned non-supersymmetric maximal enhancement. 

Whenever a non-supersymmetric enhancement can be obtained by putting a shift along $T$, the conjugacy class $\Gamma_0$ will contain norm $-1$ vectors. From Section \ref{sec:enhancements} we see that these vectors give rise to $SU(2)$ enhancements at level 2 in the right-moving sector. In this case, the result is that the full gauge symmetry is
\begin{equation} \label{E6A11SU2}
    G = \left(\frac{E_6\times SU(12)}{\mathbb{Z}_3}\right)_L \times SU(2)_R\,.
\end{equation}
On the other hand, one can add together the aforementioned norm $-1$ vectors with the charge vectors associated to the roots of the left-moving gauge group, resulting in norm $1$ vectors giving rise to massless scalars as in the previous example. To be precise we obtain a spacetime scalar transforming in the adjoint representation of the full gauge symmetry group. Any variation of the moduli whatsoever results in some of the states forming this scalar becoming tachyonic, and so this maximal enhancement is completely unstable; it sits at a knife edge with respect to every direction in moduli space. There are various maximal enhancements exhibiting $SU(2)_R$ gauge symmetries (see Table \ref{tab:9Dlist}), but the example discussed here is the only one without tachyons in nine dimensions. 

We remark that in the case of $T^8$ compactifications there are 24 special points in the classical moduli space where $\Upsilon_{8,24}$ splits orthogonally into $N_I \oplus D_8^*(-1)$ with $N_I$ one of the 24 Niemeier lattices. There is an enhancement of the right-moving sector to $SU(2)^8/\mathbb{Z}_2^7$ in accord with the fact that the worldsheet CFT must factorize into a left-moving CFT based on $N_I$ and a right-moving CFT realized with 24 real fermions and current algebra $8\, \hat A_{1,2}$ \cite{Harrison:2021gnp}. Since the $N_I$ are self-dual, there are no tachyons nor massless fermions. Moreover, in the case of the Leech lattice, which lacks roots, there is no left-moving non-Abelian gauge group and so there are no knife edges. These 24 points belong to a family of so-called MSDS models \cite{Kounnas:2008ft}; their massive spectra are degenerate since the partition function of the right-moving CFT is exactly 24.\footnote{We thank B. Percival for bringing these models to our attention.}

\section{One-loop cosmological constant}\label{s:cc}

All maximal enhancements lie at the intersection of 
seventeen walls in moduli space, and as such are invariant under the seventeen reflections corresponding to each node. This is sufficient to show that maximal enhancements extremize the one-loop cosmological constant \cite{Ginsparg:1986wr}. 
However the proof  does not tell us if they are maxima, minima or saddle points. Furthermore, the sign of the cosmological constant could be either positive or negative. Here we compute the one-loop cosmological constant in the whole seventeen-dimensional moduli space, and determine its Hessian at the eight points of tachyon-free maximal enhancement. We will show that none of them are minima.

\subsection{Computing the cosmological constant for \texorpdfstring{$T^d$}{Td} compactifications}

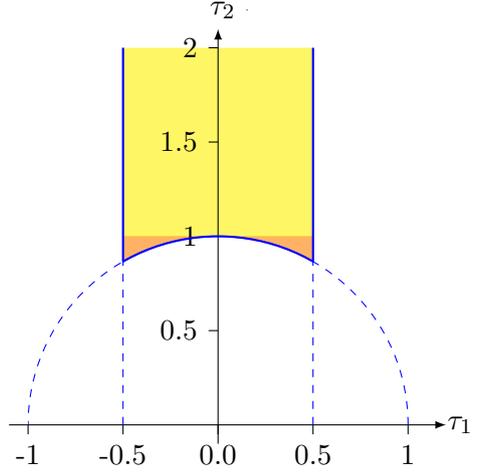
\begin{figure}
\centering
\begin{tikzpicture}[scale=2.5]
\pgfmathsetmacro{\myxlow}{-1}
\pgfmathsetmacro{\myxlowb}{-0.5}
\pgfmathsetmacro{\myxhigh}{1}
    \draw[-latex](\myxlow-0.1,0) -- (\myxhigh+0.2,0);
    \pgfmathsetmacro{\succofmyxlow}{\myxlow+0.5}
        \pgfmathsetmacro{\succofmyxlowb}{\myxlowb+0.5}
      \draw (-1,0) -- (-1,-0.05) node[below,font=\small] {-1};
    \foreach \x in {\myxlowb,\succofmyxlowb,...,\myxhigh}
    {   \draw (\x,0) -- (\x,-0.05) node[below,font=\small] {\x};}
    \foreach \y  in {0.5,1.0,...,2}
    {   \draw (0,\y) -- (-0.05,\y) node[left,font=\small] {\pgfmathprintnumber{\y}};}
    \draw[-latex](0,-0.1) -- (0,2.1);
    \begin{scope}   
    \clip (-1,0) rectangle (1,1.1);
    \draw [dashed,blue] (1,0) arc(0:180:1);       
    \end{scope}
    \begin{scope}   
    \clip (-1/2,0) rectangle (1/2,1.1);
     \draw [thick,blue] (1,0) arc(0:180:1);             
    \end{scope}
    \draw [dashed,blue] (-0.5,0) -- (-0.5,0.866);
    \draw [thick,solid,blue] (-0.5,0.866) -- (-0.5,2);
    \draw [dashed,blue] (0.5,0) -- (0.5,0.866);
    \draw [thick,solid,blue] (0.5,0.866) -- (0.5,2);
    \begin{scope}
        \begin{pgfonlayer}{background}
            \clip (-0.5,0) rectangle (0.5,2);
            \clip   (1,2) -| (-1,0) arc (180:0:1) -- cycle;            \fill[orange,opacity=0.6] (-1,-1) rectangle (1,1);
            \fill[yellow,opacity=0.6] (-1,1) rectangle (1,2);
        \end{pgfonlayer}
    \end{scope}
    {   \draw (0.15,2.2) -- (0.15,2.2) node[left,font=\small] {$\tau_2$};}
        {   \draw (1.4,0) -- (1.4,0) node[left,font=\small]{$\tau_1$};}
\end{tikzpicture}
\caption{A fundamental domain for $\tau$, divided into two regions. When integrating over the fundamental domain, the yellow region gets contributions only from  terms with no dependence on $\tau_1$.}
\label{fig:sl2z}
\end{figure}

The one-loop cosmological constant is obtained by integrating the partition function over the fundamental domain $F_0$ of the complex structure of the world-sheet two-torus (see Figure \ref{fig:sl2z}) 
\begin{equation} \label{Lambda}
\Lambda_{\rm{1-loop}}=-(4 \pi^2 \alpha')^{-\frac{10-d}{2}} \int_{F_0} \frac{d^2\tau}{2 \tau_2^2} Z(\tau) \, ,
\end{equation}
where $10-d$ is the number of external dimensions.

The partition function of the $\os$ theory compactified on $T^d$ was computed in \cite{Itoyama:2021itj} and involves sums over states in subsets of the charge lattice $\Upsilon_{d+16,d}$, 
\begin{equation} 
Z(\tau)=Z^{8-d} \left(
{\bar V}_8 \, Z_v -{\bar S}_8 \, Z_s 
- {\bar C}_8 \, Z_c +  {\bar O}_8 \, Z_0 \right)\, . 
\end{equation}
The $SO(8)$ characters $V_8,S_8,C_8,O_8$ are given in \eqref{characters}, and the individual partition functions are
\begin{equation} \label{Z}
\begin{split}
Z^{8-d} (
\tau)&=\tau _2^{-\frac{1}{2} (8-d)} \left(\eta  \bar{\eta }\right)^{-(8-d)}
\, \ , \\
Z_{v,s,c,0}(\tau) &=\frac{1}{\eta ^{16+d} \, \bar{\eta }^d}{\sum _{P\in \Gamma_{v,s,c,0}} \, q^{\frac{P_L^2}{2}} \, \bar{q}^{\frac{p_R^2}{2}}} 
\, ,
\end{split}
\end{equation}
with the Dedekind eta function given by
\begin{equation}
\eta(\tau) =q^{1/24}\overset{\infty }{\prod_{n=1}}(1-q^n) \ , \quad q=e^{2 \pi  i \tau } \ .
\end{equation}
Using $O_8$ and $V_8=S_8=C_8$ given in \eqref{characters}, we get 
\bea 
Z=\frac{1}{
2\tau _2^{\frac{8-d}{2}} 
\eta_*^{24} \bar{\eta}_*^{12}}\, \Big[
f_1(\bar q) \, (z_v-z_s-z_c)
 + f_2(\bar q) \, z_0\Big]\,,
\eea
where we have defined
\begin{equation*}
\begin{split}
\eta_*(q) &=  q^{-1/24}\eta(\tau) = \overset{\infty }{\prod_{n=1}}(1-q^n)= 1-q-q^2+ O\left(q^{5}\right)\,, \\
 f_1(\bar{q}) &= \frac{1}{\sqrt{\bar{q}}}\left(\overset{\infty }{\sum _{n=-\infty } }\bar{q}^{\frac{1}{2} \left(n+\frac{1}{2}\right)^2}\right)^4
= 16 + 64 \bar{q} + 96 \bar{q}^2 + O(\bar{q}^3)\,, \\
f_2(\bar{q}) &= \left(\overset{\infty }{\sum _{n=-\infty } }\bar{q}^{\frac{n^2}{2}}\right)^4+\left(\overset{\infty }{\sum _{n=-\infty } }(-1)^n \bar{q}^{\frac{n^2}{2}}\right)^4 = 2 + 48 \bar{q} + 48 \bar{q}^2 +  O(\bar{q}^3)\,,
\end{split}
\end{equation*}
and 
\begin{equation} \label{z}
 \begin{split}
  z_{v,s,c} & =  \frac1q \sum_{P\in \Gamma_{v,s,c}} \, q^{\frac{P_L^2}{2}} \, \bar{q}^{\frac{p_R^2}{2}}= \sum_{P \in \Gamma_{v,s,c}}  e^{ \pi i \tau_1 (\braket{Z|Z}-2)} e^{- \pi \tau_2 (P_L^2+p_R^2-2)}\,, \\
  z_{0} & =  \frac1{q\sqrt{\bar q}}  \sum_{P\in \Gamma_{o}} \, q^{\frac{P_L^2}{2}} \, \bar{q}^{\frac{p_R^2}{2}}= \sum_{P \in \Gamma_{0}} \, e^{ \pi i \tau_1 (\braket{Z|Z}-1)} e^{- \pi \tau_2 (P_L^2+p_R^2-3)}\,, 
  \end{split}
\end{equation}
where we used \eqref{innproduct}  for the difference $P_L^2-p_R^2$ that appears in the terms with $\tau_1$. 

To compute the one-loop cosmological constant \eqref{Lambda}, one has to integrate over the fundamental domain for $\tau$ (see Figure \ref{fig:sl2z}) where $|\tau|\ge 1$.  
As such, it makes sense to use the small $|q|=e^{-2\pi \tau_2}\equiv \epsilon$ expansion. Furthermore, from the sum over vectors in the lattice, the states with the smallest $|P|^2$ contribute most significantly. The lowest orders in the $\epsilon$ expansion of \eqref{Z} are 
\beq \label{expansionZ}
Z =  \left[  \left( c_{0}  +c_1 \epsilon  \right) (z_v-z_s-z_c)  +
 \left( c'_{0}  +c'_{1} \epsilon \right) z_0  \right] +{\cal O}(\epsilon^{2}) \, z \ ,
\eeq
where (ignoring the overall $1/\tau_2$) $c$ and $c'$ depend only on $\tau_1$ (and are independent of the point in moduli space). On the other hand, the sums in $z$ involve  semi-positive powers of $\epsilon$ if 
\beq
\begin{split} \label{finiteCC}
    P_L^2+p_R^2 \ge 2 \quad &\text{for} \ P\in \Gamma_{v,s,c}\,, \\
     P_L^2+p_R^2 \ge 3 \quad &\text{for} \ P\in \Gamma_{0}\, . \\
\end{split}
\eeq
Under these conditions, the contribution of {\it each state} to the cosmological constant is finite. 
These inequalities are precisely saturated by the massless bosons and fermions. On the other hand, the tachyons (in $\Gamma_0$) violate the second inequality (see \eqref{tachyons}) and thus the regions in moduli space where there are tachyons have a divergent one-loop cosmological constant.

Note however that there is a subtlety here: if the $\tau_1$ dependence of the terms that have an inverse power of $\epsilon$  is such that the integral over the yellow region of the fundamental domain in figure \ref{fig:sl2z} (where $\epsilon \to 0$) vanishes then the contribution of those states to the cosmological constant is finite, even when the conditions \eqref{finiteCC} are not met. One such state is the graviton, with $P=0$. The $\tau_1$ dependent parameters $c$ and $c'$ 
are explicitly
\begin{equation}
    \begin{split}
        c_{0}&= \frac{8}{\tau_2^{\frac{8-d}{2}}}  \, , \quad \quad
        c_{1}= \frac{64}{\tau_2^{\frac{8-d}{2}}} \,  (3 e^{2\pi i \tau_1}+ 2 \, e^{-2\pi i \tau_1})\, ,\\
        c'_{0} &= \frac{1}{\tau_2^{\frac{8-d}{2}}} \, ,   \quad \quad
        c'_{1} = \frac{12}{\tau_2^{\frac{8-d}{2}}} \,  (2 \, e^{2\pi i \tau_1} + 3\, e^{-2 i \pi \tau_1}) \ .
    \end{split}
\end{equation}
At the same time, since $\braket{Z|Z}$ is even for $Z \in \Gamma_{v,c,s}$ and odd for $Z \in \Gamma_0$, 
both $z_{v,c,s}$ and $z_0$ have integer powers of $e^{2\pi \tau_1}$. Therefore, the integral over the yellow region, with $\tau_1 \in [-1/2,1/2]$, will vanish unless the overall power is zero. States with $P=0$, which contribute to $z_v$ with a power $1/\epsilon$ have $\braket{Z|Z}=0$, which implies that the overall power of the term with $c_0$ is non-zero, and thus they do not give a divergent contribution to the cosmological constant.

Additionally because $c_0$ is independent of $\tau_1$, the only possible states leading to a divergent contribution are those with $\braket{Z|Z}=2$. On the other hand, $|P|^2=\braket{Z|Z}+2 p_R^2\ge \braket{Z|Z}$, and therefore these states satisfy the inequality \eqref{finiteCC} and are therefore not dangerous. We conclude that none of the states in $\Gamma_{v,c,s}$  gives an infinite contribution to the cosmological constant. This is not the case for states in $\Gamma_0$: out of the states that violate the inequality \eqref{finiteCC}, only the $\braket{Z|Z}=1$ are dangerous. This is precisely the situation for the tachyons but not any other state. Thus the $\tau_1$ dependence does not save the tachyons from giving an infinite contribution to the cosmological constant. On the other hand, by the same argument as above, tachyons are the only states in $\Gamma_0$ that do give a divergent contribution. We conclude that the only states in $\Upsilon_{d+16,d}$ that give a divergent contribution to the cosmological constant are the tachyons.

 So far we have talked  about the individual contribution of each state to the cosmological constant. However, the sum over all states in a given sector like say, $\Gamma_v$, could a priori diverge because the number of states grows exponentially with $|P|^2$. This is in fact not the case for all the points of maximal enhancement in $S^1$ compactifications. Furthermore, we found that the cosmological constant is {\it always finite} in any region in moduli space where there are no tachyons. We expect this to be the case for all $T^d$ compactifications because of the difference in sign from the boson and fermion contributions. 

We will now explore compactification on a circle in more detail.

\subsection{Cosmological constant in \texorpdfstring{$S^1$}{S1} compactifications}\label{ss:s1}

As mentioned in the previous section, in the absence of tachyons the largest contribution of individual states to the cosmological constant comes from massless states. Even though the contribution from all other states can be larger than that of the massless states as is the case of, for example, the point with symmetry enhancement to $E_6 \times SU(12)$, it is instructive to compute the contribution of the massless terms to the partition function, which we can do for $T^d$ compactifications for any $d$, and to the cosmological constant for the specific case $d=1$.

The massless states comprise states with $P=0$, which are massless at any point in moduli space, and states with $P\neq 0$ which become massless at particular points in moduli space. The former include the metric, B-field, dilaton, gauge bosons of $U(1)^{16+2d}$. The latter include the ten-dimensional massless fermions charged under the gauge group which acquire a mass at a generic point in moduli space. The contribution of the $P=0$ bosons to the partition function can be obtained from eqs. \eqref{z}-\eqref{expansionZ}, and is given by 
\begin{equation}
Z_{P=0}=  \frac{8}{\tau_2^{\frac{8-d}{2}}} \left[e^{-2 \pi i \tau_1}   e^{2 \pi \tau_2}+ 8(3+ 2 e^{-4 \pi i \tau_1}) \right] + {\cal O}(e^{-2\pi \tau_2}) \, .
\end{equation}
The extra massless states that appear at points of symmetry enhancement  are the vectors and spinors with 
$p_R=0$, $P_L^2=\braket{Z|Z}=2$, the scalars with $P_L^2+p_R^2=3$, $\braket{Z|Z}=1$, plus the right-moving vectors with $P_L=0$, $p_R^2=-\braket{Z|Z}=1$. Their contribution to the partition function, which we will call $Z_{\rm{enh}}$, is
\begin{align}
Z_{\rm{enh}} = &\frac1{\tau_2^{\frac{8-d}2}} \Bigg[ 8 \, (N_v-N_s-N_c) +  N_0 +
\left[ e^{-2 \pi i \tau_1}   e^{2 \pi \tau_2} +  12 (2+ 3e^{-4\pi i \tau_1})  \right]
N_0' 
\Bigg] \\  & + {\cal O}(e^{-\pi \tau_2}) \, , \nn
\end{align}
where $N_v, N_s, N_c, N_0$ and $N_0'$ are  the number of massless vectors, spinors in $s$ and $c$ representations, scalars and right-moving vectors, respectively.\footnote{We use the subscript $0$ in $N_{0'}$ because they are scalars from the left-moving point of view and belong to $\Gamma_0$.}

The total partition function is then
\beq
Z=Z_{P=0}+ Z_{\rm{enh}} + \sum {\cal O}(e^{-\pi \tau_2}) \ ,
\eeq
where we emphasized the fact that there is a sum of subleading terms which at the end of the day can give a contribution larger than the contribution from the massless states. 

The integral of $Z$ over the fundamental region depends on the dimension.
For $S^1$ compactifications we evaluate it numerically, and find that the massless states generate a one-loop cosmological constant given by:
\begin{equation} \label{Lambdamassless}
    (4 \pi^2 \alpha')^{\frac{9}{2}} \,  \Lambda_{m=0} = 28.1 + 1.1 \, (N_s + N_c - N_v ) - 0.1 N_0 + 3.4 N_0' + {\cal O}(10^{-4}) (N_f-N_b) \, .
\end{equation}
Here $N_f-N_b$ denotes schematically the number of fermions minus the number of bosons.
These numbers, as well as the contribution to the cosmological constant from the massless sector, are recorded in Table \ref{tab:CCmassless} for the eight points of maximal enhancement without tachyons. In some cases, like the enhancement to $A_1+2D_8$, this approximates the total cosmological constant reasonably well. In other cases, it does not provide a good approximation at all. In particular, for $A_{11}+E_6$, where there are no massless fermions, this contribution is negative while the true cosmological constant is positive, as we discuss below.

\begin{table}[H] 
	\begin{center}
	\renewcommand{\arraystretch}{1.2}
		\begin{tabular}
{|>{$}l<{$}|>{$}c<{$}|>{$}c<{$}|>{$}c<{$}|>{$}c<{$}|>{$}c<{$}|>{$}c<{$}|>{$}c<{$}|}
\hline
& N_v & N_s & N_c & N_0 & N'_0 & \Lambda _{m=0} & \Lambda \\ \hline
 \text{A}_1+2 \text{D}_8 & 226 & 256 & 256 & 0 & 0 & 341.6 & 431.4 \\ 
  \text{A}_1+\text{A}_2+\text{D}_6+\text{D}_8 & 180 & 192 & 192 & 0 & 0 & 251.8 & 383.5\\
 \text{A}_4+\text{D}_5+\text{D}_8 & 172 & 128 & 160 & 0 & 0 & 155.4 & 359.2\\
 \text{A}_7+2 \text{D}_5 & 136 & 0 & 170 & 0 & 0 & 65.6 & 303.8 \\
 \text{D}_8+\text{D}_9 & 256 & 128 & 288 & 256 & 0 & 168.7 & 305.0 \\
 \text{D}_4+\text{D}_5+\text{D}_8 & 176 & 208 & 128 & 256 & 0 & 168.7 & 305.0 \\
 2 \text{A}_1+\text{A}_3+2 \text{D}_6 & 136 & 128 & 168 & 256 & 0 & 168.7 & 305.0 \\
 \text{A}_{11}+\text{E}_6 & 204 & 0 & 0 & 408 & 2 & -243.7 & 180.4\\ \hline
\end{tabular}
		\caption{Cosmological constant for the eight tachyon-free maximal enhancements. We additionally give the number of massless states (excluding the graviton) and their contribution to the cosmological constant.\protect\footnotemark \ } 
	\label{tab:CCmassless}
 \end{center}\end{table}
\footnotetext{For clarity we give only the first decimal digit of the cosmological constant but we have actually computed it up to ${\cal O}(10^{-4})$.}
Given that the contribution from light states can completely change the value of the cosmological constant from the value computed using massless states only, one is forced to include many more states in the calculation, and see where one can safely truncate the sum. 
We computed the cosmological constant for $S^1$ compactifications numerically at the eight points of maximal enhancement without tachyons, and in the vicinity of these special points. The values at the points of maximal enhancement are recorded in Table \ref{tab:CCandHessians}. The cosmological constant is always positive at these points. It is surprising that for the $E_6 \times SU(12)$ enhancement, where there are no massless fermions, the cosmological constant is nevertheless positive. The reason for this is that there are very light fermions that give a very large positive contribution, uplifting the negative cosmological constant of the bosonic massless states. Since this goes against the common lore that ``the cosmological constant goes like the number of massless fermions minus the number of massless bosons," which is true close to the decompactification limits where supersymmetry is restored, but not necessarily in the bulk, we work out this example in detail in Appendix \ref{App:lightfermions}.

Our results also show that out of the eight extremal points shown in Table \ref{tab:CCmassless}, three have exactly the same value of the cosmological constant at one loop. Even though the numbers of bosons and fermions at each mass level are different, it seems that their overall partition functions coincide so they should have the same cosmological constant at all loops. Proving this statement is outside of the scope of the present work, but it would be interesting to do it. As we will see in Section \ref{ss:nonmax}, there are two more extrema of the cosmological constant with this same value (although they are not maximal enhancements by our definition).

 \subsubsection{Plots of the cosmological constant}
\label{sec:pictures}

Here we present some plots of the cosmological constant for two-dimensional slices of the seventeen-dimensional moduli space, given either by the radius and one parameter of the Wilson line, or two parameters of the Wilson lines keeping $R^2+\tfrac12 A^2$ fixed  at the value $E=1$ where there are maximal enhancements.\footnote{Maximal enhancements at $E\neq 1$ can be found from the EDD whenever the central node is deleted. However, all of them are dual to another point in the fundamental region with $E=1$.} 

There are many interesting features to notice about these plots. The most obvious ones are the reflective symmetries at the maximal enhancement points, and the fact that they are extrema of the cosmological constant. 

\begin{figure}[H]
 \centering\includegraphics[align=c,height=0.55\textwidth]{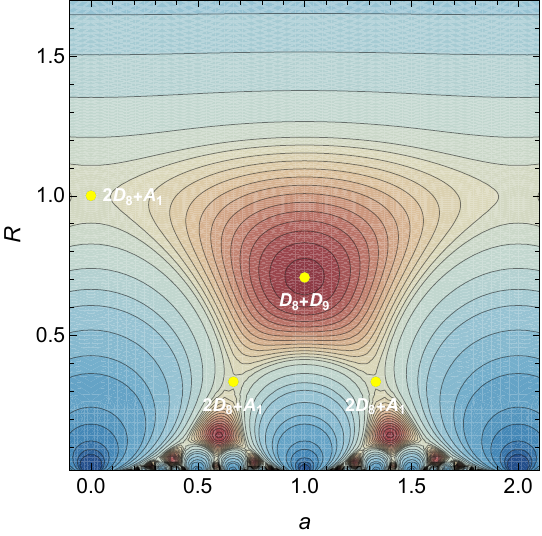}\hspace{1em}\includegraphics[align=c,height=0.45\textwidth]{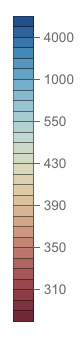}
\caption{Cosmological constant of the slice of moduli space given by the radius $R$ and Wilson line of the form $A = \left(0^{15},a\right)$. In this slice we find the enhancements $2D_8+A_1$ and $D_8+D_9$. }\label{fig:a015}
 \end{figure}

In figure \ref{fig:a015}, with one component of the Wilson line, we see the maximal enhancements $2D_8+A_1$ and $D_8+D_9$. The former appears at different points, $a=0$ and $R=1$, the standard one, and at dual points, which are in different fundamental regions. This plot already shows a surprise: while the point $2D_8 + A_1$ (i.e. with gauge symmetry $SO(16)\times SO(16) \times SU(2)$)
is a minimum with respect to the radial direction, it is a maximum if we move in the direction of the Wilson line. Therefore this point is not a local minimum, but a saddle point, contradicting the claim in  \cite{Ginsparg:1986wr} (in agreement on the other hand with the recent results of \cite{Koga:2022qch}). 

\begin{figure}[H]
 \centering
\includegraphics[align=c,height=0.55\textwidth]{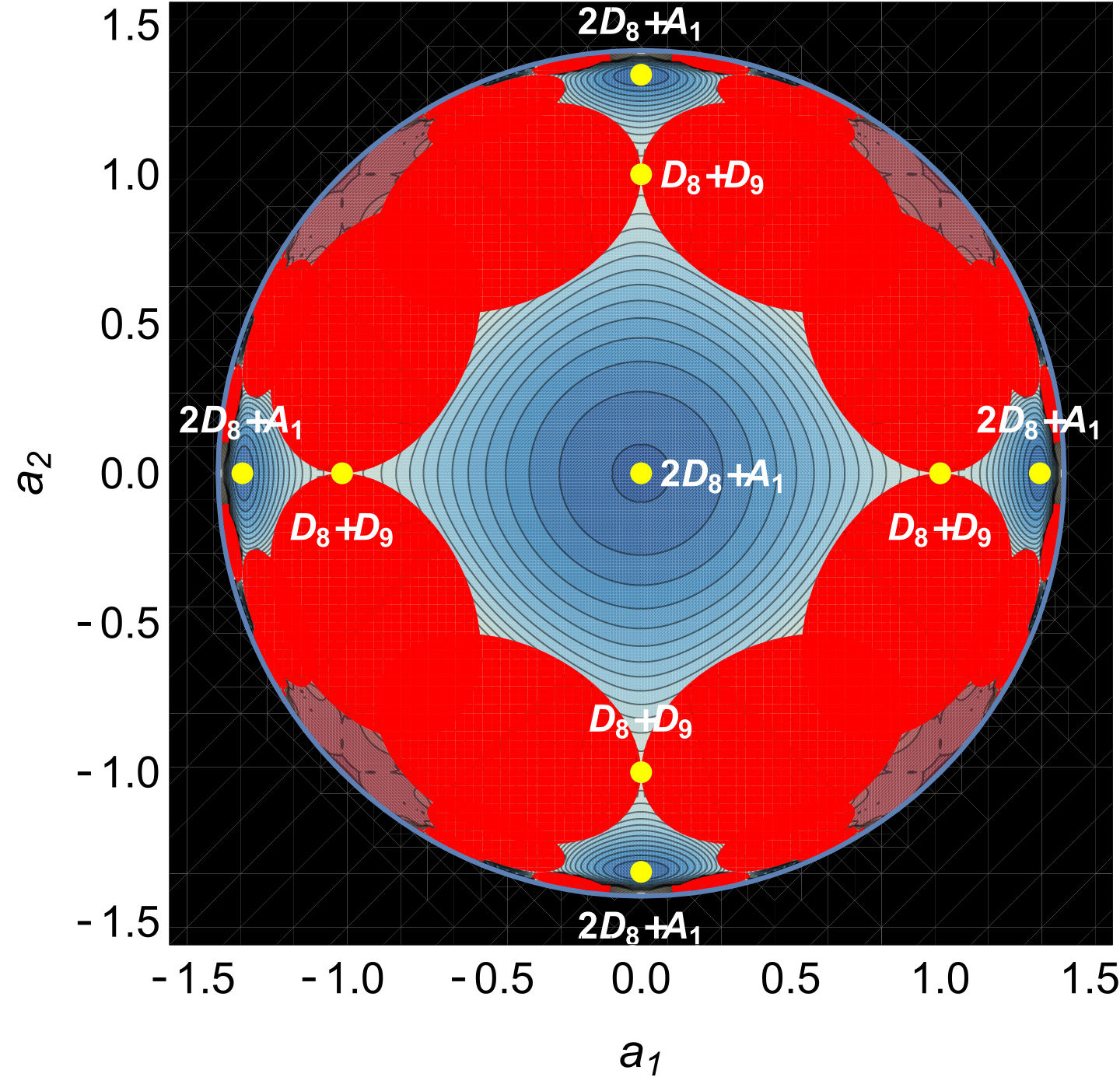}\hspace{1em}\includegraphics[align=c,height=0.4\textwidth]{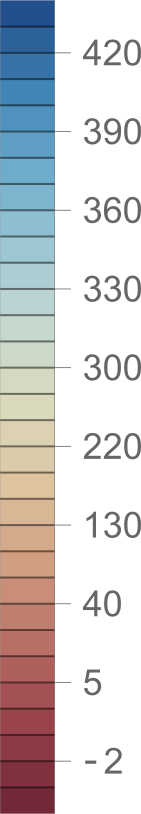}
 
 \caption{Cosmological constant of the slice of moduli space given by the radius \\ $R^2 = 1 - \tfrac12 A^2$ and Wilson line of the form $A = \left(0^{7},a_1,0^{7},a_2\right)$. In this slice we find the enhancements $2D_8+A_1$ and $D_8+D_9$.}\label{fig:a107a207}
 \end{figure}

\begin{figure}[H]
 \centering
\includegraphics[align=c,height=0.60\textwidth]{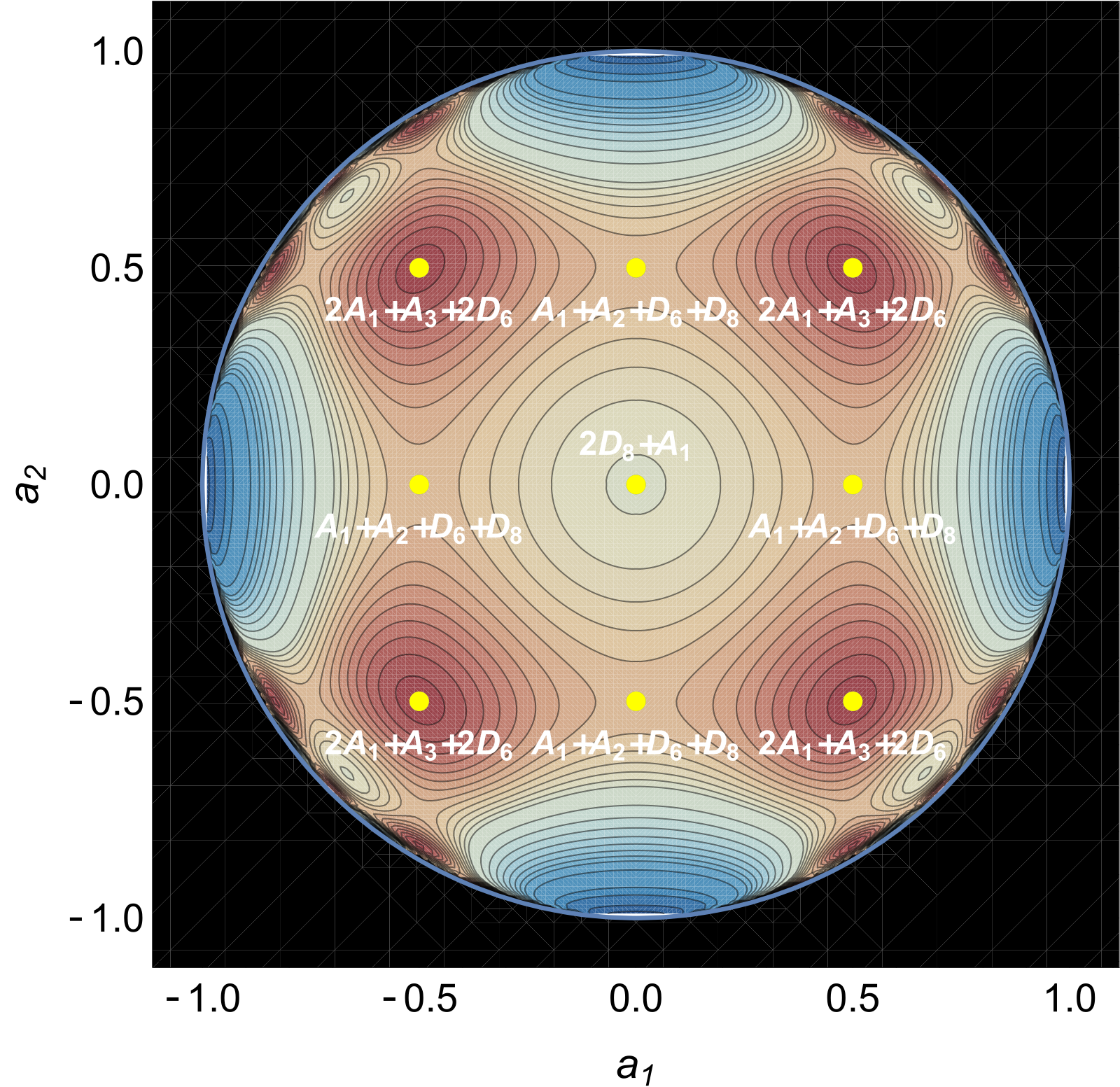}\hspace{1em}\includegraphics[align=c,height=0.4\textwidth]{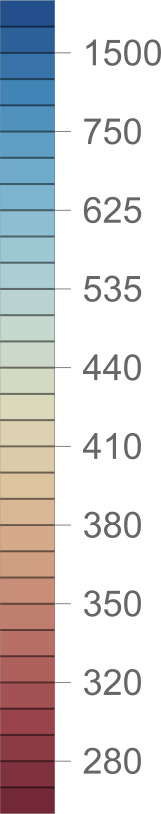}
 \caption{Cosmological constant of the slice of moduli space given by the radius \\ $R^2 = 1 - \tfrac12 A^2$ and Wilson line of the form $A = \left(0^{6},a_1^2,0^{6},a_2^2\right)$. In this slice we find the enhancements $2D_8+A_1$, $A_1+A_2+D_6+D_8$ and $2A_1+A_3+2D_6$.}\label{fig:a1206a2206}
 \end{figure}

On the other hand, the enhancement to $D_8+D_9$ in figure \ref{fig:a015}, as well as the one to $D_8+D_6+A_2+A_1$ in plot \ref{fig:a1206a2206}, appear as  local minima. We will see in the next section whether this feature persists to the whole moduli space, though we can already guess that these enhancements cannot be local minima. The fact that at these points there are massless scalars in $\Gamma_0$ (the ones we called $N_0$) implies that tachyons appear when moving infinitesimally away in certain directions, as shown in section \ref{sec:knifeedges}, and therefore these are knife edges. The knife-edge property of the first enhancement is illustrated in figure \ref{fig:a107a207}, where it is mostly surrounded by tachyons with the tachyonic regions depicted in red. In the tachyonic regions the cosmological constant is dominated by the scalars in $\Gamma_0$ that become massless at their boundaries, and slightly massive nearby. In figure \ref{fig:a107a207} we can see that tachyons appear when moving in the $a_1$ direction (where we break the $SO(16)$ symmetry), while when moving in the $a_2$ direction, corresponding to breaking the $O(18)$ symmetry, we get  a positive cosmological constant. 

\begin{figure}[H]
 \centering\includegraphics[align=c,height=0.55\textwidth]{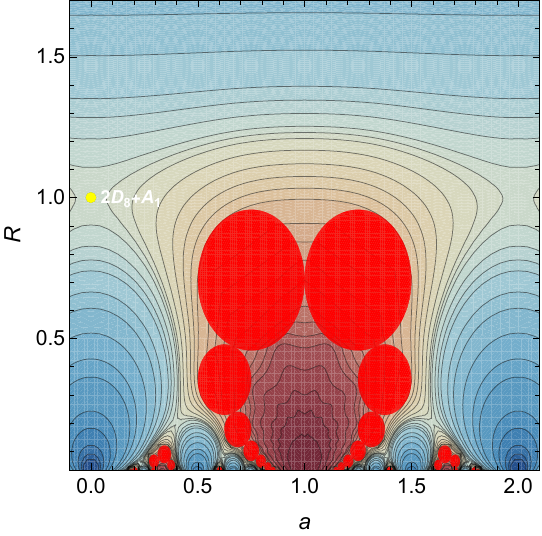}
\hspace{1em}\includegraphics[align=c,height=0.45\textwidth]{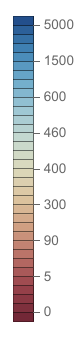}
 \caption{Cosmological constant of the slice of moduli space given by the radius $R$ and Wilson line of the form $A = \left(0^{7},a,0^{7},a\right)$. In this slice we find the enhancement $2D_8+A_1$.}\label{fig:a07a07}
 \end{figure}

On the other hand, the slice chosen in figure \ref{fig:a07a07}, where $D_8+A_1$ also appears as a local maximum, shows a tachyonic region closeby (but at a finite distance). 
This figure also shows that one can pass from negative to positive cosmological constant avoiding tachyons. 

All plots contain points that are decompactification limits (or rather, they are at $R=0$ which is T-dual to decompactification points). Figures \ref{fig:a07a07} and \ref{fig:a107a207} contain, respectively, the infinite distance points   $R=0$, $a=1$ and $a_1=a_2=1$ which happens at $R=0$. When approaching these points the 
quantity $\Lambda \times R$, which one identifies as the ten-dimensional cosmological constant, tends to zero in agreement with the fact that we recover the ten-dimensional supersymmetric $SO(32)$ theory. Recall there are other infinite distance limits that correspond to ten-dimensional non-supersymmetric heterotic theories, see discussion around figure \ref{fig:infdist}. Close to the points of supersymmetry restoration we get a very small cosmological constant which seems negative when approaching the limit at constant Wilson line $a=1$ in figure \ref{fig:a07a07} and also in figure \ref{fig:a107a207}, where the infinite distance limit is taken by keeping $E=1$ fixed. In the latter case, when moving away from the decompactification point, one always reaches a tachyonic region where the cosmological constant is  $-\infty$; after which it becomes positive. It would be very interesting to see if all the regions of negative cosmological constant are of this type, i.e. if the magnitude of the cosmological constant is very small and the regions  are ``close" to a decompactification point. 

In the decompactification limits with zero Wilson lines, we recover the cosmological constant of the ten-dimensional $O(16) \times O(16)$
theory after multiplying (dividing) $\Lambda$ by $R$ for small (big) radius. With generic Wilson lines and outside of tachyonic regions, the cosmological constant tends to $+\infty$. It would be interesting to see what happens with $\Lambda \times R$ in this case.

 We leave for future work a detailed analysis of the behavior of the cosmological constant close to decompactification limits, as well in the regions where it is negative.

\subsection{Maximal enhancements: maxima, minima or saddle points?}\label{sec:maximalextrema}
\label{maximal-saddle?}

We now want to determine if the extrema corresponding to the eight points of maximal enhancement without tachyons are maxima, minima or saddle points. To this end, we compute the Hessian of the cosmological constant at these points.

To compute the Hessian, it is convenient to write the sums \eqref{z} appearing in the 
cosmological constant solely in terms of $p_R$  as follows, 
\begin{equation}
\begin{split}
 z_{v,s,c} &= \sum_{Z \in \Gamma_{v,s,c}} \, e^{\pi i \tau_1 (\braket{Z|Z}-2)} e^{- \pi \tau_2 (2p_R^2+\braket{Z|Z}-2) }\, , \\ 
 z_{0} &= \sum_{Z \in \Gamma_{0}} \, e^{\pi i \tau_1 (\braket{Z|Z}-1)} e^{- \pi \tau_2 (2p_R^2+\braket{Z|Z}-3) } \, ,
   \end{split}
\end{equation}
 where we used \eqref{innproduct} to solve for $P_L$. This expression is easier to handle when taking derivatives with respect to the moduli because all the moduli-dependence is encoded in the norm of the $d$-dimensional vector $p_R$, which we 
rewrite here for the $d=1$ case that we will analyse:
\beq
p_R^2= \frac{1}{2R^2} \left(n-E\, w - \pi \cdot A \right)^2=\frac{1}{2R^2} \left(n-(R^2+ \frac12 A^2)\, w - \pi \cdot A \right)^2 \ .
\eeq

\subsubsection{Extremal points}\label{sec:extremal} 

The fact that maximal enhancements extremize the cosmological constant was proven in \cite{Ginsparg:1986wr}, but for what follows it is instructive to rederive these results in a  more explicit way. 

Recall all the moduli-dependence is encoded in the sums $z_{v,c,0}$ defined in \eqref{z}, so it is enough to show that the gradient of these sums with respect to the moduli vanishes. This gradient is given by
\begin{equation} \label{nablaz}
    \nabla z= g(\tau) \sum_{Z}  e^{\pi i \tau \braket{Z|Z}} e^{-2 \pi \tau_2 p_R^2} \, \, p_R \nabla p_R  
\end{equation}
where $g(\tau)$ are functions independent of the  moduli and $Z$. These functions will not play a role in this section, but we give them for future reference 
\beq \label{g}
g_{v,s,c}(\tau)=-4 \pi \tau_2 e^{-2\pi i \tau} \, , \qquad g_{0}(\tau)=-4 \pi \tau_2 e^{-\pi i \tau} e^{2\pi \tau_2}.
\eeq
We combine the moduli in a ket
\begin{equation}
    \ket{{\cal E}}= (-1,E,A)
\end{equation}
with the same $O(17,1)$ inner product as $\ket{Z}$, given in \eqref{innproduct}.\footnote{Actually this argument is for generic $d$, but to avoid confusion with $d$ indices we just write it for the circle.} Note that 
\begin{equation}
    -\braket{{\cal E}|{\cal E}}\equiv |\CE|^2 = 2 R^2 \ , \quad  \braket{{\cal E}|Z}=n-E\, w - \pi \cdot A=  |\CE| \,  p_R \, .
\end{equation}
Defining
\begin{equation}
    \ket{\nabla} = \frac{\partial}{\partial \bra{\CE}}=\left(\partial_{\cancel{E}},\partial_E,\partial_{A^1},...,\partial_{A^{16}} \right)
\end{equation}
where we have defined a spurious coordinate $\cancel{E}$ in the kets : $ \ket{{\cal E}}= (\cancel{E},E,A)$, which is set to $-1$. After some manipulation, we can show that\footnote{Note that the matrix ${\ket{\CE}\bra{\CE}}$ is computed using the $O(17,1)$ invariant, namely  
\begin{equation} \label{CECE}
    {\ket{\CE}\bra{\CE}}=\left(\begin{matrix}   \cancel{E} E  &\cancel{E}^2 & \cancel{E} A \\ E^2  &  E \cancel{E} & {E} A \\ A^T {E} & A^T \cancel{E} & A^T A \\ \end{matrix} \right) \, .
\end{equation}}
\bea 
\bra{\nabla p_R}  = \frac1{|\CE|}  \bra{Z} \left(\mathbb{I}+ \frac1{|\CE|^2}    {\ket{\CE}\bra{\CE}} \right) \ .
\eea 
Points of maximal enhancement have $E=1$ and rational Wilson lines, and therefore the moduli vector $\ket{E}$ can be embedded in the charge lattice by  rescaling by an integer $n$. One can then use this lattice vector $n \ket{\CE}$ to perform a Weyl reflection and get another lattice vector
\bea \label{WeylnE}
\ket{Z'} = \ket{Z} - 2 \tfrac{n\braket{\CE|Z}}{n^2\braket{\CE|\CE}} n\ket{\CE} = \ket{Z} - 2 \tfrac{\braket{\CE|Z}}{\braket{\CE|\CE}}\ket{\CE} \ .
\eea 
The reflected vector satisfies
\begin{equation} \label{reflected}
    \braket{Z'|Z'}=\braket{Z|Z} \ , \quad \braket{\CE|Z'}=-\braket{\CE|Z}\, \Rightarrow \, p_R'=-p_R \ , \quad \ket{\nabla p'_R}=\ket{\nabla p'_R} \ .
\end{equation}
Looking back at $\nabla z$ given in \eqref{nablaz}, we see that the contribution from $Z'$ cancels that of $Z$, and thus we conclude $\nabla z=0$, which means 
\beq
\nabla {\Lambda}|_{\rm max \, enh}=0 \ .
\eeq

 Note
that in the proof we did not need to perform the integral over the fundamental domain of the genus one surface that gives the one-loop contribution. This implies that the fact that maximal enhancements are extremal is true at all loops, as already pointed out in \cite{Ginsparg:1986wr}. Also note that in this proof we have restricted to reflections in the T-duality group, but as shown in \cite{Ginsparg:1986wr}, the proof extends beyond reflections to also include outer automorphisms. There is, for example, a point in the moduli space of $T^8$ compactifications associated to the Leech lattice which is fully stabilized by outer automorphisms; later in this section, we will give some examples involving outer automorphisms in $S^1$ compactifications.

\subsubsection{Hessian} \label{sec:hessian}

The Hessian of the cosmological constant is given by the matrix $\ket{\nabla}\bra{\nabla} \Lambda$.
As pointed out before, the moduli dependence of the cosmological constant appears only in $z$. Eq. \eqref{nablaz} says that 
\begin{equation}
    \nana z =g(\tau) \sum_Z e^{\pi i \tau \braket{Z|Z}} \ket{\nabla} \left(e^{-2 \pi \tau_2 p_R^2} \, \, p_R \bra{\nabla p_R}  \right) \ ,
\end{equation}
where 
\begin{align}\label{nanazterms}
\ket{\nabla} \left(e^{-2 \pi  \tau_2 p_R^2}  p_R\bra{\nabla p_R} \right)  
= 
e^{-2 \pi  \tau_2 p_R^2} \left((1-4 \pi  \tau_2 p_R^2)
 \ket{\nabla p_R}
\bra{\nabla p_R}  + 
  p_R \ket{\nabla} \bra{\nabla} p_R  \right) \ . 
\end{align}  
Using $\ket{\nabla} (\braket{\CE|\CE}) = 2 \ket{\CE}$ we compute the second term:
\beq
p_R \ket{\nabla} \bra{\nabla} p_R  = \frac{p_R}{|\CE|^2}\left[
  \tfrac{ 
\ket{\CE}\bra{Z}+  \ket{Z}\bra{\CE}}{{|\CE|}}
+ p_R  \left(\mathbb{I} + 3\tfrac{\ket{\CE}\bra{\CE}}{|\CE|^2}\right)\right] \ .
\eeq 
 
The first term in \eqref{nanazterms} gives
\beq
\ket{\nabla p_R}\bra{\nabla p_R}
= \frac{1}{|\CE|^2} \Big[
\ket{Z}\bra{Z}
+ p_R \tfrac{\ket{\CE}\bra{Z} + \ket{Z}\bra{\CE}}{|\CE|}
+ p_R^2 \tfrac{\ket{\CE}\bra{\CE}}{|\CE|^2}
\Big] \ .
\eeq

Putting all the pieces together we get\footnote{Recall that $\cancel{E}$ is a spurious coordinate, so in these matrices we have to remove the first column and second row (see eq. \eqref{CECE}), such that the Hessian is a 17 $\times$ 17 matrix given by 
\begin{equation} \label{Hessmatrix}
{\ket{\nabla}\bra{\nabla}}=\left(\begin{matrix}   \nabla_E^2 & \nabla_E \nabla_A \\  \\ \nabla_A^T  \nabla_{E} & (\nabla^T  \nabla)_{A} \\ \end{matrix} \right)\,.
\end{equation}
}
\begin{align} \label{HessianmatrixLambda}
    \nana z= & \frac{g(\tau)}{2R^2} \sum_Z e^{\pi i \tau \braket{Z|Z}} e^{-2 \pi \tau_2 p_R^2} \, \times  
    \Big[(1-4 \pi  \tau_2 p_R^2)
\ket{Z} \bra{Z} +
p_R^2\,  \mathbb{I}
+ \frac{2 \pi  \tau_2 p_R^4}{R^2}\,  \ket{\CE}\bra{\CE} + y \Big] 
\,,
\end{align}
with the sum over $y=(2 - 4 \pi \tau_2 p_R^2)p_R \left(
\ket{\CE}\bra{\nabla p_R} +   
\ket{\nabla p_R}\bra{\CE}\right)$ vanishing at points of maximal enhancement.

Therefore the Hessian of the cosmological constant is,
\footnote{\label{foot:Hessianknife} Note that here we are computing the Hessian of the partition function and then doing the integral over the fundamental domain. These two operations commute if the integral of the partition function converges, which is not the case in the presence of (nearby) tachyons. }
\begin{align}
    (4 \pi^2 \alpha')^{\frac{9}{2}} \,  \nana \Lambda= \int_{F_0} \frac{d^2 \tau}{4 
\tau_2^{\frac{12-d}{2}} 
\eta_*^{24} \bar{\eta}_*^{12}}\, \Big[ &
f_1(\bar q) \, \nana (z_s+z_c-z_v)
 - f_2(\bar q) \, \nana z_0\Big] \, .\nn
\end{align}
As for the cosmological constant itself, we see again that terms with increasing $p_R$ contribute less significantly to its Hessian, which is computed using a power expansion in $q$. We will illustrate this explicitly in the next subsection for the simplest point of maximal enhancement.

\subsubsection{Example $SO(16)\times SO(16) \times SU(2)$}

We illustrate the computation of the Hessian with the simplest example of maximal enhancement to $2D_{8}+A_1$ at $A=0$, $E=R=1$. We have
\begin{equation*}
    \nana z= \frac{g(\tau)}{2} \sum_{(w,n,\pi) \in \Gamma} e^{\pi i \tau (2w n+\pi^2)} e^{-2 \pi \tau_2 p_R^2} \Big[(1-4 \pi  \tau_2 p_R^2)
\left(\begin{smallmatrix} 
 w^2 & 0  \\
 0 & \pi^T\pi
\end{smallmatrix}\right) +
p_R^2\,  \mathbb{I}
+ \frac{2 \pi  \tau_2 p_R^4}{R^2}\,  \left(\begin{smallmatrix} 
 1 & 0  \\
 0 & 0 
\end{smallmatrix}\right) \Big] \, ,
\end{equation*}
with $p_R^2=\tfrac12 (w-n)^2$. Here we have used that in the off-diagonal terms of $\ket{Z} \bra{Z}$, equal to $w \pi$, the contribution from a vector with a given $w$ cancels with that of  opposite $w$ and same $\pi$.  The sum over vectors in the lattice can be split into a sum over $\pi\in \Gamma_{v,s,c,0} $ (we refer here to the sixteen-dimensional part only), and a sum over $(w,n) \in \Gamma_{1,1}$. For the former, we define the theta series of a lattice $\Gamma$ by
\bea 
\theta_{\Gamma}(q) \equiv \sum_{\pi \in \Gamma} q^{\frac{\pi^2}{2}} \ . 
\eea 
In \eqref{Thetas}, we give the leading terms of the $q$ series expansion for the various $\theta$ series for the 16-dimensional part of $\Gamma_{v,c,s,0}$ defined in \eqref{conjdef}.  
For computing the terms with $\pi^T \pi$, we use the following observation: for lattices symmetric both under permutation of directions and under reflections on any pair of directions,  one has 
\beq
\begin{split}
\sum_{\pi \in \Gamma \, {\rm with} \, \pi^2=\pi_0^2}  \pi^T \pi & = \frac{N }{r} \, \pi_0^2 \, \mathbb{I}_{r} \, , \\
\sum_{\pi \in \Gamma \, {\rm with} \, \pi^2=\pi_0^2} \pi^T \pi \, q^{\frac{\pi^2}{2}}  &= \frac{N }{r}\,  \pi_0^2 \, q^{\frac{\pi_0^2}{2}} \, \mathbb{I}_{r} = \frac{2N }{r}\,  q \frac{\partial}{\partial q}  q^{\frac{\pi_0^2}{2}} \, \mathbb{I}_{r} \, ,
\end{split}
\eeq
where $N$ is the number of terms in the sum, and $r$ is the rank of the lattice. These relations imply
\bea 
\sum_{\pi \in \Gamma} \pi^T \pi q^{\frac{\pi^2}{2}} = q\frac{\mathbb{I}_{16}}{8}    \frac{\partial \theta_{\Gamma}(q)}{\partial \text{log}(q)} \ .
\eea 
For the sums over momenta and winding we define the functions
\beq \label{sigma}
\sigma_k(q,\bar q) \equiv \sum_{w,n} w^{2k}{\bar q}^{\frac{(w-n)^2}{4}} q^{\frac{(w+n)^2}{4}} \, .
\eeq
The power series expansions for the cases $k=0,1$ that appear in the Hessian are given in \eqref{powersigma}.  In terms of $\sigma_k$, we see that
\begin{equation}
\begin{split}    
    \nana z_{v,s,c,0}= \frac{g_{v,s,c,0}(q,\bar q)}{2} \,   \Bigg[&\tfrac18\left(\begin{smallmatrix} 
 0 & 0 \\
 0 & \mathbb{I}_{16}
\end{smallmatrix}\right)
\tfrac{\partial \theta_{v,s,c,0}}{\partial \text{log}(q)}
(1- 8 \pi  \tau_2 \tfrac{\partial}{\partial \text{Log}(\bar{q})})
\sigma_0(q,\bar q)
 \nonumber\\ 
+ \theta_{v,s,c,0} 
\bigg[ 
\left(\begin{smallmatrix} 
 1 & 0 \\
 0 & 0 
\end{smallmatrix}\right)
(1- 8 \pi  \tau_2 \tfrac{\partial}{\partial \text{log}(\bar{q})})
\sigma_1(q,\bar q)  
&+ 
\left(
2\left(\begin{smallmatrix} 
 0 & 0 \\
 0 & \mathbb{I}_{16} 
\end{smallmatrix}\right)
\tfrac{\partial}{\partial \text{log}(\bar{q})}
+  
8 \pi  \tau_2
\left(\begin{smallmatrix} 
 1 & 0  \\
 0 & 0 
\end{smallmatrix}\right)
\tfrac{\partial^2}{\partial \text{log}(\bar{q})^2}
\right)
\sigma_0(q,\bar q)
\bigg] \Bigg] \, , \label{hess1}
\end{split}
\end{equation}
where we have replaced $e^{-2\pi \tau_2 p_R^2}=(q \bar q)^{\tfrac12 p_R^2}$, and  $p_R^2 \,e^{-2\pi \tau_2 p_R^2}=2\frac{\partial}{\partial \text{log}(\bar q)} e^{-2\pi \tau_2 p_R^2}$, and we used the explicit form of the function $g(\tau)$ given in \eqref{g}.

Inserting the power expansion of this result into \eqref{HessianmatrixLambda} and performing  the integral over the fundamental domain numerically, we find
\bea
H_{\Lambda} \simeq 
(4 \pi^2 \alpha')^{-\frac{9}{2}}\left( \begin{matrix}
  416 & 0 \\
  0 & -306 \, \mathbb{I}_{16}
  \end{matrix}\right) \, .
\eea
Surprisingly, this extremum is not a local minimum, as one might reasonably have thought, but is instead a saddle point. In this particularly simple case, where the cosmological constant is most significantly sourced by the massless states, one can get some intuition about what is going on. 

When moving in the radial direction, while keeping the Wilson line turned off, only the $SU(2)$ bosons get a mass, and none of the spinors. Then $N_v$ decreases while $N_s$ and $N_c$ are constant, and   
therefore \eqref{Lambdamassless} says that the cosmological constant increases, compatible with the Hessian in the radial direction being positive. On the other hand, when turning on a Wilson line that breaks all of the $SO(16)\times SO(16)$ symmetry, all bosons except the graviton, B-field and dilaton become massive, and thus the first term in \eqref{Lambdamassless} stays the same, while the next term decreases, and thus the cosmological constant decreases. 

\subsubsection{Hessian for all maximal enhancement points}
\label{sec:Hessianmaxenh}

Computing the Hessian of the cosmological constant for other points of maximal enhancement is more cumbersome. We will not present the technical details here, but we just give in Table \ref{tab:CCandHessians} the eigenvalues of the Hessian at all points of maximal enhancement without tachyons.

As we can see from that table, {\it none} of the points of maximal enhancement is a local {\it minumum}. 

We initially concentrate on the first four enhancements, since, as discussed earlier, the last four are knife edges. 
The first three points are saddle points, while the fourth one is a local maximum. All three saddle points are minima in the radial direction, while they are a maximum in most of the directions involving varying the Wilson line.

With regard to the last four points, one should take the Hessian  in some of the directions in moduli space with a grain of salt: at these points there are no tachyons, but since they have $N_0\neq 0$ massless scalars, tachyons appear when moving away from those critical points in specific directions (see section \ref{sec:knifeedges}), and the cosmological constant in those regions jumps to $- \infty$  (the minus sign is because the leading order divergent term for tachyons has a negative sign). They are therefore knife edges.

The reason why we see a finite value for the Hessian in all directions for these knife edges is that in the directions where tachyons arise, the cosmological constant is not a continuous function, and thus its derivatives are ill-defined. What we are computing is the integral over the fundamental domain of the Hessian of the partition function. We expect this to coincide with the Hessian of the cosmological constant only when the integral of the partition function converges (see footnote \ref{foot:Hessianknife}).

\begin{table}[H] 
	\begin{center}\renewcommand{\arraystretch}{1.2}
		\begin{tabular}
{|@{\hskip 0.1cm}>{$}l<{$}@{\hskip 0.1cm}|@{\hskip 0.1cm}>{$}c<{$}@{\hskip 0.1cm}|@{\hskip 0.1cm}>{$}r<{$}@{\hskip 0.1cm}|@{\hskip 0.1cm}>{$}c<{$}@{\hskip 0.1cm}|@{\hskip 0.1cm}>{$}r<{$}@{\hskip 0.1cm}|}
\hline
	\qquad \qquad  \qquad \text{Group} & R^2 & \text{Wilson line} & \Lambda & \nabla_i \nabla_j \Lambda  \qquad \quad \\ \hline
 \left[Spin(16)^2\right]/{\mathbb{Z}_2}\times SU(2) & 1 & 0^{16} & 431.4 & \text{-}306^{16},831 \\ \hline 
 \left[Spin(16)\times Spin(12) \times SU(2)\right]/\mathbb{Z}_2\times SU(3) & \tfrac34 & 0^{14},\tfrac12^2 & 383.5 & \text{-}307^{15},544^2 \\ \hline 
 Spin(16)\times Spin(10) \times SU(5) & \tfrac58 & 0^{13},\tfrac12^3 & 359.2 & \text{-}569^{5},\text{-}256^8,355^4 \\ \hline
 \left[Spin(10)^2\times SU(8)\right]/{\mathbb{Z}_4} & \tfrac14 & 0^{4}, \tfrac12^4,\tfrac14^8 & 303.8 & \text{-}195^{17} \\\hline 
 Spin(18)\times Spin(16) & \tfrac12 & 0^{15}, 1 & 305.0& \text{-}1283^{8},588^9 \\ \hline 
 \left[Spin(16)\times Spin(10)\times Spin(8)\right] / {\mathbb{Z}_2} & \tfrac12 & 0^{12}, \tfrac12^4 & 305.0 & \text{-}1283^{4},\text{-}347^8,588^5 \\ \hline 
 \left[Spin(12)^2\times SU(4)\times SU(2)^2\right] / {\mathbb{Z}_2^2} & \tfrac12 & 0^{6}, \tfrac12^2,0^6,\tfrac12^2 & 305.0 & \text{-}1283^{2},\text{-}347^{12},588^3 \\\hline
 \left[E_6 \times SU(12)\right] / {\mathbb{Z}_3} & \tfrac18 & 0^{3}, \tfrac12^5,\text{-}\tfrac14,\tfrac14^7 & 180.4 & \text{-}72^{17} \\\hline 
\end{tabular}\caption[Caption for LOF]
  {One loop cosmological constant in units of $(4 \pi^2 \alpha')^{-\frac92}$ and the eigenvalues of the Hessian matrix \eqref{Hessmatrix} multiplied by $R^2$  for the eight tachyon-free maximal enhancements.\protect\footnotemark} \label{tab:CCandHessians}
\end{center}\end{table}
\footnotetext{The multiplicative factor of $R^2$ gives the Hessian in an orthonormal frame with respect to the classical metric on moduli space. In the circle reduction of the heterotic  string, the derivatives in an orthonormal frame are given by (see e.g. \cite{Collazuol:2022jiy}) $\nabla_\rho = \frac{\sqrt{7}}4 \, R \, \nabla_R \ , \ \nabla_{\tilde A}= R \, \nabla_A$.} 
  
\subsubsection{Refined de Sitter swampland conjecture} \label{sec:desitter}

Having computed the Hessian of the cosmological constant, or more precisely, the Hessian of the one-loop potential energy, it is straightforward to see whether the refined de Sitter conjecture is satisfied~\cite{Andriot:2018wzk,Ooguri:2018wrx}. This conjecture states that at extremal points with positive cosmological constant,
\beq \label{dSconj}
\frac{{\rm min} (\nabla_i \nabla_j V)}{V} \le - c\,,  
\eeq
with $c$ some order one constant, and where the left-hand side means the minimum of the Hessian in an orthonormal frame.
We record in Table \ref{tab:checkConjecture} these ratios for the four unstable points of maximal enhancement which are not knife edges; these are the first four entries of Table \ref{tab:CCandHessians}. The table shows that the refined de Sitter conjecture is satisfied at all points of maximal enhancement in circle compactifications of the non-supersymmetric heterotic string with $c=0.64$. 
\begin{table}[H] 
	\begin{center}
 \renewcommand{\arraystretch}{1.2} 
 \begin{tabular}
 {|>{$}l<{$}|>{$}c<{$}|>{$}c<{$}|>{$}c<{$}|}
\hline
\qquad \qquad \qquad	 \text{Group} &   \Lambda & \text{min} (\nabla_i \nabla_j V)/V\\ \hline
\left[Spin(16)^2\right]/{\mathbb{Z}_2}\times SU(2) & 431.4 & -0.71 \\ \hline 
 \left[Spin(16)\times Spin(12) \times SU(2)\right]/\mathbb{Z}_2\times SU(3) & 383.5 & -0.80 \\ \hline 
 Spin(16)\times Spin(10) \times SU(5) & 359.2 & -1.58 \\ \hline
 \left[Spin(10)^2\times SU(8)\right]/{\mathbb{Z}_4} & 303.8 & 
 -0.64
 \\\hline 
\end{tabular}
\caption{Minimum eigenvalue of the Hessian of the potential divided by the potential at the four points of maximal enhancement that are not knife-edges.}\label{tab:checkConjecture}
\end{center}\end{table}

\subsection{Extrema without maximal enhancement}\label{ss:nonmax}

As discussed in Section \ref{sec:extremal}, maximal enhancement points are guaranteed to be extrema of the cosmological constant. However, the converse is not necessarily true: there can in principle exist extremal points  that do not have maximal enhancement. We actually found four such points,
proving that maximal enhancement is a sufficient but not necessary condition. These points, which we list in Table \ref{tab:nonmaxExtrema}, are points of ``almost maximal enhancement," meaning that they are obtained from the EDD \eqref{diag1} by keeping only 16 instead of 17 nodes. 

Note that we are not strictly proving that there are extrema of the cosmological constant which fall outside of the proof relating them to maximal enhancements. Technically speaking, this proof works for any point which is completely fixed by the T-duality group, but it is not true that every fixed point of T-duality has gauge symmetry enhancement.  For example consider the outer automorphism of the charge lattice which exchanges the $SO(16)$ factors in the ten-dimensional gauge group. It fixes a region in moduli space where the first and the last eight components of the Wilson line are equal; however, this condition does not lead to any enhancement in the spectrum. In general, involutions in the T-duality group may fix regions with enhanced gauge bosons or tachyons if they are inner automorphisms, or with no enhancement at all if they are outer automorphisms.

It turns out that the fourth entry in Table \ref{tab:nonmaxExtrema} is completely fixed under T-duality. Deleting the nodes $(0), (0'), (8)$ and $(8')$ in the EDD \eqref{diag1} leaves us with 16 nodes associated to inner automorphisms and so a rank 16 gauge symmetry enhancement. By considering the aforementioned outer automorphisms exchanging the values of the Wilson line entries, it is easy to see that the values of the moduli in Table \ref{tab:nonmaxExtrema} are fixed. It is likely that the three remaining cases behave in a similar way, lying at point-like boundaries in the moduli space. It would be interesting to check this further.

\begin{table}[H] 
	\begin{center}
	\renewcommand{\arraystretch}{1.2}
		\begin{tabular}
{|>{$}l<{$}|>{$}c<{$}|>{$}c<{$}|>{$}c<{$}|>{$}c<{$}|}
\hline
	\text{Algebra} & R^2 & \text{Wilson line} & \Lambda & \nabla_i \nabla_j \Lambda \\ \hline
 \text{D}_8 + \text{D}_5 + \text{A}_3 & 3/8 & \left(0^{5},\tfrac12^3,0^8\right) & 367.1 & \text{-}338^{14},424^3 
\\ \hline 
 \text{D}_8 + \text{D}_4 + \text{A}_4 & 2/5 & \left(0^{3},\tfrac45^5,0^8\right) & 359.2 & 
\text{-}569,\text{-}412^{8},\text{-}256^4,355^4 
\\ \hline 
 4\text{D}_4 & 1/2 & \left(0^{4},\tfrac12^4,0^4, \tfrac12^4\right) & 305.0 & \text{-}347^{16}, 588
\\ \hline 
 2\text{A}_7 + 2\text{A}_1 & 1/2 & \left(\tfrac14^{16}\right) & 305.0 & \text{-}1283,\text{-}347^{14}, 588^2 
\\ \hline 
\end{tabular}
		\caption{Points in moduli space that do not correspond to maximal enhancements, yet are  extrema of the cosmological constant.}
	\label{tab:nonmaxExtrema}
 \end{center}\end{table}

As commented in Section \ref{ss:s1}, two of the points in Table \ref{tab:nonmaxExtrema} have the same value of $\Lambda$ as three of the maximal enhancements in Table \ref{tab:CCmassless}. There could be some deep reason why this is so, and if so, it would be very interesting to understand it. 

\section{Building \texorpdfstring{$\AdS_3$}{AdS3} vacua} \label{sec:AdS}

We can now examine the implications of our results for the construction of stabilized string vacua. Typically in discussions of non-supersymmetric string theory, the dilaton is left unfixed with the expectation or hope that it will be stabilized at weak coupling by non-perturbative physics. The focus is usually on the conformal field theory moduli of the worldsheet theory. The first thing we note is that the Minkowski background $\R^9 \times S^1$ is now unstable even at the level of these CFT Narain moduli because no extremum of the $1$-loop potential energy is a minimum.

Now one might say that this observation simply places the CFT moduli on the same footing as the dilaton, which also has no extremum in Minkowski space. Fortunately stabilizing the dilaton is a problem that can be solved for an $\AdS_3$ spacetime using ingredients that admit a tractable worldsheet description~\cite{Baykara:2022cwj}.

 In light of our realization that there are no minima for the CFT moduli, we will revisit that $\AdS_3$ construction. Specifically we consider
\begin{align}
    {\rm AdS}_3 \times S^3 \times \hat{S}^3 \times S^1\, ,
\end{align}
with $(n_1, n_5, \hat{n}_5)$ quanta of $H_3$-flux threading $\AdS_3 \times S^3 \times \hat{S}^3$, respectively. In this model the dilaton is stabilized at a tunably small value of the string coupling $g_s$ determined by the flux quanta. Ignoring the $1$-loop string potential for the moment, we note that
\begin{align}\label{roughgs}
g_s^4 \sim \frac{n_5^2\hn_5^2(\left|n_5\right|+\left|\hn_5\right|)}{n_1^2}.    
\end{align}
To make $g_s$ small, we choose  $n_1$ large with $(n_5, \hn_5)$ fixed. The spacetime cosmological constant is determined by the $D=3$ potential energy, 
\begin{equation}\label{thecc}
    \Lambda_{\AdS}= \frac{1}{2} V_{D=3} =   -\frac{1}{\alpha'} \left( \frac{1}{|n_5|}+\frac{1}{|\hn_5|} \right) \, .
\end{equation}
Note that this is a classical potential energy, which did not appear in our preceding discussion of $\R^9 \times S^1$. These results are modified by the $1$-loop potential energy on $\R^9 \times S^1$, which is a key ingredient in this construction.

The analysis in \cite{Baykara:2022cwj} examined how the values of $g_s$ and $\Lambda$ change when the $1$-loop potential is taken into account. For example, the classical string coupling of \C{roughgs} diverges as $n_1 \rightarrow 0$. The $1$-loop potential, however, prevents such a divergence even when $n_1=0$. Let us denote $1$-loop corrected quantities with a loop subscript. The $1$-loop corrected string coupling scales like $g_{s, \circ} \sim 1/\sqrt{n}$ for $n=n_5=\hn_5$ and $n_1=0$. This case corresponds to an intrinsically quantum string vacuum that only exists because of a balance between a tree-level potential and a $1$-loop potential.

What changes because the extrema on $S^1$ are not minima but saddle points? There now exist CFT moduli from the $S^1$ which become tachyonic at $1$-loop with masses determined by the Hessian of the potential energy studied in Section~\ref{s:cc}. Unlike flat space, tachyons in $\AdS$ are not necessarily problematic. If the tachyon is above the BF bound~\cite{Breitenlohner:1982bm}, the background is still perturbatively stable. 

\subsubsection*{\ul{\it $n_1>>1$ }}

There are two regimes that we can examine. In the first regime, $n_1$ is very large compared to $n_5$ and $\hn_5$ ensuring that $g_s$ is very small. In this limit, the $1$-loop potential is a small perturbation on the classical string background. 

Using the one-loop quantum corrected AdS cosmological constant, $\Lambda_{\AdS,\circ}$, we can approximate the one-loop corrected BF bound for $\AdS_3$ at small $g_s$~\cite{Baykara:2022cwj}:
\begin{align} \label{BFbound}
    m_{BF,\circ}^2:=\Lambda_{\AdS,\circ} = \Lambda_{\AdS} + \frac{1}{4 \alpha'} \lambda g_s^2 + O(g_s^4)\,.
\end{align}
Here quantities on the right-hand side without a loop subscript are tree-level values and $\lambda$ is a constant that is essentially the value of the $D=3$ $1$-loop potential evaluated at a chosen critical point:
\begin{align} \label{V1loop}
    V^{\rm{1-loop}} = 2 \lambda \frac{g_s^2}{\alpha'}\, .
\end{align}
In the notation of  Section \ref{s:cc} and Tables~\ref{tab:CCmassless}, \ref{tab:CCandHessians} and \ref{tab:checkConjecture}, $\lambda = \frac{\Lambda}{(2\pi)^9}$. 
We need to worry about the most tachyonic mode from the $S^1$ critical point and impose the condition that it is a safe tachyon above the BF bound \C{BFbound}. 

There are two issues to take into account. First we need to know the value of the largest negative eigenvalue of the Hessian for the $1$-loop potential studied in Section \ref{sec:hessian}.  Second we need to make sure the tachyonic mode is normalized so that the kinetic term is canonical. The kinetic terms can be evaluated at tree-level using the metric on Narain moduli space, which is a locally homogeneous space, because we are in a regime of small $g_{s}$. 
These are exactly the same issues that arose in checking the de Sitter swampland conjecture in Section \ref{sec:desitter}. We can conveniently parametrize the mass of the most tachyonic mode in terms of $c$ of \C{dSconj} as follows,
\begin{align}
    m^2_{\rm tachyon} = - c V^{\rm{1-loop}} \, ,
\end{align}
which gives the constraint:
\begin{align} \label{BF1}
    \left| \frac{m^2_{\rm tachyon}}{m_{BF,\circ}^2} \right| < 1\quad \rightarrow \quad \left| \frac{c}{\frac{1}{2 \lambda g_s^2} \left( \frac{1}{|n_5|}+\frac{1}{|\hn_5|} \right) - 1} \right| <1 
\end{align}
Taking a sufficiently large $n_1$ always guarantees that this constraint is satisfied because $g_s$ can be made as small as one desires. Our first conclusion is that this background is perturbatively stable for fixed $(n_5, \hn_5)$ and large enough $n_1$. 

However, there is now a critical value of $g_s^2$ where the BF bound is violated. Without worrying about precise numerical factors, this is roughly given by 
\begin{align}
    g_s^2 \sim \left( \frac{1}{|n_5|}+\frac{1}{|\hn_5|} \right) \quad \rightarrow \quad n_1 \sim \frac{(n_5 \hn_5)^2}{\sqrt{|n_5| + |\hn_5|}}\, .
\end{align}
This instability can appear in a regime of very weak coupling! However we need to be careful because with this critical scaling for $g_s$, the $1$-loop potential energy \C{V1loop} is comparable to the tree-level potential \C{thecc}, and we might worry about using the expansion \C{BFbound}.

\subsubsection*{\ul{\it The $n_1=0$ case }}

To simplify the discussion, let us take $n_5 = \hn_5 =n$. Unfortunately, there are no closed form expressions for the general case and numerical techniques are needed. The exception is the intrinsically quantum case with $n_1=0$ to which we now turn. In this case, one finds that~\cite{Baykara:2022cwj}
\begin{align}
    m_{BF,\circ}^2:=\Lambda_{\AdS,\circ} = - \frac{12}{5 \sqrt{5}} \frac{1}{\alpha' |n|}\, ,
\end{align}
and that, 
\begin{align}
    g_{s, \circ}^2 = \frac{24}{5 \sqrt{5}} \frac{1}{\lambda |n|}\, , \qquad V^{\rm 1-loop} = 2 \lambda \frac{g_{s, \circ}^2}{\alpha'} \, . 
\end{align}
This gives the constraint, 
\begin{align}
  4 \,  c  < 1\, .
\end{align}
This differs in spirit from the refined de Sitter swampland conjecture in the sense that it is a sharp statement about whether the AdS vacuum is perturbatively stable or not. 

It does not appear that any of the critical points we found satisfies this criterion. The smallest value of $c$, which is determined by the largest negative eigenvalue of the Hessian, is $c \sim 0.64$. The intrinsically quantum vacuum is a very special case because our a priori expectation is that any $\AdS_3$ vacuum with $n_1>0$ will decay non-perturbatively down to this case by discharging the spacetime electric $H_3$-flux via string nucleation. So finding perturbatively stable examples with $n_1=0$ is important in the hunt for non-supersymmetric examples of AdS/CFT. \\

\section{Summary and discussion of results}
 \label{sec:discussion}

In this paper we used lattice embedding techniques to characterize all points of gauge symmetry enhancement and their massless spectra in compactifications of the $\os$  string on the circle. We constructed an extended Dynkin diagram (EDD) that encodes all the information about gauge symmetry enhancements. The full diagram contains 22 nodes, corresponding to the generators of the 22 Weyl reflections of the T-duality group; see Figure \ref{diagcomplete}. Two of them are nodes of norm 1 which correspond to ``extremal" tachyons with the largest absolute value of the mass, namely $m^2=-2$, which generate reflections that exchange the (s) and (c) spinor classes. Because of these two nodes, the EDD becomes non-planar.
Each node represents a boundary of 
moduli space; for example, the central node corresponds to $R^2+\tfrac12 A^2=E=1$, see Table \ref{tab:boundaries}. The fundamental region has the form of a chimney with four horizontal walls, corresponding to the central, the spinor weight and the two tachyonic nodes   
delimiting the fundamental region in moduli space (see Figure \ref{fig:chimney}). 

All non-Abelian left-moving gauge groups of rank $k$ can be obtained from the diagram by deleting $22-k$ nodes, such that the resulting diagram is ADE, plus eventually a tachyonic node. When this happens, the  algebra which seems to be $B_n$ is actually to be understood as $D_n$ plus an extremal tachyon. The hypersurface in moduli space where the enhancements arise is found by satisfying the equations of the remaining nodes. The equations corresponding to the two tachyonic nodes are only satisfied at the same time at a decompactification point shown in red in Figure \ref{fig:chimney}. Therefore, for any enhancement at a finite distance point it suffices to work with the single-tachyon planar diagram in Figure \ref{diag2} and delete $21-k$ nodes. 
The non-supersymmetric setting also allows for an enhancement of the right-moving $U(1)_R$ to $SU(2)_R$.\\  

We found a total of 107 maximal enhancements,\footnote{This list includes  enhancements with the tachyonic node disconnected from the rest, which actually corresponds to a $U(1)$ gauge group plus an extremal tachyon. According to the EDD, these enhancements are on the same footing as the maximal ones.}  of which only eight are tachyon-free. The complete list of maximal enhancements is given in Table \ref{tab:9Dlist}, while the tachyon-free algebras are reported in Table \ref{tab:9Dtachyonfree} (see also Table  \ref{tab:CCandHessians} for the actual groups and their global structure). Four of the maximal enhancements have massless scalars that signal the presence of tachyons when moving infinitesimally in certain directions in moduli space. Only the $E_6 \times SU(12)$ point has an enhancement of the right-moving gauge symmetry. Note that this tachyon-free point is actually surrounded by tachyons, which we showed is an expected feature. It is also the only tachyon-free maximal enhancement with no massless fermions. 

The EDD additionally gives the affine algebras that arise at the eight decompactification limits to the two supersymmetric and six non-supersymmetric heterotic theories (see Figure \ref{fig:infdist}), which arise as $R\to 0$ or $R\to \infty$ with various Wilson lines. 

Maximal enhancements are particularly interesting points in moduli space because they extremize the cosmological constant to all loops, which one can show using the fact that they are fixed points under the full reflective symmetries. We numerically computed the one-loop cosmological constant at these points and at their vicinity. We found that all eight tachyon-free maximal enhancements have a positive cosmological constant, but are in fact unstable points. The four enhancements with massless scalars are knife edges. When one moves infinitesimally in moduli space in directions where these scalars become tachyonic, the cosmological constant jumps to minus infinity. Of the other four points, one is a local maximum while the remaining three are saddle points. We give the Hessian of the one-loop potential at these points in Table \ref{tab:CCandHessians}.  For the purpose of illustration, we gave the contribution of the massless states to the cosmological constant. This showed that massless states do not necessarily provide the dominant contribution to the vacuum energy. The extreme case is the $E_6 \times SU(12)$ enhancement where the cosmological constant is positive despite the absence of any massless fermions. 

We  computed the minimum value of the Hessian of the potential divided by the potential, and obtained an order one constant, in accord with the refined de Sitter swampland conjecture. We also presented a few points in moduli space that are extrema of the cosmological constant, yet do not possess maximal enhancement, showing that the latter is a sufficient but not necessary condition.

Plots of the cosmological constant for two-dimensional slices in moduli space are given in Figures \ref{fig:a015} to \ref{fig:a107a207}, where the unstable nature of the critical points can clearly be seen.  In one of the plots, we can see a decompactification limit (or rather, its T-dual limit $R\to 0$), where the cosmological constant multiplied by the radius goes to zero, signaling a decompactification to one of the supersymmetric theories. The paths towards supersymmetric theories that do not cross tachyonic regions are particularly interesting. This is because  the cosmological constant can be exponentially suppressed along paths where there is bose-fermi
degeneracy at the massless level~\cite{Itoyama:2019yst,Itoyama:2020ifw,Koga:2022qch}.  These regions merit further exploration. 

It would be very interesting to see if one can find actual local minima of the potential energy in compactifications on higher-dimensional tori. We leave this for future work. There are many other compelling avenues to explore. As one example:  constructing non-supersymmetric theories with rank reduction.  It is interesting to note that the $\os$ theory combines the $\mathbb{Z}_2$ gauge factors appearing in the supersymmetric cousins $Spin(32)/\Z_2$ and $E_8 \times E_8 \rtimes \Z_2$. Each of these $\Z_2$ factors can be used to further orbifold the parent theory to generate a reduced rank theory. 

The outer automorphism can now be used to construct a non-supersymmetric analog of the CHL string in nine dimensions, perhaps giving rise to the theory recently constructed in \cite{Nakajima:2023zsh}.\footnote{This automorphism was used in ten dimensions together with the spacetime fermion number in a manner analogous to the construction of the non-supersymmetric $E_8$ string from the $E_8 \times E_8 \rtimes \mathbb{Z}_2$ string in \cite{Forgacs:1988iw}. Curiously it again leads to the $E_8$ string.} The non-trivial fundamental group could be exploited in eight dimensions to construct a `holonomy double' flat connection in analogy to the compactification without vector structure of the $Spin(32)/\mathbb{Z}_2$ string~\cite{Witten:1997bs}. 

Finally there are many issues to explore around the $\AdS_3$ construction and constructing non-supersymmetric $\AdS$ more generally. Here we mention two directions: the first is taking the $\AdS_3 \times S^3 \times S^3$ geometry into account when computing the spacetime potential energy, rather than just studying $\R^9\times S^1$. 
Specifically whether the pathologies we see in the potential energy, like the knife-edges, are smoothed by quantum gravity. The second direction is exploring the bound we see on intrinsically quantum $\AdS$ solutions. Is this bound satisfied by any critical points found in toroidal compactifications? Are there any completely stable non-supersymmetric examples of AdS/CFT?

\section*{Acknowledgements}

We would like to thank Paul Ginsparg, Yuichi Koga, Wolfgang Lerche,  Dieter Lust, Miguel Montero, Benjamin Percival, Ignacio Ruiz, Cumrun Vafa and Irene Valenzuela for interesting discussions and useful comments. We also thank Markus Dierigl and Pavel Tumarkin for useful correspondence. S.~S. is supported in part by NSF Grant No. PHY2014195. MG and HP are partly supported 
by the ERC Consolidator Grant 772408-Stringlandscape. BF is supported in part by the ERC Starting Grant QGuide-101042568 - StG 2021.

\newpage
\appendix

\section{All maximal enhancement groups}
\label{app:results}
Here we present the list of maximal enhancements in  $S^1$ compactifications of the $\os$ heterotic string. 

\begin{table}[H] 
	\begin{center}
 \footnotesize 
\renewcommand{\arraystretch}{0.9}
		\begin{tabular}
			{|@{\hskip 0.2cm}
            >{$}c<{$}@{\hskip 0.2cm}|
			@{\hskip 0.2cm}>{$}r<{$}@{\hskip 0.2cm}|
			@{\hskip 0.2cm}>{$}c<{$}@{\hskip 0.1cm}|
			@{\hskip 0.1cm}>{\begin{scriptsize}$}c<{$\end{scriptsize}}@{\hskip 0.1cm}|
            >{\begin{scriptsize}$}r<{$\end{scriptsize}}
                @{\hskip 0.1cm}
                |
                @{\hskip 0.2cm}
                >{\begin{scriptsize}$}c<{$\end{scriptsize}}
                @{\hskip 0.2cm}
                |
                @{\hskip 0.2cm}
                >{\begin{scriptsize}$}c<{$\end{scriptsize}}
                @{\hskip 0.2cm}
                |
				@{\hskip 0.2cm}
                >{\begin{scriptsize}$}r<{$\end{scriptsize}}
                @{\hskip 0.2cm}
                |
                @{\hskip 0.1cm}
                >{\begin{scriptsize}$}c<{$\end{scriptsize}}
                @{\hskip 0.1cm}
                |} 
			\hline
			\# & \text{L} & \text{H} & \text{k} & m^2 & N_b & N_f & \text{Wilson line} & R^2 \\
			\hline	  
 \hline 1 & 2 {\color{red}\text{A}}_1+3 \text{A}_5 & \mathbb{Z}_6 & 
\begin{array}{@{}c@{}c@{}c@{}c@{}c@{}}
 1 & 1 & 1 & 4 & 1 \\
\end{array}
 & \text{-}2,\text{-}\frac{2}{3} & 94 & 160 & \frac{1}{5}(0^3,1^4,4^1,0^2,2^6) & \frac{3}{25} \\
 \hline 2 & \text{A}_1+2 {\color{red}\text{A}}_1+\text{A}_3+\text{A}_5+\text{A}_6 & \mathbb{Z}_2 & 
\begin{array}{@{}c@{}c@{}c@{}c@{}c@{}c@{}}
 1 & 1 & 1 & 2 & 3 & 0 \\
\end{array}
 & \text{-}2,\text{-}\frac{5}{6} & 90 & 128 & \frac{1}{18}(0^2,3^5,15^1,0^2,7^6) & \frac{7}{54} \\
 \hline 3 & 2 \text{A}_1+2 {\color{red}\text{A}}_1+2 \text{A}_3+\text{A}_7 & \mathbb{Z}_2 \mathbb{Z}_4 & 
\begin{array}{@{}c@{}c@{}c@{}c@{}c@{}c@{}c@{}}
 1 & 1 & 1 & 1 & 0 & 0 & 4 \\
 0 & 0 & 1 & 1 & 1 & 1 & 6 \\
\end{array}
 & \text{-}2,\text{-}1 & 88 & 96 & \frac{1}{6}(0^2,1^5,5^1,0^2,2^4,3^2) & \frac{1}{9} \\
 \hline 4 & 2 {\color{red}\text{A}}_1+\text{A}_2+\text{A}_5+\text{A}_8 & \mathbb{Z}_3 & 
\begin{array}{@{}c@{}c@{}c@{}c@{}c@{}}
 0 & 0 & 1 & 4 & 6 \\
\end{array}
 & \text{-}2,\text{-}1 & 112 & 80 & \frac{1}{8}(0^2,3^6,1^7,7^1) & \frac{9}{64} \\
 \hline 5 & \text{A}_1+2 {\color{red}\text{A}}_1+\text{A}_2+\text{A}_3+\text{A}_9 & \mathbb{Z}_2 & 
\begin{array}{@{}c@{}c@{}c@{}c@{}c@{}c@{}}
 1 & 1 & 1 & 0 & 0 & 5 \\
\end{array}
 & \text{-}2,\text{-}\frac{7}{6} & 114 & 48 & \frac{1}{16}(0^2,3^4,8^2,\text{-}6^1,6^7) & \frac{15}{128} \\
 \hline 6 & 2 {\color{red}\text{A}}_1+\text{A}_5+\text{A}_{10} & 1 & 
\begin{array}{@{}c}
 \text{} \\
\end{array}
 & \text{-}2,\text{-}\frac{17}{33} & 144 & 80 & \frac{1}{18}(0^2,7^6,\text{-}3^1,3^6,15^1) & \frac{11}{108} \\
 \hline 7 & 2 {\color{red}\text{A}}_1+2 \text{A}_2+\text{A}_{11} & \mathbb{Z}_6 & 
\begin{array}{@{}c@{}c@{}c@{}c@{}c@{}}
 1 & 1 & 1 & 1 & 10 \\
\end{array}
 & \text{-}2,\text{-}\frac{4}{3} & 148 & 0 & \frac{1}{8}(0^3,2^3,4^2,\text{-}3^1,3^7) & \frac{3}{32} \\
 \hline 8 & \text{A}_1+2 {\color{red}\text{A}}_1+\text{A}_3+\text{A}_{11} & \mathbb{Z}_4 & 
\begin{array}{@{}c@{}c@{}c@{}c@{}c@{}}
 0 & 1 & 1 & 1 & 9 \\
\end{array}
 & \text{-}2,\text{-}\frac{2}{3} & 150 & 48 & \frac{1}{6}(0^2,1^5,5^1,0^2,2^5,4^1) & \frac{1}{12} \\
 \hline 9 & 2 {\color{red}\text{A}}_1+\text{A}_2+\text{A}_{13} & 1 & 
\begin{array}{@{}c}
 \text{} \\
\end{array}
 & \text{-}2,\text{-}\frac{17}{21} & 192 & 0 & \frac{1}{16}(0^2,5^5,11^1,2^7,14^1) & \frac{21}{256} \\
 \hline 10 & 2 {\color{red}\text{A}}_1+\text{A}_{15} + \text{A}_1^{(\text{R})} & \mathbb{Z}_4 & 
\begin{array}{@{}c@{}c@{}c@{}}
 1 & 1 & 4 \\
\end{array}
 & \text{-}2 & 244 & 0 & \frac{1}{6}(0^2,2^5,4^1,\text{-}1^1,1^6,5^1) & \frac{1}{18} \\
 \hline 11 & 2 \text{A}_1+{\color{red}\text{D}}_1+2 \text{A}_7 & \mathbb{Z}_2 \mathbb{Z}_4 & 
\begin{array}{@{}c@{}c@{}c@{}c@{}c@{}}
 1 & 1 & 1 & 0 & 4 \\
 0 & 0 & 1 & 2 & 6 \\
\end{array}
 & \text{-}2,0 & 118 & 0 & \frac{1}{4}(0^2,1^5,3^1,\text{-}1^1,1^5,2^2) & \frac{1}{8} \\
 \hline 12 & {\color{red}\text{D}}_1+2 \text{A}_8 & \mathbb{Z}_3 & 
\begin{array}{@{}c@{}c@{}c@{}}
 0 & 3 & 6 \\
\end{array}
 & \text{-}2,0 & 146 & 0 & \frac{1}{10}(\text{-}3^1,3^7,2^7,8^1) & \frac{9}{50} \\
 \hline 13 & \text{A}_1+{\color{red}\text{D}}_1+\text{A}_6+\text{A}_9 & \mathbb{Z}_2 & 
\begin{array}{@{}c@{}c@{}c@{}c@{}}
 1 & 1 & 0 & 5 \\
\end{array}
 & \text{-}2,\text{-}\frac{4}{7} & 136 & 0 & \frac{1}{22}(0^2,6^5,16^1,\text{-}7^1,7^7) & \frac{35}{242} \\
 \hline 14 & {\color{red}\text{D}}_1+\text{A}_5+\text{A}_{11} & \mathbb{Z}_6 & 
\begin{array}{@{}c@{}c@{}c@{}}
 1 & 2 & 10 \\
\end{array}
 & \text{-}2,\text{-}1 & 164 & 0 & \frac{1}{3}(0^3,1^4,2^1,\text{-}1^1,1^7) & \frac{1}{9} \\
 \hline 15 & \text{A}_1+{\color{red}\text{D}}_1+\text{A}_{15} & \mathbb{Z}_4 & 
\begin{array}{@{}c@{}c@{}c@{}}
 0 & 1 & 4 \\
\end{array}
 & \text{-}2 & 244 & 0 & \frac{1}{4}(0^2,1^5,3^1,\text{-}1^1,1^6,3^1) & \frac{1}{16} \\
 \hline 16 & {\color{red}\text{D}}_1+\text{A}_{16} & 1 & 
\begin{array}{@{}c}
 \text{} \\
\end{array}
 & \text{-}2 & 274 & 0 & \frac{1}{22}(\text{-}7^1,7^7,\text{-}6^1,6^6,16^1) & \frac{17}{242} \\
 \hline 17 & \text{A}_7+2 \text{D}_5 & \mathbb{Z}_4 & 
\begin{array}{@{}c@{}c@{}c@{}}
 2 & 1 & 3 \\
\end{array}
 &  & 136 & 170 & \frac{1}{2}(0^5,1^3,0^5,1^3) & \frac{1}{4} \\
 \hline 18 & \text{A}_4+\text{A}_8+{\color{red}\text{D}}_5 & 1 & 
\begin{array}{@{}c}
 \text{} \\
\end{array}
 & \text{-}2,\text{-}\frac{8}{9} & 132 & 0 & \frac{1}{38}(0^3,4^4,34^1,9^5,19^3) & \frac{45}{722} \\
 \hline 19 & \text{A}_1+\text{A}_2+\text{A}_9+{\color{red}\text{D}}_5 & \mathbb{Z}_2 & 
\begin{array}{@{}c@{}c@{}c@{}c@{}}
 1 & 0 & 5 & 2 \\
\end{array}
 & \text{-}2,\text{-}\frac{4}{5} & 138 & 0 & \frac{1}{22}(0^3,6^5,\text{-}9^1,9^5,11^2) & \frac{15}{242} \\
 \hline 20 & \text{A}_1+\text{A}_{11}+{\color{red}\text{D}}_5 & \mathbb{Z}_2 & 
\begin{array}{@{}c@{}c@{}c@{}}
 0 & 6 & 2 \\
\end{array}
 & \text{-}2,\text{-}\frac{2}{3} & 174 & 0 & \frac{1}{7}(0^3,2^5,\text{-}3^1,3^7) & \frac{3}{49} \\
 \hline 21 & \text{A}_{12}+{\color{red}\text{D}}_5 & 1 & 
\begin{array}{@{}c}
 \text{} \\
\end{array}
 & \text{-}2,\text{-}\frac{8}{13} & 196 & 0 & \frac{1}{22}(0^3,6^5,\text{-}9^1,9^6,13^1) & \frac{13}{242} \\
 \hline 22 & \text{A}_7+{\color{red}\text{D}}_5+\text{D}_5 & \mathbb{Z}_2 & 
\begin{array}{@{}c@{}c@{}c@{}}
 4 & 2 & 2 \\
\end{array}
 & \text{-}2,\text{-}1 & 136 & 0 & \frac{1}{8}(0^5,4^3,\text{-}3^1,3^7) & \frac{1}{16} \\
 \hline 23 & \text{A}_1+2 {\color{red}\text{A}}_1+\text{A}_3+\text{A}_5+\text{D}_6 & \mathbb{Z}_2^2 & 
\begin{array}{@{}c@{}c@{}c@{}c@{}c@{}c@{}c@{}}
 0 & 0 & 0 & 2 & 3 & 1 & 0 \\
 1 & 1 & 1 & 0 & 3 & 1 & 1 \\
\end{array}
 & \text{-}2,\text{-}\frac{1}{2} & 108 & 176 & \frac{1}{6}(0^6,3^2,\text{-}2^1,2^5,3^2) & \frac{1}{6} \\
 \hline 24 & 2 {\color{red}\text{A}}_1+\text{A}_4+\text{A}_5+\text{D}_6 & \mathbb{Z}_2 & 
\begin{array}{@{}c@{}c@{}c@{}c@{}c@{}c@{}}
 1 & 1 & 0 & 3 & 0 & 1 \\
\end{array}
 & \text{-}2,\text{-}\frac{1}{3} & 114 & 208 & \frac{1}{10}(0^6,5^2,\text{-}3^1,3^4,5^3) & \frac{3}{20} \\
 \hline 25 & 2 {\color{red}\text{A}}_1+\text{A}_2+\text{A}_7+\text{D}_6 & \mathbb{Z}_2 & 
\begin{array}{@{}c@{}c@{}c@{}c@{}c@{}c@{}}
 1 & 1 & 0 & 4 & 1 & 1 \\
\end{array}
 & \text{-}2,\text{-}\frac{2}{3} & 126 & 128 & \frac{1}{8}(0^6,4^2,\text{-}3^1,3^7) & \frac{3}{16} \\
 \hline 26 & 2 {\color{red}\text{A}}_1+\text{A}_9+\text{D}_6 & \mathbb{Z}_2 & 
\begin{array}{@{}c@{}c@{}c@{}c@{}c@{}}
 0 & 0 & 5 & 1 & 0 \\
\end{array}
 & \text{-}2,\text{-}\frac{1}{5} & 154 & 128 & \frac{1}{6}(0^6,3^2,\text{-}2^1,2^6,4^1) & \frac{5}{36} \\
 \hline 27 & \text{A}_1+\text{A}_5+\text{D}_5+\text{D}_6 & \mathbb{Z}_2 & 
\begin{array}{@{}c@{}c@{}c@{}c@{}c@{}}
 1 & 3 & 2 & 1 & 1 \\
\end{array}
 & \text{-}\frac{1}{2} & 132 & 224 & \frac{1}{2}(0^6,1^2,0^5,1^3) & \frac{3}{8} \\
 \hline 28 & 2 \text{A}_1+\text{A}_3+2 \text{D}_6 & \mathbb{Z}_2^2 & 
\begin{array}{@{}c@{}c@{}c@{}c@{}c@{}c@{}c@{}}
 0 & 0 & 2 & 1 & 0 & 1 & 0 \\
 1 & 1 & 0 & 0 & 1 & 0 & 1 \\
\end{array}
 & 0 & 136 & 296 & \frac{1}{2}(0^6,1^2,0^6,1^2) & \frac{1}{2} \\
 \hline 29 & 2 {\color{red}\text{A}}_1+\text{A}_3+2 \text{D}_6 & \mathbb{Z}_2^2 & 
\begin{array}{@{}c@{}c@{}c@{}c@{}c@{}c@{}c@{}}
 0 & 0 & 2 & 1 & 0 & 0 & 1 \\
 1 & 1 & 0 & 0 & 1 & 1 & 0 \\
\end{array}
 & \text{-}2,0 & 136 & 256 & \frac{1}{4}(0^6,2^2,\text{-}1^1,1^3,2^4) & \frac{1}{8} \\
 \hline 30 & \text{A}_1+\text{D}_4+2 \text{D}_6 & \mathbb{Z}_2^2 & 
\begin{array}{@{}c@{}c@{}c@{}c@{}c@{}c@{}c@{}}
 0 & 0 & 1 & 1 & 0 & 0 & 1 \\
 1 & 1 & 0 & 1 & 1 & 1 & 0 \\
\end{array}
 & \text{-}1 & 146 & 320 & \frac{1}{2}(0^6,1^2,0^4,1^4) & \frac{1}{4} \\
 \hline 31 & \text{A}_1+\text{A}_4+\text{A}_6+{\color{red}\text{D}}_6 & 1 & 
\begin{array}{@{}c}
 \text{} \\
\end{array}
 & \text{-}2,\text{-}\frac{9}{7},\text{-}\frac{3}{5} & 124 & 128 & \frac{1}{24}(0^3,2^4,22^1,7^6,12^2) & \frac{35}{576} \\
 \hline 32 & 2 \text{A}_1+\text{A}_2+\text{A}_7+{\color{red}\text{D}}_6 & \mathbb{Z}_2 & 
\begin{array}{@{}c@{}c@{}c@{}c@{}c@{}c@{}}
 1 & 1 & 0 & 4 & 1 & 1 \\
\end{array}
 & \text{-}2,\text{-}\frac{5}{4},\text{-}\frac{2}{3} & 126 & 128 & \frac{1}{14}(0^2,1^5,13^1,4^6,7^2) & \frac{3}{49} \\
 \hline 33 & 2 \text{A}_1+\text{A}_9+{\color{red}\text{D}}_6 & \mathbb{Z}_2 & 
\begin{array}{@{}c@{}c@{}c@{}c@{}c@{}}
 1 & 0 & 5 & 1 & 1 \\
\end{array}
 & \text{-}2,\text{-}\frac{6}{5},\text{-}\frac{1}{5} & 154 & 128 & \frac{2}{9}(0^2,1^6,\text{-}2^1,2^7) & \frac{5}{81} \\
 \hline 34 & \text{A}_1+\text{A}_{10}+{\color{red}\text{D}}_6 & 1 & 
\begin{array}{@{}c}
 \text{} \\
\end{array}
 & \text{-}2,\text{-}\frac{13}{11} & 172 & 128 & \frac{1}{14}(0^2,3^6,\text{-}6^1,6^6,8^1) & \frac{11}{196} \\
 \hline 35 & \text{A}_1+\text{A}_5+\text{D}_5+{\color{red}\text{D}}_6 & \mathbb{Z}_2 & 
\begin{array}{@{}c@{}c@{}c@{}c@{}c@{}}
 1 & 3 & 2 & 1 & 1 \\
\end{array}
 & \text{-}2,\text{-}\frac{4}{3},\text{-}\frac{1}{2} & 132 & 128 & \frac{1}{6}(0^5,3^3,\text{-}2^1,2^5,3^2) & \frac{1}{24} \\
 \hline 36 & \text{A}_4+\text{D}_5+\text{D}_8 & 1 & 
\begin{array}{@{}c}
 \text{} \\
\end{array}
 &  & 172 & 288 & \frac{1}{2}(0^{13},1^3) & \frac{5}{8} \\
 \hline 37 & \text{D}_4+\text{D}_5+\text{D}_8 & \mathbb{Z}_2 & 
\begin{array}{@{}c@{}c@{}c@{}c@{}c@{}}
 1 & 1 & 2 & 1 & 0 \\
\end{array}
 & 0 & 176 & 336 & \frac{1}{2}(0^{12},1^4) & \frac{1}{2} \\
 \hline 38 & {\color{red}\text{D}}_4+\text{D}_5+\text{D}_8 & \mathbb{Z}_2 & 
\begin{array}{@{}c@{}c@{}c@{}c@{}c@{}}
 1 & 1 & 2 & 0 & 1 \\
\end{array}
 & \text{-}2,0 & 176 & 256 & \frac{1}{6}(0^4,1^3,5^1,1^4,3^4) & \frac{1}{18} \\
 \hline 39 & \text{A}_1+2 {\color{red}\text{A}}_1+\text{D}_6+\text{D}_8 & \mathbb{Z}_2^2 & 
\begin{array}{@{}c@{}c@{}c@{}c@{}c@{}c@{}c@{}}
 0 & 1 & 1 & 1 & 1 & 0 & 1 \\
 1 & 0 & 0 & 1 & 0 & 1 & 0 \\
\end{array}
 & \text{-}2,\text{-}1 & 178 & 256 & \frac{1}{4}(0^4,1^3,3^1,0^2,1^2,2^4) & \frac{1}{16} \\
 \hline 40 & \text{A}_1+\text{A}_2+\text{D}_6+\text{D}_8 & \mathbb{Z}_2 & 
\begin{array}{@{}c@{}c@{}c@{}c@{}c@{}c@{}}
 1 & 0 & 0 & 1 & 1 & 0 \\
\end{array}
 &  & 180 & 384 & \frac{1}{2}(0^{14},1^2) & \frac{3}{4} \\
 \hline 41 & \text{A}_1+2 \text{D}_8 & \mathbb{Z}_2 & 
\begin{array}{@{}c@{}c@{}c@{}c@{}c@{}}
 0 & 1 & 0 & 1 & 0 \\
\end{array}
 &  & 226 & 512 & (0^{16}) & 1 \\
 \hline 42 & {\color{red}\text{D}}_1+2 \text{D}_8 + \text{A}_1^{(\text{R})}& \mathbb{Z}_2^2 & 
\begin{array}{@{}c@{}c@{}c@{}c@{}c@{}}
 0 & 0 & 1 & 1 & 0 \\
 1 & 1 & 0 & 1 & 1 \\
\end{array}
 & \text{-}2,0 & 226 & 0 & \frac{1}{6}(0^4,2^3,4^1,\text{-}1^1,1^3,3^4) & \frac{1}{18} \\
 \hline 43 & \text{A}_4+\text{A}_5+{\color{red}\text{D}}_8 & 1 & 
\begin{array}{@{}c}
 \text{} \\
\end{array}
 & \text{-}2,\text{-}\frac{1}{3} & 162 & 256 & \frac{1}{17}(0^3,1^4,16^1,6^8) & \frac{15}{289} \\

\hline\end{tabular}
 \end{center}\end{table}
\begin{table}[H] 
\vspace{-0.6in}
\begin{center}\footnotesize 
\renewcommand{\arraystretch}{0.9}
		\begin{tabular}
		{|@{\hskip 0.2cm}
            >{$}c<{$}@{\hskip 0.2cm}|
			@{\hskip 0.2cm}>{$}r<{$}@{\hskip 0.2cm}|
			@{\hskip 0.2cm}>{$}c<{$}@{\hskip 0.1cm}|
			@{\hskip 0.1cm}>{\begin{scriptsize}$}c<{$\end{scriptsize}}@{\hskip 0.1cm}|
            >{\begin{scriptsize}$}r<{$\end{scriptsize}}
                @{\hskip 0.1cm}
                |
                @{\hskip 0.2cm}
                >{\begin{scriptsize}$}c<{$\end{scriptsize}}
                @{\hskip 0.2cm}
                |
                @{\hskip 0.2cm}
                >{\begin{scriptsize}$}c<{$\end{scriptsize}}
                @{\hskip 0.2cm}
                |
				@{\hskip 0.2cm}
                >{\begin{scriptsize}$}r<{$\end{scriptsize}}
                @{\hskip 0.2cm}
                |
                @{\hskip 0.1cm}
                >{\begin{scriptsize}$}c<{$\end{scriptsize}}
                @{\hskip 0.1cm}
                |} 
			\hline
			\# & \text{L} & \text{H} & \text{k} & m^2 & N_b & N_f & \text{Wilson line} & R^2 \\
			\hline	
 \hline 44 & \text{A}_1+\text{A}_2+\text{A}_6+{\color{red}\text{D}}_8 & 1 & 
\begin{array}{@{}c}
 \text{} \\
\end{array}
 & \text{-}2,\text{-}\frac{2}{7} & 162 & 256 & \frac{1}{20}(0^2,1^5,19^1,7^8) & \frac{21}{400} \\
 \hline 45 & \text{A}_1+\text{A}_8+{\color{red}\text{D}}_8 & 1 & 
\begin{array}{@{}c}
 \text{} \\
\end{array}
 & \text{-}2,\text{-}\frac{2}{9} & 186 & 256 & \frac{2}{13}(\text{-}3^1,3^7,1^8) & \frac{9}{169} \\
 \hline 46 & \text{A}_9+{\color{red}\text{D}}_8 & 1 & 
\begin{array}{@{}c}
 \text{} \\
\end{array}
 & \text{-}2,\text{-}\frac{1}{5} & 202 & 256 & \frac{1}{15}(0^3,3^4,12^1,0^3,7^4,8^1) & \frac{1}{45} \\
 \hline 47 & \text{A}_4+\text{D}_5+{\color{red}\text{D}}_8 & 1 & 
\begin{array}{@{}c}
 \text{} \\
\end{array}
 & \text{-}2,\text{-}\frac{2}{5} & 172 & 256 & \frac{1}{10}(0^5,5^3,\text{-}3^1,3^4,5^3) & \frac{1}{40} \\
 \hline 48 & \text{A}_1+2 {\color{red}\text{A}}_1+\text{A}_5+\text{D}_9 & \mathbb{Z}_2 & 
\begin{array}{@{}c@{}c@{}c@{}c@{}c@{}}
 1 & 1 & 1 & 3 & 2 \\
\end{array}
 & \text{-}2,\text{-}\frac{4}{3} & 180 & 224 & \frac{1}{3}(0^7,3^1,0^2,1^6) & \frac{1}{6} \\
 \hline 49 & {\color{red}\text{D}}_1+\text{A}_7+\text{D}_9 & \mathbb{Z}_2 & 
\begin{array}{@{}c@{}c@{}c@{}}
 1 & 4 & 2 \\
\end{array}
 & \text{-}2,\text{-}1 & 202 & 0 & \frac{1}{4}(0^7,4^1,\text{-}1^1,1^7) & \frac{1}{4} \\
 \hline 50 & \text{A}_4+{\color{red}\text{D}}_4+\text{D}_9 & 1 & 
\begin{array}{@{}c}
 \text{} \\
\end{array}
 & \text{-}2,\text{-}\frac{2}{5} & 188 & 288 & \frac{1}{5}(0^7,5^1,0^3,2^5) & \frac{1}{10} \\
 \hline 51 & \text{D}_8+\text{D}_9 & 1 & 
\begin{array}{@{}c}
 \text{} \\
\end{array}
 & 0 & 256 & 416 & (0^{15},1^1) & \frac{1}{2} \\
 \hline 52 & {\color{red}\text{D}}_8+\text{D}_9 & 1 & 
\begin{array}{@{}c}
 \text{} \\
\end{array}
 & \text{-}2,0 & 256 & 256 & \frac{1}{3}(0^7,3^1,1^8) & \frac{1}{18} \\
 \hline 53 & 2 \text{A}_1+2 {\color{red}\text{A}}_1+\text{A}_3+\text{D}_{10} & \mathbb{Z}_2^2 & 
\begin{array}{@{}c@{}c@{}c@{}c@{}c@{}c@{}c@{}}
 0 & 1 & 0 & 0 & 2 & 1 & 0 \\
 1 & 0 & 1 & 1 & 0 & 0 & 1 \\
\end{array}
 & \text{-}2,\text{-}\frac{3}{2},0 & 200 & 208 & \frac{1}{4}(0^7,4^1,0^2,1^4,2^2) & \frac{1}{8} \\
 \hline 54 & \text{A}_1+{\color{red}\text{D}}_1+\text{A}_5+\text{D}_{10} & \mathbb{Z}_2^2 & 
\begin{array}{@{}c@{}c@{}c@{}c@{}c@{}}
 0 & 0 & 3 & 1 & 0 \\
 1 & 1 & 0 & 0 & 1 \\
\end{array}
 & \text{-}2,\text{-}\frac{4}{3} & 214 & 0 & \frac{1}{6}(0^7,6^1,\text{-}1^1,1^5,3^2) & \frac{1}{6} \\ \hline 55 & \text{A}_1+\text{A}_2+{\color{red}\text{D}}_4+\text{D}_{10} & \mathbb{Z}_2 & 
\begin{array}{@{}c@{}c@{}c@{}c@{}c@{}c@{}}
 1 & 0 & 1 & 1 & 0 & 1 \\
\end{array}
 & \text{-}2,\text{-}\frac{2}{3} & 212 & 320 & \frac{1}{6}(0^7,6^1,0^3,2^3,3^2) & \frac{1}{12} \\
 \hline 56 & \text{A}_1+\text{D}_6+\text{D}_{10} & \mathbb{Z}_2 & 
\begin{array}{@{}c@{}c@{}c@{}c@{}c@{}}
 0 & 1 & 0 & 1 & 0 \\
\end{array}
 & \text{-}1 & 242 & 304 & \frac{1}{2}(0^7,2^1,0^6,1^2) & \frac{1}{4} \\
 \hline 57 & \text{A}_1+{\color{red}\text{D}}_6+\text{D}_{10} & \mathbb{Z}_2 & 
\begin{array}{@{}c@{}c@{}c@{}c@{}c@{}}
 1 & 1 & 1 & 0 & 1 \\
\end{array}
 & \text{-}2,\text{-}1 & 242 & 128 & \frac{1}{4}(0^7,4^1,1^6,2^2) & \frac{1}{16} \\
 \hline 58 & \text{A}_1+2 {\color{red}\text{A}}_1+\text{A}_2+\text{D}_{12} & \mathbb{Z}_2 & 
\begin{array}{@{}c@{}c@{}c@{}c@{}c@{}c@{}}
 0 & 1 & 1 & 0 & 1 & 0 \\
\end{array}
 & \text{-}2,\text{-}\frac{5}{3},\text{-}\frac{2}{3} & 276 & 192 & \frac{1}{6}(0^7,6^1,0^2,1^3,3^3) & \frac{1}{12} \\
 \hline 59 & {\color{red}\text{D}}_1+\text{A}_4+\text{D}_{12} & \mathbb{Z}_2 & 
\begin{array}{@{}c@{}c@{}c@{}c@{}}
 1 & 0 & 1 & 0 \\
\end{array}
 & \text{-}2,\text{-}\frac{8}{5} & 286 & 0 & \frac{1}{10}(0^7,10^1,\text{-}1^1,1^4,5^3) & \frac{1}{10} \\
 \hline 60 & \text{A}_1+{\color{red}\text{D}}_4+\text{D}_{12} & \mathbb{Z}_2 & 
\begin{array}{@{}c@{}c@{}c@{}c@{}c@{}}
 0 & 1 & 1 & 1 & 0 \\
\end{array}
 & \text{-}2,\text{-}1 & 290 & 384 & \frac{1}{4}(0^7,4^1,0^3,1^2,2^3) & \frac{1}{16} \\
 \hline 61 & \text{D}_5+\text{D}_{12} & 1 & 
\begin{array}{@{}c}
 \text{} \\
\end{array}
 & \text{-}\frac{3}{2} & 304 & 240 & \frac{1}{2}(0^7,2^1,0^5,1^3) & \frac{1}{8} \\
 \hline 62 & {\color{red}\text{D}}_5+\text{D}_{12} + \text{A}_1^{(\text{R})} & \mathbb{Z}_2 & 
\begin{array}{@{}c@{}c@{}c@{}}
 2 & 1 & 0 \\
\end{array}
 & \text{-}2,0 & 304 & 0 & \frac{1}{6}(0^7,6^1,1^5,3^3) & \frac{1}{18} \\
 \hline 63 & \text{D}_5+{\color{red}\text{D}}_{12} & 1 & 
\begin{array}{@{}c}
 \text{} \\
\end{array}
 & \text{-}2,\text{-}\frac{3}{2},0 & 304 & 0 & \frac{1}{6}(0^5,3^3,\text{-}2^1,2^6,4^1) & \frac{1}{72} \\
 \hline 64 & {\color{red}\text{D}}_4+\text{D}_{13} & 1 & 
\begin{array}{@{}c}
 \text{} \\
\end{array}
 & \text{-}2,0 & 336 & 416 & \frac{1}{3}(0^7,3^1,0^3,1^4,2^1) & \frac{1}{18} \\
 \hline 65 & \text{A}_4+{\color{red}\text{D}}_{13} & 1 & 
\begin{array}{@{}c}
 \text{} \\
\end{array}
 & \text{-}2,\text{-}\frac{8}{5},\text{-}\frac{2}{5} & 332 & 0 & \frac{1}{24}(0^3,2^4,22^1,7^7,17^1) & \frac{5}{288} \\
 \hline 66 & \text{A}_1+2 {\color{red}\text{A}}_1+\text{D}_{14} & \mathbb{Z}_2 & 
\begin{array}{@{}c@{}c@{}c@{}c@{}c@{}}
 1 & 0 & 0 & 1 & 0 \\
\end{array}
 & \text{-}2,\text{-}1 & 370 & 224 & \frac{1}{4}(0^7,4^1,0^2,1^5,3^1) & \frac{1}{16} \\
 \hline 67 & \text{A}_1+\text{A}_2+{\color{red}\text{D}}_{14} & 1 & 
\begin{array}{@{}c}
 \text{} \\
\end{array}
 & \text{-}2,\text{-}\frac{5}{3},\text{-}\frac{2}{3} & 372 & 0 & \frac{1}{14}(0^2,1^5,13^1,4^7,10^1) & \frac{3}{196} \\
 \hline 68 & {\color{red}\text{D}}_1+\text{D}_{16}+ \text{A}_1^{(\text{R})} & \mathbb{Z}_2 & 
\begin{array}{@{}c@{}c@{}c@{}}
 0 & 0 & 1 \\
\end{array}
 & \text{-}2,0 & 482 & 0 & \frac{1}{6}(0^7,6^1,\text{-}1^1,1^6,5^1) & \frac{1}{18} \\
 \hline 69 & \text{A}_1+{\color{red}\text{D}}_{16} & 1 & 
\begin{array}{@{}c}
 \text{} \\
\end{array}
 & \text{-}2,\text{-}1 & 482 & 0 & \frac{2}{9}(\text{-}2^1,2^7,\text{-}1^1,1^7) & \frac{1}{81} \\
 \hline 70 & {\color{red}\text{D}}_{17} + \text{A}_1^{(\text{R})}& 1 & 
\begin{array}{@{}c}
 \text{} \\
\end{array}
 & \text{-}2,0 & 544 & 0 & \frac{1}{14}(\text{-}6^1,6^6,8^1,\text{-}3^1,3^7) & \frac{1}{98} \\
 \hline 71 & \text{A}_1+2 \text{A}_5+\text{E}_6 & \mathbb{Z}_3 & 
\begin{array}{@{}c@{}c@{}c@{}c@{}}
 0 & 2 & 4 & 1 \\
\end{array}
 & \text{-}\frac{2}{3} & 134 & 80 & \frac{1}{4}(0^5,2^3,1^6,2^2) & \frac{3}{16} \\
 \hline 72 & \text{A}_4+\text{A}_7+\text{E}_6 & 1 & 
\begin{array}{@{}c}
 \text{} \\
\end{array}
 & \text{-}\frac{13}{20} & 148 & 70 & \frac{1}{10}(0^5,5^3,2^5,5^3) & \frac{3}{20} \\
 \hline 73 & \text{A}_{11}+\text{E}_6 + \text{A}_1^{(\text{R})}& \mathbb{Z}_3 & 
\begin{array}{@{}c@{}c@{}}
 4 & 1 \\
\end{array}
 & 0 & 204 & 0 & \frac{1}{4}(0^5,2^3,1^7,3^1) & \frac{1}{8} \\
 \hline 74 & \text{A}_6+{\color{red}\text{D}}_5+\text{E}_6 & 1 & 
\begin{array}{@{}c}
 \text{} \\
\end{array}
 & \text{-}2,\text{-}\frac{8}{7} & 154 & 0 & \frac{1}{26}(0^5,4^2,22^1,7^5,13^3) & \frac{21}{338} \\
 \hline 75 & \text{A}_1+\text{A}_4+{\color{red}\text{D}}_6+\text{E}_6 & 1 & 
\begin{array}{@{}c}
 \text{} \\
\end{array}
 & \text{-}2,\text{-}\frac{7}{5},\text{-}\frac{1}{3} & 154 & 128 & \frac{1}{16}(0^5,2^2,14^1,5^6,8^2) & \frac{15}{256} \\
 \hline 76 & {\color{red}\text{D}}_4+\text{D}_7+\text{E}_6 & 1 & 
\begin{array}{@{}c}
 \text{} \\
\end{array}
 & \text{-}2,\text{-}\frac{1}{2} & 180 & 224 & \frac{1}{4}(0^7,3^1,0^3,2^5) & \frac{3}{32} \\
 \hline 77 & \text{A}_3+\text{D}_8+\text{E}_6 & 1 & 
\begin{array}{@{}c}
 \text{} \\
\end{array}
 & \text{-}\frac{1}{2} & 196 & 224 & \frac{1}{2}(0^{11},1^5) & \frac{3}{8} \\
 \hline 78 & \text{A}_3+{\color{red}\text{D}}_8+\text{E}_6 & 1 & 
\begin{array}{@{}c}
 \text{} \\
\end{array}
 & \text{-}2,\text{-}\frac{1}{2} & 196 & 256 & \frac{1}{11}(0^5,1^2,10^1,4^8) & \frac{6}{121} \\
 \hline 79 & {\color{red}\text{D}}_{11}+\text{E}_6 & 1 & 
\begin{array}{@{}c}
 \text{} \\
\end{array}
 & \text{-}2,\text{-}\frac{4}{3} & 292 & 0 & \frac{1}{16}(0^5,2^2,14^1,5^7,11^1) & \frac{3}{128} \\
 \hline 80 & \text{A}_1+2 {\color{red}\text{A}}_1+\text{A}_3+\text{A}_4+\text{E}_7 & \mathbb{Z}_2 & 
\begin{array}{@{}c@{}c@{}c@{}c@{}c@{}c@{}}
 1 & 1 & 1 & 2 & 0 & 1 \\
\end{array}
 & \text{-}2,\text{-}\frac{2}{5} & 164 & 272 & \frac{1}{10}(0^6,5^2,0^1,4^5,5^2) & \frac{1}{10} \\
 \hline 81 & 2 \text{A}_1+\text{A}_3+\text{A}_5+\text{E}_7 & \mathbb{Z}_2 & 
\begin{array}{@{}c@{}c@{}c@{}c@{}c@{}}
 1 & 1 & 0 & 3 & 1 \\
\end{array}
 & \text{-}\frac{4}{3},\text{-}\frac{1}{2} & 172 & 176 & \frac{1}{6}(0^6,3^2,2^6,3^2) & \frac{1}{6} \\
 \hline 82 & 2 {\color{red}\text{A}}_1+\text{A}_3+\text{A}_5+\text{E}_7 & \mathbb{Z}_2 & 
\begin{array}{@{}c@{}c@{}c@{}c@{}c@{}}
 1 & 1 & 0 & 3 & 1 \\
\end{array}
 & \text{-}2,\text{-}\frac{1}{2} & 172 & 304 & \frac{1}{8}(0^6,4^2,0^1,3^4,4^3) & \frac{3}{32} \\
 \hline 83 & \text{A}_1+\text{A}_4+\text{A}_5+\text{E}_7 & 1 & 
\begin{array}{@{}c}
 \text{} \\
\end{array}
 & \text{-}\frac{7}{5},\text{-}\frac{1}{3} & 178 & 152 & \frac{1}{10}(0^6,5^2,3^5,5^3) & \frac{3}{20} \\
 \hline 84 & 2 {\color{red}\text{A}}_1+\text{A}_2+\text{A}_6+\text{E}_7 & 1 & 
\begin{array}{@{}c}
 \text{} \\
\end{array}
 & \text{-}2,\text{-}\frac{2}{7} & 178 & 224 & \frac{1}{14}(0^6,7^2,0^1,6^7) & \frac{3}{28} \\
 \hline 85 & \text{A}_1+\text{A}_2+\text{A}_7+\text{E}_7 & \mathbb{Z}_2 & 
\begin{array}{@{}c@{}c@{}c@{}c@{}}
 1 & 0 & 4 & 1 \\
\end{array}
 & \text{-}\frac{5}{4} & 190 & 182 & \frac{1}{6}(0^5,3^3,0^2,2^3,3^3) & \frac{1}{12} \\
 \hline 86 & 2 {\color{red}\text{A}}_1+\text{A}_8+\text{E}_7 & 1 & 
\begin{array}{@{}c}
 \text{} \\
\end{array}
 & \text{-}2,\text{-}\frac{2}{9} & 202 & 224 & \frac{1}{10}(0^6,5^2,0^1,4^6,6^1) & \frac{9}{100} \\
 \hline 87 & \text{A}_1+\text{A}_9+\text{E}_7 & 1 & 
\begin{array}{@{}c}
 \text{} \\
\end{array}
 & \text{-}\frac{6}{5} & 218 & 112 & \frac{1}{6}(0^6,3^2,2^7,4^1) & \frac{5}{36} \\
 \hline 88 & \text{A}_5+{\color{red}\text{D}}_5+\text{E}_7 & \mathbb{Z}_2 & 
\begin{array}{@{}c@{}c@{}c@{}}
 3 & 2 & 1 \\
\end{array}
 & \text{-}2,\text{-}\frac{4}{3} & 196 & 0 & \frac{1}{10}(0^6,2^1,8^1,3^5,5^3) & \frac{3}{50} \\
 \hline 89 & 2 {\color{red}\text{A}}_1+\text{A}_2+\text{D}_6+\text{E}_7 & \mathbb{Z}_2 & 
\begin{array}{@{}c@{}c@{}c@{}c@{}c@{}c@{}}
 1 & 1 & 0 & 1 & 0 & 1 \\
\end{array}
 & \text{-}2,\text{-}\frac{2}{3} & 196 & 352 & \frac{1}{6}(0^6,3^2,0^1,2^3,3^4) & \frac{1}{12} \\
 \hline 90 & \text{A}_1+\text{A}_3+\text{D}_6+\text{E}_7 & \mathbb{Z}_2 & 
\begin{array}{@{}c@{}c@{}c@{}c@{}c@{}}
 1 & 2 & 1 & 1 & 1 \\
\end{array}
 & \text{-}\frac{3}{2},0 & 200 & 248 & \frac{1}{2}(0^6,1^2,0^3,1^5) & \frac{1}{8} \\
 \hline 91 & {\color{red}\text{D}}_4+\text{D}_6+\text{E}_7 & \mathbb{Z}_2 & 
\begin{array}{@{}c@{}c@{}c@{}c@{}c@{}}
 1 & 1 & 0 & 1 & 1 \\
\end{array}
 & \text{-}2,\text{-}1 & 210 & 192 & \frac{1}{4}(0^6,1^1,3^1,0^3,2^5) & \frac{1}{16} \\
 \hline 92 & \text{A}_1+\text{A}_3+{\color{red}\text{D}}_6+\text{E}_7 & \mathbb{Z}_2 & 
\begin{array}{@{}c@{}c@{}c@{}c@{}c@{}}
 1 & 2 & 1 & 1 & 1 \\
\end{array}
 & \text{-}2,\text{-}\frac{3}{2},0 & 200 & 128 & \frac{1}{6}(0^6,1^1,5^1,2^6,3^2) & \frac{1}{18} \\
 \hline 93 & \text{A}_1+2 {\color{red}\text{A}}_1+\text{D}_7+\text{E}_7 & \mathbb{Z}_2 & 
\begin{array}{@{}c@{}c@{}c@{}c@{}c@{}}
 1 & 1 & 1 & 2 & 1 \\
\end{array}
 & \text{-}2,0 & 216 & 336 & \frac{1}{2}(0^7,1^1,0^2,1^6) & \frac{1}{8} \\
 \hline 94 & 2 \text{A}_1+\text{D}_8+\text{E}_7 & \mathbb{Z}_2 & 
\begin{array}{@{}c@{}c@{}c@{}c@{}c@{}}
 1 & 0 & 1 & 0 & 1 \\
\end{array}
 & \text{-}1 & 242 & 304 & \frac{1}{2}(0^{10},1^6) & \frac{1}{4} \\
 \hline 95 & \text{A}_2+{\color{red}\text{D}}_8+\text{E}_7 & 1 & 
\begin{array}{@{}c}
 \text{} \\
\end{array}
 & \text{-}2,\text{-}\frac{2}{3} & 244 & 256 & \frac{1}{8}(0^6,1^1,7^1,3^8) & \frac{3}{64} \\
 \hline 96 & {\color{red}\text{D}}_{10}+\text{E}_7 & 1 & 
\begin{array}{@{}c}
 \text{} \\
\end{array}
 & \text{-}2,\text{-}1 & 306 & 0 & \frac{1}{6}(0^6,1^1,5^1,2^7,4^1) & \frac{1}{36} \\
 \hline 97 & \text{A}_1+2 {\color{red}\text{A}}_1+2 \text{E}_7 & \mathbb{Z}_2 & 
\begin{array}{@{}c@{}c@{}c@{}c@{}c@{}}
 0 & 1 & 1 & 1 & 1 \\
\end{array}
 & \text{-}2,\text{-}1 & 258 & 448 & \frac{1}{4}(0^6,2^2,0^1,1^2,2^5) & \frac{1}{16} \\
 \hline 98 & \text{A}_1+\text{A}_2+2 \text{E}_7 & 1 & 
\begin{array}{@{}c}
 \text{} \\
\end{array}
 & \text{-}\frac{5}{3} & 260 & 224 & \frac{1}{6}(0^6,3^2,1^3,3^5) & \frac{1}{12} \\
 \hline 99 & {\color{red}\text{A}}_3+2 \text{E}_7 + \text{A}_1^{(\text{R})} & \mathbb{Z}_2 & 
\begin{array}{@{}c@{}c@{}c@{}}
 2 & 1 & 1 \\
\end{array}
 & \text{-}2,0 & 264 & 0 & \frac{1}{6}(0^6,2^1,4^1,1^3,3^5) & \frac{1}{18} \\
 \hline 100 & {\color{red}\text{D}}_4+\text{D}_5+\text{E}_8 & 1 & 
\begin{array}{@{}c}
 \text{} \\
\end{array}
 & \text{-}2,\text{-}\frac{3}{2},0 & 304 & 160 & \frac{1}{4}(0^5,1^2,3^1,0^3,2^5) & \frac{1}{32} \\
 \hline
 \end{tabular}
 \end{center}\end{table}
\begin{table}[H] 
\vspace{-0.6in}
\begin{center}\footnotesize 
\renewcommand{\arraystretch}{0.9}
		\begin{tabular}
		{|@{\hskip 0.2cm}
            >{$}c<{$}@{\hskip 0.2cm}|
			@{\hskip 1.6cm}>{$}r<{$}@{\hskip 0.2cm}|
			@{\hskip 0.2cm}>{$}c<{$}@{\hskip 0.1cm}|
			@{\hskip 0.8cm}>{\begin{scriptsize}$}c<{$\end{scriptsize}}@{\hskip 0.1cm}|
            >{\begin{scriptsize}$}r<{$\end{scriptsize}}
                @{\hskip 0.1cm}
                |
                @{\hskip 0.2cm}
                >{\begin{scriptsize}$}c<{$\end{scriptsize}}
                @{\hskip 0.2cm}
                |
                @{\hskip 0.2cm}
                >{\begin{scriptsize}$}c<{$\end{scriptsize}}
                @{\hskip 0.2cm}
                |
				@{\hskip 1.1cm}
                >{\begin{scriptsize}$}r<{$\end{scriptsize}}
                @{\hskip 0.2cm}
                |
                @{\hskip 0.1cm}
                >{\begin{scriptsize}$}c<{$\end{scriptsize}}
                @{\hskip 0.1cm}
                |} 
			\hline
			\# & \text{L} & \text{H} & \hspace{-0.5cm}\text{k} & m^2 & N_b & N_f & \text{Wilson line} & R^2 \\
			\hline	
 \hline 101 & \text{A}_4+{\color{red}\text{D}}_5+\text{E}_8 & 1 & 
\begin{array}{@{}c}
 \text{} \\
\end{array}
 & \text{-}2,\text{-}\frac{8}{5},\text{-}\frac{2}{5} & 300 & 0 & \frac{1}{14}(0^7,10^1,5^5,7^3) & \frac{5}{98} \\
 \hline 102 & \text{A}_1+\text{A}_2+{\color{red}\text{D}}_6+\text{E}_8 & 1 & 
\begin{array}{@{}c}
 \text{} \\
\end{array}
 & \text{-}2,\text{-}\frac{5}{3},\text{-}\frac{2}{3} & 308 & 128 & \frac{1}{8}(0^7,6^1,3^6,4^2) & \frac{3}{64} \\
 \hline 103 & \text{A}_1+{\color{red}\text{D}}_8+\text{E}_8 & 1 & 
\begin{array}{@{}c}
 \text{} \\
\end{array}
 & \text{-}2,\text{-}1 & 354 & 256 & \frac{2}{5}(0^7,2^1,1^8) & \frac{1}{25} \\
 \hline 104 & {\color{red}\text{D}}_9+\text{E}_8 + \text{A}_1^{(\text{R})} & 1 & 
\begin{array}{@{}c}
 \text{} \\
\end{array}
 & \text{-}2,0 & 384 & 0 & \frac{1}{8}(0^7,6^1,3^7,5^1) & \frac{1}{32} \\
 \hline 105 & {\color{red}\text{A}}_3+\text{E}_6+\text{E}_8 & 1 & 
\begin{array}{@{}c}
 \text{} \\
\end{array}
 & \text{-}2,\text{-}\frac{4}{3} & 324 & 0 & \frac{1}{10}(0^7,6^1,3^3,5^5) & \frac{3}{50} \\
 \hline 106 & 2 {\color{red}\text{A}}_1+\text{E}_7+\text{E}_8 & 1 & 
\begin{array}{@{}c}
 \text{} \\
\end{array}
 & \text{-}2,\text{-}1 & 370 & 224 & \frac{1}{4}(0^7,2^1,1^2,2^6) & \frac{1}{16} \\
 \hline 107 & {\color{red}\text{D}}_1+2 \text{E}_8 + \text{A}_1^{(\text{R})} & 1 & 
\begin{array}{@{}c}
 \text{} \\
\end{array}
 & \text{-}2,0 & 482 & 0 & \frac{1}{6}(0^7,2^1,1^1,3^7) & \frac{1}{18} \\
\hline\end{tabular}
		\caption{Maximal enhancements in the $\os$ heterotic string compactified on a circle and the moduli where they appear. $L$ is the gauge algebra, $H$ the fundamental group and $k$ its generators. Under $m^2$, we list the non-positive squared masses appearing as states in $\Gamma_0$, which are associated to either tachyons or scalars leading to instabilities. $N_b=N_v$ is the number of massless gauge bosons, while $N_f=N_s+N_c$ is the number of massless fermions. The color red indicates that there are tachyons in the vector representation of the associated $\mathfrak{so}_{2n}$ gauge factor.
        }
	\label{tab:9Dlist}
 \end{center}\end{table}

\section{Further technical details}
\label{appendix}

\subsection{\texorpdfstring{$SO(8)$}{SO(8)} characters}
The $SO(8)$ characters are written in terms of theta functions with characteristics, defined by
\begin{equation}\vartheta \left[
\begin{array}{c}
 a \\
 b \\
\end{array}
\right](0,\tau )=\overset{\infty }{\sum _{n=-\infty } }e^{2 \pi  b i (a+n)+\pi  i \tau  (a+n)^2} \ .
\end{equation}
The $SO(8)$ characters are given by
\begin{equation} \label{characters}
\begin{split}
    {\bar O}_8 &= \frac{1}{2 \bar{\eta }^4}\left(\bar{\vartheta }^4\left[
\begin{array}{c}
 0 \\
 0 \\
\end{array}
\right](0,\tau )+\bar{\vartheta }^4\left[
\begin{array}{c}
 0 \\
 \frac{1}{2} \\
\end{array}
\right](0,\tau )\right)  \\
& =  \frac{1}{2 \bar{\eta }^4}\left(\left(\overset{\infty }{\sum _{n=-\infty } }\bar{q}^{\frac{n^2}{2}}\right)^4+\left(\overset{\infty }{\sum _{n=-\infty } }(-1)^n \bar{q}^{\frac{n^2}{2}}\right)^4\right) \, ,\\
{\bar V}_8 &= \frac{1}{2 \bar{\eta }^4}\left(\bar{\vartheta }^4\left[
\begin{array}{c}
 0 \\
 0 \\
\end{array}
\right](0,\tau )-\bar{\vartheta }^4\left[
\begin{array}{c}
 0 \\
 \frac{1}{2} \\
\end{array}
\right](0,\tau )\right)
= \frac{1}{2 \bar{\eta }^4} \bar{\vartheta }^4\left[
\begin{array}{c}
 \frac{1}{2} \\
 0 \\
\end{array}
\right](0,\tau ) \\
&= \frac{1}{2 \bar{\eta }^4} \left(\overset{\infty }{\sum _{n=-\infty } } \bar{q}^{\frac12\left(n + \frac12\right)^2}\right)^4 \, ,\\
{\bar S}_8 &= \frac{1}{2 \bar{\eta }^4}\left(\bar{\vartheta }^4\left[
\begin{array}{c}
 \frac{1}{2} \\
 0 \\
\end{array}
\right](0,\tau )+\bar{\vartheta }^4\left[
\begin{array}{c}
 \frac{1}{2} \\
 \frac{1}{2} \\
\end{array}
\right](0,\tau )\right) 
= \frac{1}{2 \bar{\eta }^4} \bar{\vartheta }^4\left[
\begin{array}{c}
 \frac{1}{2} \\
 0 \\
\end{array}
\right](0,\tau ) \\
&= \frac{1}{2 \bar{\eta }^4} \left(\overset{\infty }{\sum _{n=-\infty } } \bar{q}^{\frac12\left(n + \frac12\right)^2}\right)^4 \, , \nn 
\end{split}\end{equation}
\begin{equation}\begin{split}
   {\bar C}_8 &= \frac{1}{2 \bar{\eta }^4}\left(\bar{\vartheta }^4\left[
\begin{array}{c}
 \frac{1}{2} \\
 0 \\
\end{array}
\right](0,\tau )-\bar{\vartheta }^4\left[
\begin{array}{c}
 \frac{1}{2} \\
 \frac{1}{2} \\
\end{array}
\right](0,\tau )\right)
= \frac{1}{2 \bar{\eta }^4} \bar{\vartheta }^4\left[
\begin{array}{c}
 \frac{1}{2} \\
 0 \\
\end{array}
\right](0,\tau ) \\
&= \frac{1}{2 \bar{\eta }^4} \left(\overset{\infty }{\sum _{n=-\infty } } \bar{q}^{\frac12\left(n + \frac12\right)^2}\right)^4 \, . \\
\end{split}
\end{equation}
The $q$-expansion of the theta series for the lattices $\Gamma_{v,s,c,0}$ defined in \eqref{conjdef} is given by,
\beq \label{Thetas}
\begin{split}
\theta_{v} &= 1+224 q+31200 q^2+522880 q^3+ O(q^4) \, ,\\
\theta_{s} = \theta_{c} &= 256 q+30720 q^2+527360 q^3+ O(q^4)\, ,\\
\theta_{0} &= 4096 q^{\tfrac32} + 147456 q^{\tfrac52} + O(q^{\tfrac72})\, ,
\end{split}
\eeq
while the expansion of the functions $\sigma_{0,1}$ defined in \eqref{sigma} is given by, 
\begin{align} \label{powersigma}
   \sigma_0(q,\bar q)  = & 1 + 2 \bar{q} + 4 \bar{q}^{\frac14} q^{\frac14} + 4 \bar{q}^{\frac94} q^{\frac14} + 2 q + 4 \bar{q} q + 
 4 \bar{q}^{\frac14} q^{\frac94} + 4 \bar{q}^{\frac94} q^{\frac94} \cr & + O(q^{\frac{13}{4}}) + O({\bar q}^{\frac{13}{4}}) \, , \nn \\ 
\sigma_1(q,\bar q) 
 = & 2 \bar{q} + 2 \bar{q}^{\frac14} q^{\frac14} + 10 \bar{q}^{\frac94} q^{\frac14} + 2 q + 8 \bar{q} q + 
 10 \bar{q}^{\frac14} q^{\frac94} + 18 \bar{q}^{\frac94} q^{\frac94} \\ & + O(q^{\frac{13}{4}}) + O({\bar q}^{\frac{13}{4}}) \, . \nn
\end{align}

\subsection{Light fermions in \texorpdfstring{$E_6 \times SU(12)$}{E6 x SU(12)} enhancement}
\label{App:lightfermions}

The enhancement to $E_6 \times SU(12)$ is the best counterexample to the common lore that the cosmological constant is roughly the number of massless fermions  minus the number of massless bosons. This enhancement has no massless fermions, yet the cosmological constant is positive. The reason is that there are many -- more precisely 2176 -- very light fermions ($m^2=\frac12$), half in class (s) and half in class (c) that give a contribution to the cosmological constant that is in order of magnitude about 1.5 times that of the massless states.  Below we list these spinors, which have all  $p_R^2 = \frac{1}{4}$ and $p_L^2 = \frac{9}{4}$.

There are 1088 class (s) spinors, 
\bea 
\left(-w, n, \pi_1,\dots,\pi_8, -\pi_9,\pi_{10},\dots,\pi_{16} \right)
= \begin{cases}
\pm\left(\underline{k,0},{0^3},{0^5},\underline{-\tfrac12^{3-2k},\tfrac12^{5+2k}}\right) \quad & \rightarrow \quad 144 \\
\pm\left( \underline{k,1},{0^3},\underline{0,1^{2k}},\underline{-\tfrac12,\tfrac12^7}\right) \quad & \rightarrow \quad 400 \\
\pm\left({2,2},\underline{0,0,\pm 1},1^5,\underline{\tfrac12^7,\tfrac32}\right) \quad & \rightarrow \quad 96 \\
\pm\left(\underline{k-1,0},\underline{\pm\tfrac12^3}_{\text{even}},\underline{-\tfrac12^{4-2k}},\tfrac12^{1+2k},{0^8}\right) \quad & \rightarrow \quad 176 \\
\pm\left({1,1},\underline{\pm\tfrac12^3}_{\text{even}},{\tfrac12^5},\underline{0^6,1^2}\right) \quad & \rightarrow \quad 224 \\
\pm\left((1+2k)^2,\underline{\pm\tfrac12^3}_{\text{odd}},\underline{\tfrac12^{4-4k},\tfrac32^{1+4k}},{k^8}\right) \quad & \rightarrow \quad 48 \\
\end{cases}\nonumber
\eea 
and 1088 class (c) spinors:
\bea 
\left(-w, n, \pi_1,\dots,\pi_8, -\pi_9,\pi_{10},\dots,\pi_{16} \right)
= \begin{cases}
\pm\left(0,0,\underline{0^7,-1},\underline{0^7,1}\right) \quad & \rightarrow \quad 128 \\
\pm\left(2k,2k,\underline{0^{2+k},1^{1-k}},k^5,\underline{0^{7-4k},1^{1+4k}}\right) \quad & \rightarrow \quad 160 \\
\pm\left(\underline{k,1},0^3,\underline{0^{4-2k},1^{1+2k}},\underline{0^7,1}\right) \quad & \rightarrow \quad 352 \\
\pm \left(\underline{k+1 ,2},(\pm\tfrac12)^3_{\text{even}},\underline{\tfrac12^{4-2k},\tfrac32^{2k+1}},\tfrac12^8\right) \quad & \rightarrow \quad 176 \\
\pm \left({1,1},(\pm\tfrac12)^3_{\text{even}},\underline{-\tfrac12,\tfrac12^4},
\tfrac12^8\right) \quad & \rightarrow \quad 40 \\
\pm \left({1,1},(\pm\tfrac12)^3_{\text{odd}},\tfrac12^5,
\underline{-\tfrac12^2,\tfrac12^7}\right) \quad & \rightarrow \quad 224 \\
\pm \left({1,1},(\pm\tfrac12)^3_{\text{odd}},\tfrac12^5,
\tfrac12^8\right) \quad & \rightarrow \quad 8 \nonumber
\end{cases}
\eea 
where the underline means any permutation of the entries. On the right side we count the total number of states for each $Z$.

Each state has a contribution to $z_s$ or $z_c$ of $q^{\tfrac18}\bar{q}^{\tfrac18} = 
e^{-\frac{\pi \tau_2}{2}}$.
After integrating over the fundamental region, we get a contribution to $\Lambda$ of $0.177$ for each state. The $2176$ total fermionic states at this mass level increase $\Lambda$ by $385.2$, dominating over the massless states with negative contribution of $-243.7$. As a result we get a positive cosmological constant even though there are no massless fermions. Note that there is still a piece of the cosmological constant coming from heavier states because the contribution of the massless states plus these fermionic states equals 141.5, while the total cosmological constant at this enhancement point is 180.4.

\newpage
\bibliographystyle{JHEP}
\bibliography{refs}

\end{document}